\documentclass[galaxies,accept,submit,pdftex,moreauthors]{Definitions/mdpi}

\firstpage{1} 
\pubvolume{11}
\issuenum{1}
\articlenumber{28}
\pubyear{2023}
\copyrightyear{2023}
\externaleditor{Academic Editor: {Firstname Lastname}}
\datereceived{9 November 2022} 
\daterevised{23 January 2023} 
\dateaccepted{28 January 2023} 
\datepublished{10 February 2023} 
\hreflink{https://doi.org/10.3390/\linebreak galaxies11010028} 

\usepackage{eht}

\Title{Enabling Transformational ngEHT Science via the Inclusion of 86\,GHz Capabilities}
\TitleCitation{Enabling Transformational ngEHT Science via the Inclusion of 86\,GHz Capabilities}

\Author{%
Sara~Issaoun
 $^{1,2,}$*\orcidA{},
Dominic~W.~Pesce $^{1,3}$\orcidB{},
Freek~Roelofs $^{1,3}$\orcidC{},
Andrew~Chael $^{2,4}$\orcidD{},
Richard~Dodson $^{5}$\orcidE{},
Mar\'ia~J.~Rioja $^{5,6,7}$,
Kazunori~Akiyama $^{3,8,9}$\orcidF{},
Romy~Aran $^{1,10}$,
Lindy~Blackburn $^{1,3}$\orcidG{},
Sheperd~S.~Doeleman $^{1,3}$\orcidH{},
Vincent~L.~Fish $^{8}$\orcidM{},
Garret~Fitzpatrick $^{1}$,
Michael~D.~Johnson $^{1,3}$\orcidI{},
Gopal~Narayanan $^{11}$\orcidJ{},
\mbox{Alexander~W.~Raymond $^{1,3}$\orcidK{}} and Remo~P.~J.~Tilanus $^{12,13,14,15}$\orcidL{}} 

\AuthorCitation{Issaoun, S.; Pesce, D.W.; Roelofs, F.; Chael, A.; Dodson, R.; Rioja, M.J.; Akiyama, K.; Aran, R.; Blackburn, L.; Doeleman, S.S.; et al.}

\address{%
$^{1}$ \quad Center for Astrophysics $|$
Harvard \& Smithsonian, 60 Garden Street, Cambridge, MA 02138, USA \\
$^{2}$ \quad NASA Hubble Fellowship Program, Einstein Fellow \\
$^{3}$ \quad Black Hole Initiative at Harvard University, 20 Garden Street, Cambridge, MA 02138, USA\\
$^{4}$ \quad Princeton Center for Theoretical Science, Jadwin Hall, Princeton University, Princeton, NJ 08544, USA\\
$^{5}$ \quad International Centre for Radio Astronomy Research, M468, The University of Western Australia, 35 Stirling Hwy, Crawley, WA 6009, Australia\\
$^{6}$ \quad CSIRO Astronomy and Space Science, P.O. Box 1130, Bentley WA 6102, Australia\\
$^{7}$ \quad Observatorio Astron\'omico Nacional (IGN), Alfonso XII, 3 y 5, 28014 Madrid, Spain\\
$^{8}$ \quad Massachusetts Institute of Technology, Haystack Observatory, 99 Millstone Rd, Westford, MA 01886, USA \\
$^{9}$ \quad National Astronomical Observatory of Japan, 2-21-1 Osawa, Mitaka 181-8588, Tokyo, Japan\\
$^{10}$\quad Harvard College, Harvard University, Cambridge, MA 02138, USA\\ 
$^{11}$\quad Department of Astronomy, University of Massachusetts, Amherst, MA 01003, USA\\
$^{12}$\quad Steward Observatory and Department of Astronomy, University of Arizona, 933 N. Cherry Ave., \mbox{Tucson, AZ 85721, USA}\\
$^{13}$\quad Department of Astrophysics, Institute for Mathematics, Astrophysics and Particle Physics (IMAPP), Radboud University, P.O. Box 9010, 6500 GL Nijmegen, The Netherlands\\
$^{14}$\quad Leiden Observatory, Leiden University, Postbus 2300, 9513 RA Leiden, The Netherlands\\
$^{15}$\quad Netherlands Organisation for Scientific Research (NWO), Postbus 93138, \mbox{2509 AC Den Haag, The Netherlands}
}

\corres{Correspondence: sara.issaoun@cfa.harvard.edu}

\abstract{%
We present a case for significantly enhancing the utility and efficiency of the ngEHT by incorporating an additional 86\,GHz observing band.  In contrast to 230 or 345\,GHz, weather conditions at the ngEHT sites are reliably good enough for 86\,GHz to enable year-round observations.  Multi-frequency imaging that incorporates 86\,GHz observations would sufficiently augment the $(u,v)$ coverage at 230 and 345\,GHz to permit detection of the M87 jet structure without requiring EHT stations to join the array.  The general calibration and sensitivity of the ngEHT would also be enhanced by leveraging frequency phase transfer techniques, whereby simultaneous observations at 86\,GHz and higher-frequency bands have the potential to increase the effective coherence times from a few seconds to tens of minutes. When observation at the higher frequencies is not possible, there are opportunities for standalone 86\,GHz science, such as studies of black hole jets and spectral lines. Finally, the addition of 86\,GHz capabilities to the ngEHT would enable it to integrate into a community of other VLBI facilities---such as the GMVA and ngVLA---that are expected to operate at 86\,GHz but not at the higher ngEHT observing frequencies.
}

\begin{document}

\nocite{M87PaperI}
\nocite{M87PaperII}
\nocite{M87PaperIII}
\nocite{M87PaperIV}
\nocite{M87PaperV}
\nocite{M87PaperVI}
\nocite{M87PaperVII}
\nocite{M87PaperVIII}

\section{Introduction}
Building upon the success of the Event Horizon Telescope (EHT; \cite{M87PaperI,M87PaperII,M87PaperIII,M87PaperIV,M87PaperV,M87PaperVI,M87PaperVII,M87PaperVIII,SgrAPaperI,SgrAPaperII,SgrAPaperIII,SgrAPaperIV,SgrAPaperV,SgrAPaperVI}), the next-generation EHT (ngEHT) is a proposed global very long baseline interferometry (VLBI) telescope network that aims to carry out horizon-scale observations of M87 and Sgr A* at (sub)millimeter wavelengths \citep{Doeleman_2019}.  By adding $\sim$10 new VLBI stations to the EHT array and increasing the overall array sensitivity, the ngEHT will be able to achieve high-fidelity imaging and even movie-making capabilities. The primary scientific goals of the ngEHT require an angular resolution of $\lesssim$20\,$\upmu$as
and thus motivates observing at the highest VLBI frequencies, currently 230 and 345\,GHz (see Bustamante et al. \cite{Bustamante_2023} for dual-band receiver details).  However, the design specifications for the ngEHT have yet to be finalized, and the addition of an 86\,GHz band is under consideration.  Adding 86\,GHz capabilities to the ngEHT would provide numerous prospects for improving the performance of the array, expanding its science applications and permitting it to operate jointly with other existing or near-future facilities.  In this work, we explore a number of these motivating factors and present a case for including 86\,GHz capabilities as part of the ngEHT array's design.

\section{Science Drivers}\label{sec:drivers}

\subsection{Black Hole Shadow and Jet Physics}

\begin{figure}[H]
        \begin{tabular}{lll}
             & \hspace{0.13\linewidth} ngEHTa & \hspace{0.11\linewidth} ngEHTa + EHT \\
            \raisebox{0.15\linewidth}[0pt][0pt]{\rotatebox{90}{230 GHz}} &\includegraphics[width=0.35\linewidth]{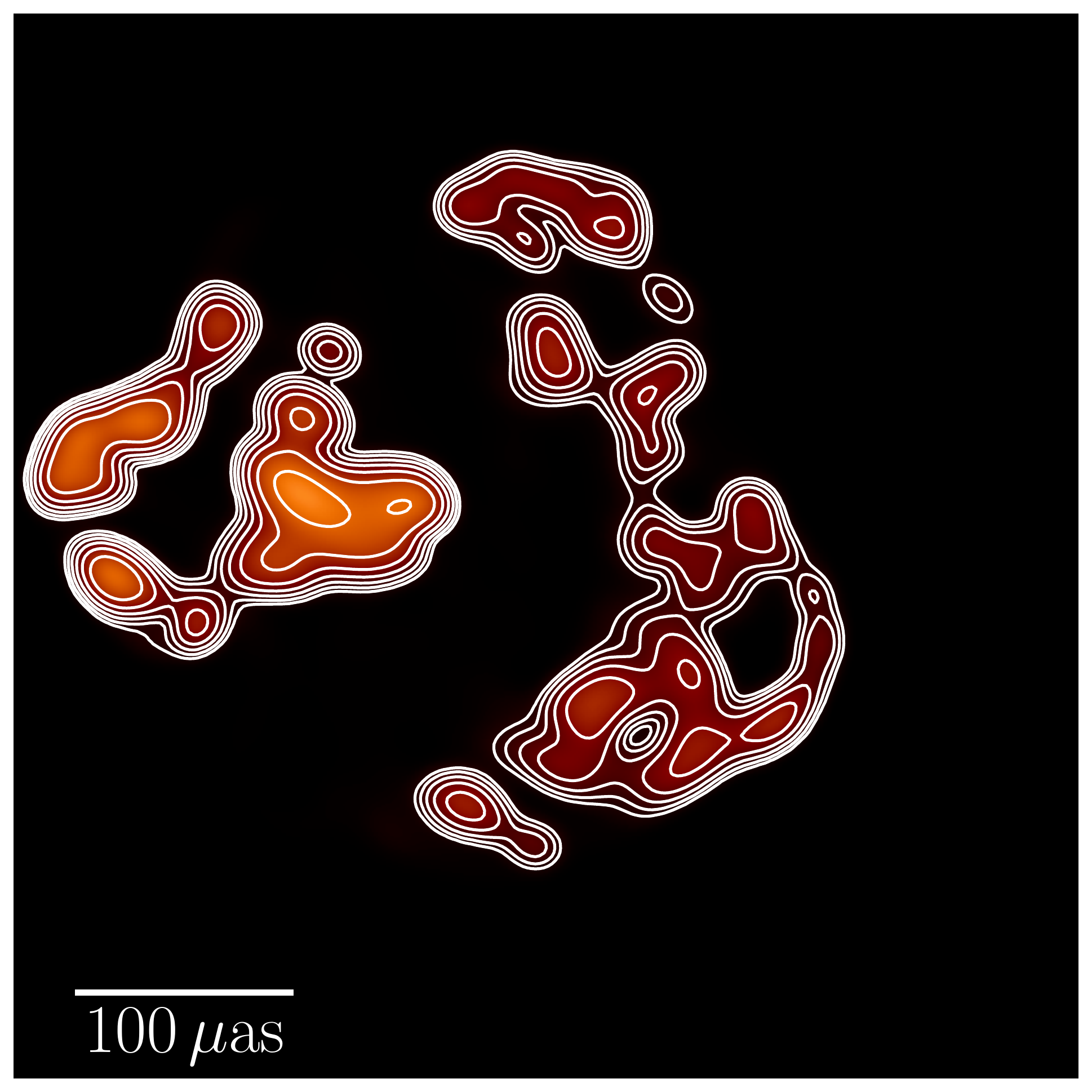} & \hspace{-0.16cm} \includegraphics[width=0.35\linewidth]{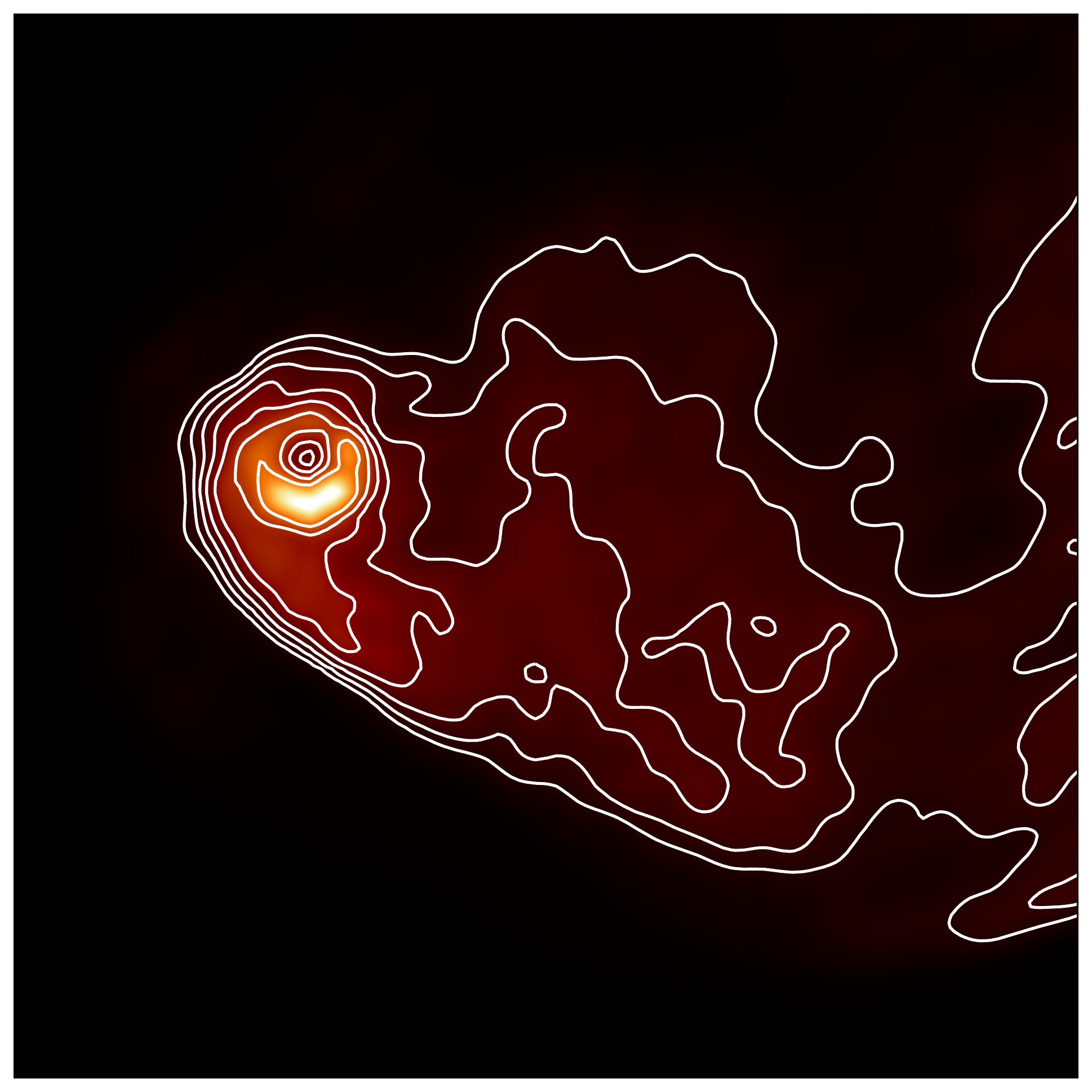} \\
            \raisebox{0.13\linewidth}[0pt][0pt]{\rotatebox{90}{86+230 GHz}} & \includegraphics[width=0.35\linewidth]{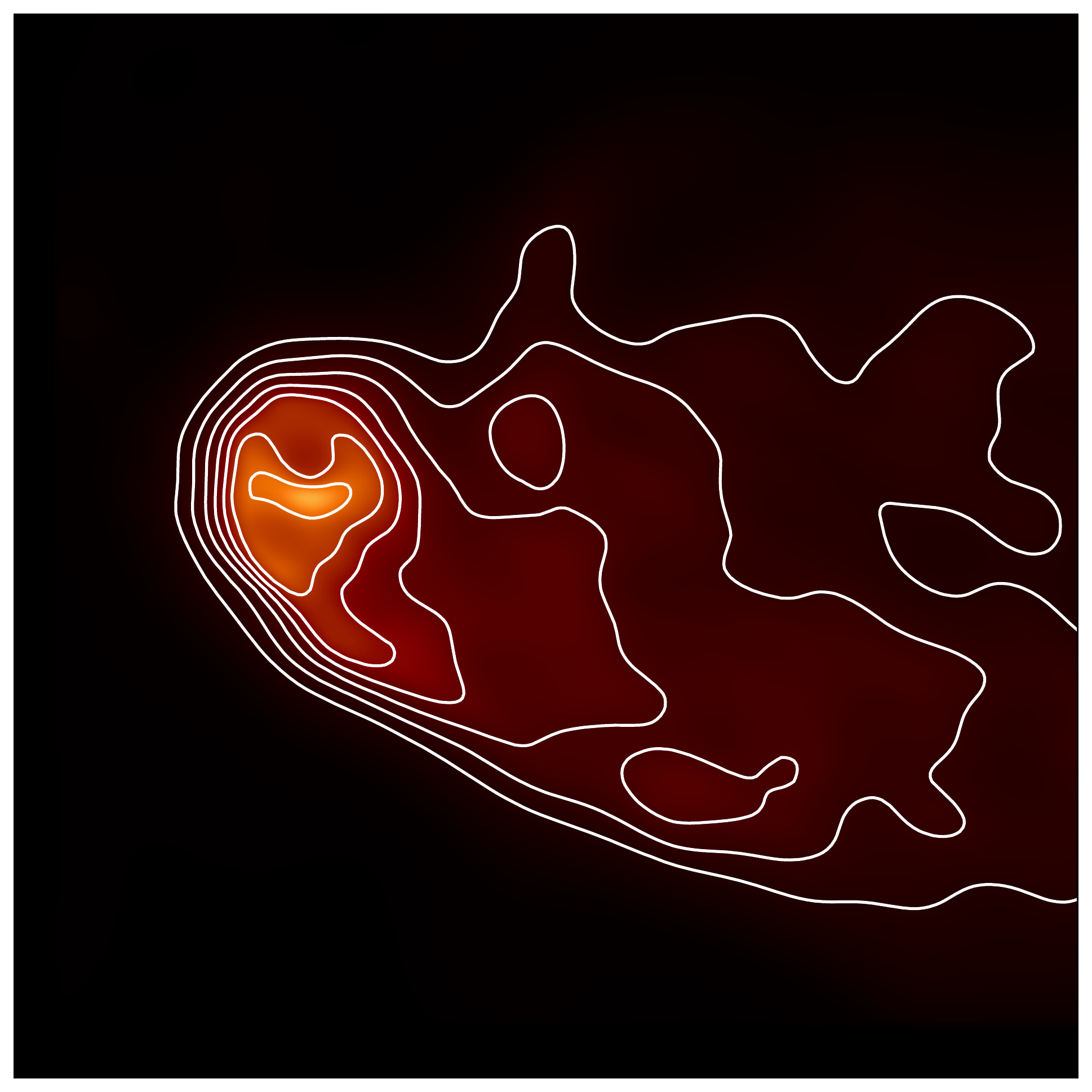} & \hspace{-0.16cm} 
    \includegraphics[width=0.35\linewidth]{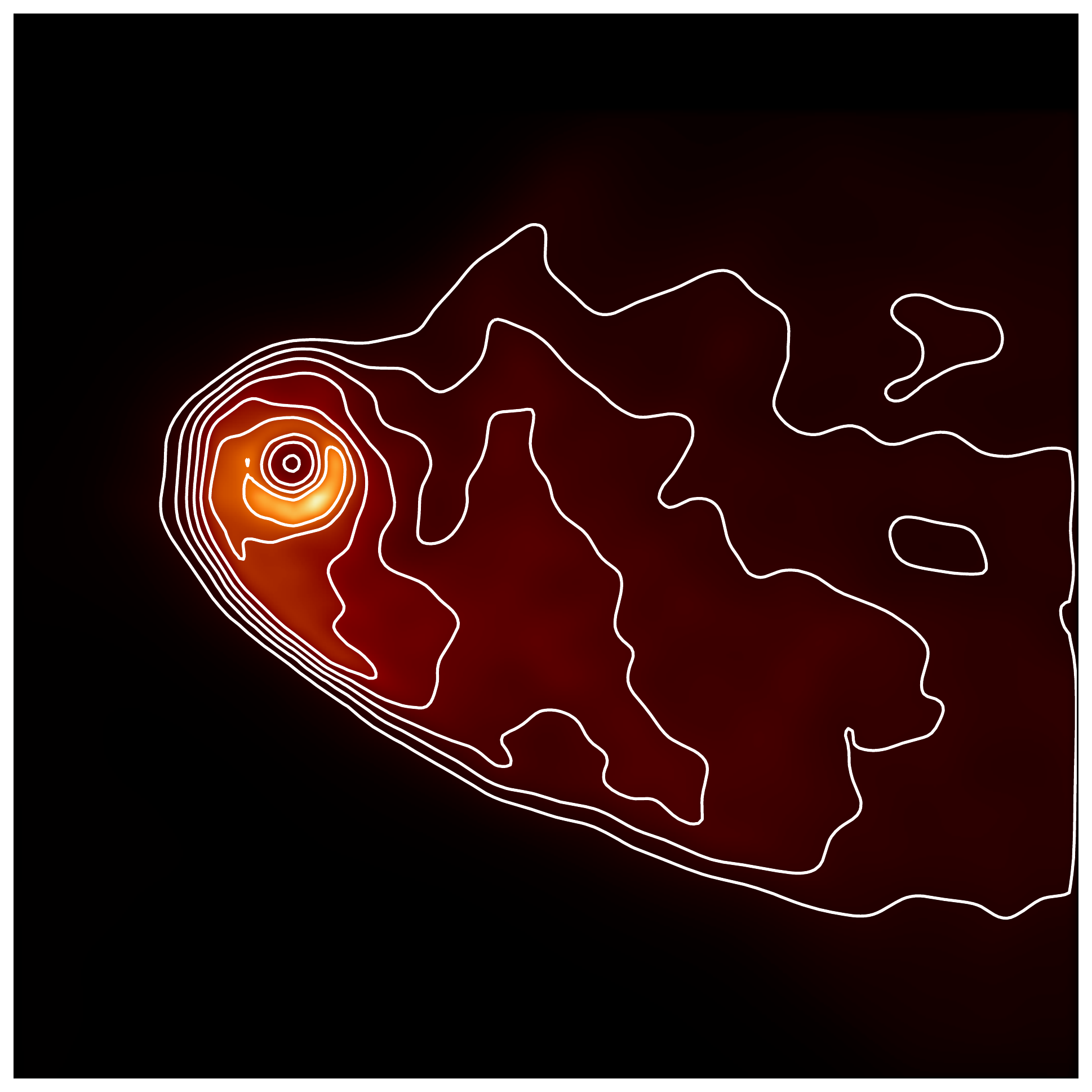} \\
\raisebox{0.1\linewidth}[0pt][0pt]{\rotatebox{90}{86+230+345 GHz}} & \includegraphics[width=0.35\linewidth]{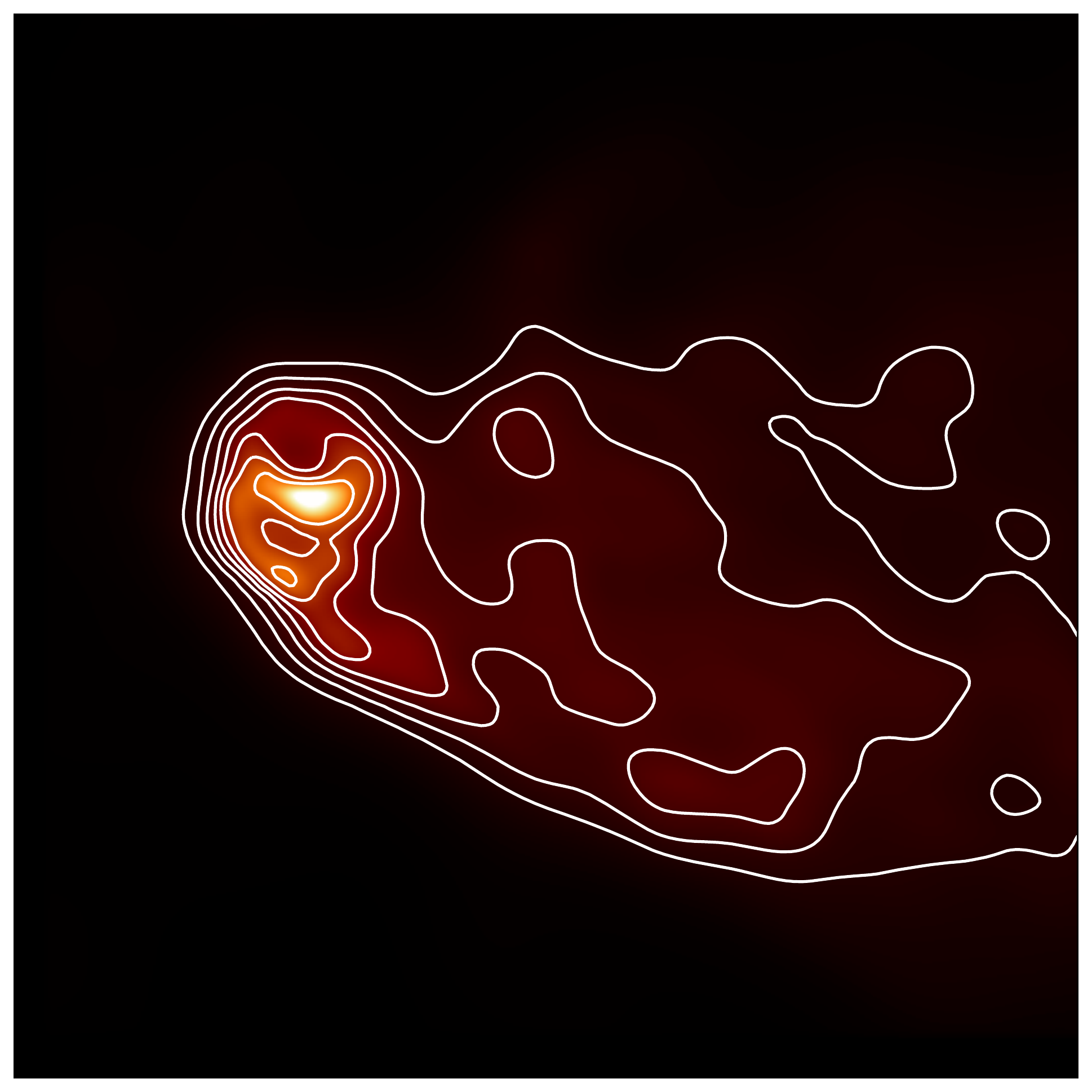} & \hspace{-0.16cm} 
    \raisebox{-0.015\linewidth}[0pt][0pt]{\includegraphics[width=0.42\linewidth]{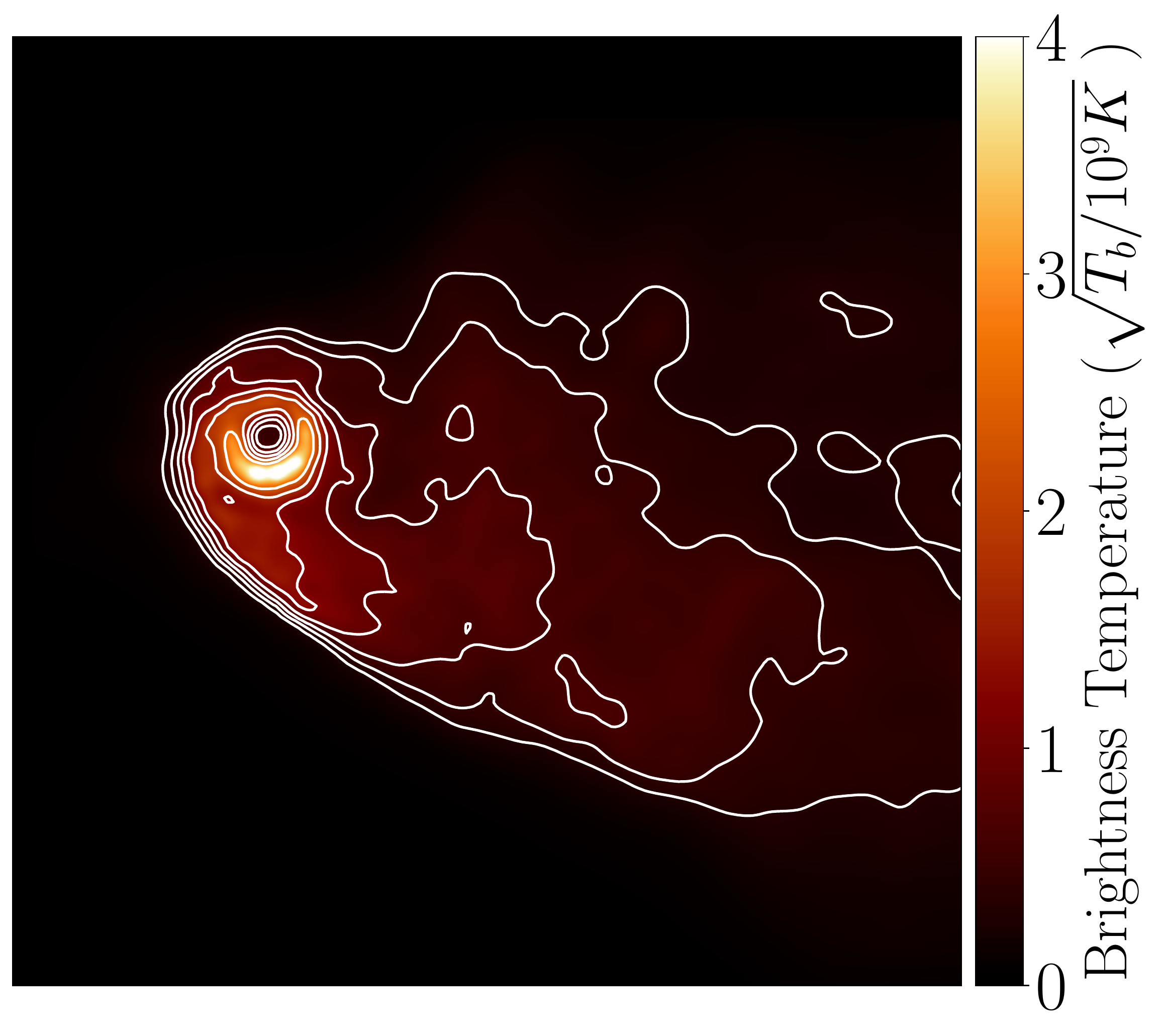}}
    \end{tabular}
    \caption{Demonstration of multi-frequency capabilities with simulated observations of the M87 jet at 230\,GHz reconstructed with and without 86\,GHz complementary observations. The left column shows reconstructions using the new ngEHTa configuration stations alone (see Section \ref{sec:Array} for array specifications), and the right column shows reconstructions with the full ngEHT array, including the core EHT stations. Top: Simulated reconstructions of 230\,GHz only observations. The contours are spaced logarithmically, starting at 1\% of the peak value and increasing by factors of 2. Middle: Simulated reconstructions at 230\,GHz using  multi-frequency 86 and 230\,GHz observations. Bottom: Simulated reconstructions at 230\,GHz using multi-frequency 86, 230 and 345\,GHz observations. This demonstration offers a compelling view of the imaging advantages of adding 86\,GHz receivers to the ngEHT, enabling flexible high-fidelity imaging of the black hole shadow and jet during times when the core EHT sites would not be readily available.}
    \label{fig:ngEHTdemo}
\end{figure}

\begin{figure}[H]
\begin{center}
\includegraphics[width=\textwidth]{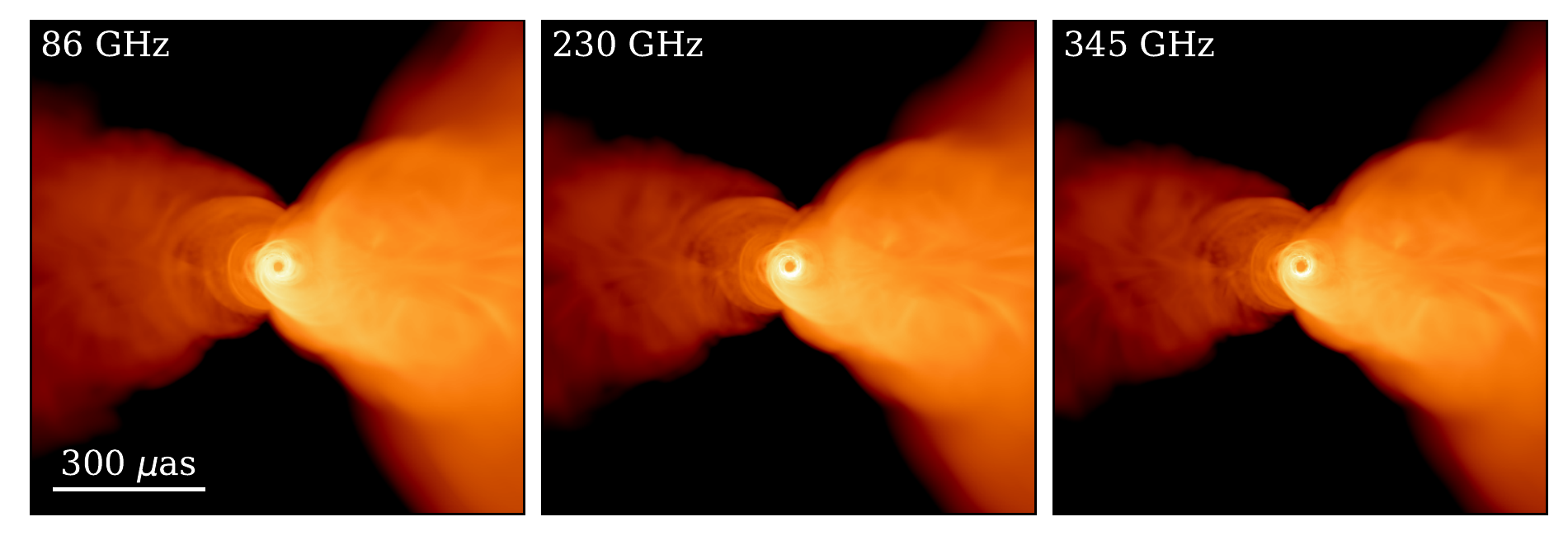}
\includegraphics[width=\textwidth]{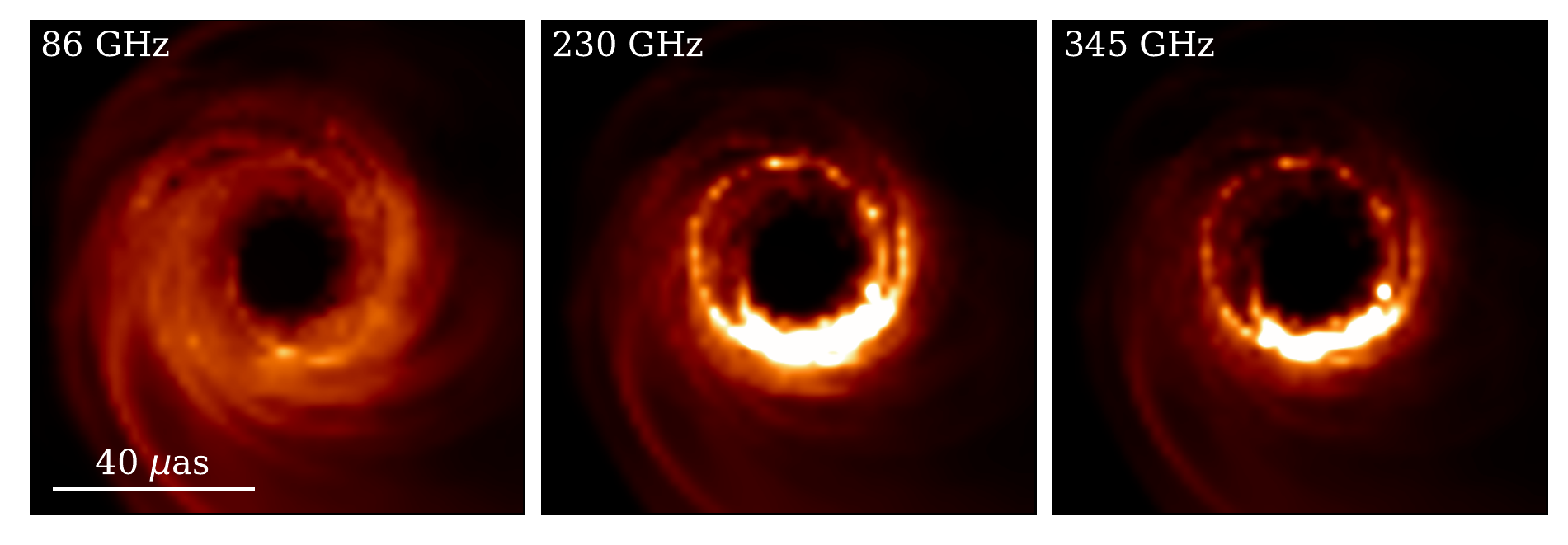}
\end{center}
\caption{Input images used for the M87 synthetic data in this work, produced from the GRMHD simulations in
 \cite{Chael_2019}.  The upper panels show a 1\,mas field of view with a logarithmic colorscale to highlight the extended jet emission, and the lower panels show the central 100\,$\upmu$as around the black hole using a linear color scale to more easily see the photon ring region.}\label{fig:m87model}
\end{figure}

Observations of horizon-scale targets at 86\,GHz supplement a primary science driver of the ngEHT: connecting dynamics and properties of black hole shadows with the creation and launching of astrophysical jets.
The horizon-scale jet emission is typically brighter at 86\,GHz than at 230 or 345\,GHz, owing to the negative spectral index and increased optical depth at 86\,GHz.
The inference of jet structure from 86\,GHz observations to 230 and 345\,GHz images, with the use of multi-frequency imaging techniques \citep{Chael_2022}, will enable the recovery of faint large-scale jet emissions in horizon-scale images with higher fidelity compared to high-frequency imaging alone. In Figure \ref{fig:ngEHTdemo}, we show example reconstructions of simulated emission from M87 at 230\,GHz with and without multi-frequency information from 86\,GHz (and 345\,GHz) observations. Imaging with the full EHT+ngEHT array was done using visibility amplitudes and closure phases, and ngEHT-only reconstructions also made use of closure amplitudes to improve convergence. The ground-truth models are shown in Figure \ref{fig:m87model}. These reconstructions assume that the telescopes are well pointed, well focused, and amplitude-calibrated, another avenue in which 86\,GHz capabilities can improve the overall array performance. Relative registration of images across frequencies was performed via the algorithm, although additional observing techniques can provide that information more robustly (see Sections \ref{sec:astrometry} and \ref{sec:phase_trans}). While the best imaging results are obtained using the full core EHT and new ngEHT stations with multi-frequency imaging, the addition of 86\,GHz information dramatically improves 230\,GHz images with the new ngEHT stations alone (or in conjunction with a single sensitive core-EHT site, such as the LMT or ALMA). The inclusion of 86\,GHz observing guarantees good coverage and station operation the entire year; see Section \ref{sec:logistics}. This would, for example, enable images and movies of the shadow and jet in M87 with high cadence over long periods of time, see Figure \ref{fig:ngEHTdemo}.

The 128\,Gbps recording rate of the ngEHT brings about a factor of 32 increase compared to the current 86\,GHz VLBI recording rate of 4\,Gbps (currently limited by the recording capability of the Very Long Baseline Array). The increase in sensitivity brought by the ngEHT at 86\,GHz would greatly improve polarimetric imaging of jet structure and inner accretion flows, particularly for low-polarization sources such as M87 and Sgr A*. Polarimetric imaging is essential for the discrimination between magnetic field configurations~\citep{M87PaperVII,M87PaperVIII}, thereby tightening our constraints on black hole astrophysical models. 

For Sgr~A*, the ngEHT will be sufficiently sensitive for detecting refractive scattering in the interstellar medium toward the Galactic Center, for which the effects are most dominant on long VLBI baselines. Detections of refractive scattering in the millimeter regime were detected on north--south baselines, where the diffractive scattering is weakest \citep{Issaoun_2019,Issaoun_2021}, and detections in the centimeter regime were detected on east--west baselines \citep{Gwinn_2014}. While these detections allowed us to eliminate the more extreme models of magnetic field wander (i.e., variations transverse to the line of sight across a range of viewing angles), further constraints of the model parameter space require a wider coverage of position angles at many refractive timescales. A higher baseline sensitivity and increased observing cadence at 86\,GHz would provide detections of refractive scattering along baselines in all directions, which could definitively discriminate between different models for the underlying magnetic field wander in the interstellar medium that predicts varying levels of refractive noise along different baseline directions \citep{Psaltis_2018}. 

Joint modeling of the screen across the three simultaneous observing bands will also improve the scattering mitigation of Sgr A* observations.

\vspace{-4pt}

Furthermore, the addition of simultaneous 86\,GHz observing in the 230 and 345\,GHz configuration will provide wide frequency coverage for rotation measure and Faraday rotation studies, core shift studies, and time-lag measurements in the event of flares. Due to the variability of Sgr A*, simultaneity of 86\,GHz observations is essential for relative astrometry between the frequency bands, and the registration of the three bands for accurate spectral index and rotation measure mapping. 

\subsection{Telescope Calibration and Astrometry}\label{sec:astrometry}

At high frequencies, single-frequency phase referencing using nearby calibrators is challenging due to the varying atmospheric conditions between the calibrator and the main target. Even at 86\,GHz, atmospheric coherence limits the switching time between the target and the calibrator reference to 15 s or less. While a single demonstration at 86\,GHz exists \citep{Porcas_2002}, single-frequency phase referencing at high frequencies is, in general, only conceivable using simultaneous observations in multiple directions, either with multiple beams or more likely paired antennas; ngVLA would have this capability by having multiple antennas per site location. Single-frequency astrometry at 230 and 345\,GHz is thus difficult without a proper phase reference position on the sky.

Frequency phase transfer (FPT) is a powerful tool for 230\,GHz calibration in marginal weather conditions. The FPT approach was initially developed for (sub)mm compact arrays (e.g., \cite[]{Asaki_1996}) and first used in VLBI by Middelberg et al. \cite{Middelberg_2005} to extend the coherence time at 86\,GHz, using the VLBA. The work of Rioja and Dodson enabled bona-fide astrometry (source frequency phase referencing; SFPR) in addition (\citet{Dodson_2009} and \citet{Rioja_2011} using the VLBA up to 86\,GHz) and has been demonstrated up to 130\,GHz using the Korean VLBI Network (KVN \cite[]{Han_2013,Rioja_2015}). With FPT techniques applied to simultaneous KVN observations at 22, 43, 87, and 130\,GHz, coherence times at 130\,GHz were increased from tens of seconds to $\sim$20 min. Adding SFPR using a separate reference source, the coherence time was further increased to many hours (20\% loss after eight hours of integration). FPT has been successfully applied to other KVN observations as well. Application to MOnitoring of GAmma-ray Bright AGN (iMOGABA) observations led to the imaging of several sources at 86 and 129\,GHz that were not detected without FPT~\citep{Algaba2015}. Zhao et al. \cite{Zhao2019} applied the technique to simultaneous 22 and 43\,GHz observations with the KVN and VERA in Japan (combined as KaVA), increasing the coherence time at 43\,GHz from $\sim$1 min to tens of minutes. As demonstrated by Zhao et al. \cite{Zhao_2018}, a second round of FPT applied to the residuals from the first round (FPT-square), was successful at taking out ionospheric effects without the need to observe a reference source as in the SFPR technique, increasing coherence times from 20 min to more than eight hours at 86\,GHz.

As shown in Section \ref{sec:Array}, 86\,GHz observing is possible throughout the entire year at all sites. On observing days with marginal or poor 230\,GHz weather, the phase stability at the lower frequency of 86\,GHz would allow us to solve for phase offsets on a particular station at the lower frequency and transfer them to a higher frequency (230 and/or 345\,GHz). This technique requires at minimum simultaneous 86 and 230\,GHz observing to be effective, due to the short coherent time at the higher frequencies. Another requirement is that sources at 86\,GHz be bright and compact enough to be detectable on all baselines, which is not a limiting factor for most sources apart from Sgr~A*, which is scatter-broadened at this frequency \citep{Lee_2008,Hada_2011,Hada_2016,Kim_2018}. Finally, an integer frequency ratio is preferred between the different frequency bands in order to optimally use these techniques and should allow bonafide astrometry between the bands (i.e., between 86 and 230 and/or 345\,GHz). Astrometry at these frequencies is expected to yield residual systematic errors of $\sim$3\,$\upmu$as \citep{Rioja_2011,Rioja_2020,Rioja_2023,Jiang_2023}, though we note that such analyses will additionally need to account for frequency-dependent structural changes such as the well-known core-shift effect (e.g., \cite[]{Marcaide_1984}). These specifications are discussed in more detail in Section \ref{sec:phase_trans}.

With simultaneous receivers at 86\,GHz, the ngEHT will have a frequency overlap with a large network of well-located VLBI stations that enable astrometric observations. Using SFPR, it would be possible to astrometrically connect these to the highest frequencies and should be able to provide relative astrometry at 345\,GHz. With certain array configurations, this should be possible with just ngEHT observations. The frequency phase transfer technique allows for correction of the fast tropospheric variations and increases the coherence at the higher frequencies. This enables the switching time between the target and calibrator to be longer (of order $\sim$ minutes) and allows for the calibrator to be at a larger distance (up to 10$^{\circ}$--20$^{\circ}$,  as demonstrated with the KVN). This technique, called "source frequency phase referencing," allows for relative astrometry to directly register observations of the target at the three observed frequencies with respect to the calibrator. 
Explanations of the various flavors of frequency phase transfer and their applications to the ngEHT are provided in a complementary publication by Rioja et al. \cite{Rioja_2023}.

In addition to scientific input, 86\,GHz capabilities enhance technical specifications of the ngEHT array. Individual station calibration will be improved with the ability to point and focus at a lower frequency, where calibrator sources have higher flux density and the atmosphere is more stable. The calibrator sky is limited at 230 and 345\,GHz, and the catalog of sources available at 86\,GHz, both for continuum and spectral-line pointing, is significantly larger. Calibration operations at 86\,GHz would enable station participation in marginal 230\,GHz weather conditions.

\subsection{Stand-Alone 86\,GHz Science}

Over the past few decades, VLBI observations at 86\,GHz have provided high-quality images of AGN sources, spatially and temporally resolving their innermost structure in total intensity and polarization, thereby providing new insights into the origins, collimation, and dynamics of relativistic jets (e.g., \cite[]{Boccardi_2017}).

For example, recent observations with the Global Millimeter VLBI Array (GMVA) have imaged the core region of 3C84 and OJ287 \citep{Kim2019, Paraschos2021, Oh2022, Gomez2021}, measured jet collimation profiles of various AGNs (e.g., \cite[]{Boccardi2021, Casadio2021}), provided a survey of AGN core brightness temperature measurements \citep{Nair2019}, and imaged jet dynamics associated with gamma ray flares and source variability across the electromagnetic spectrum  (e.g., \cite[]{Rani2014, Rani2015, Casadio2019, Schultz2020, Traianou2020}). Multi-wavelength (mm-) VLBI is also useful for studying the dynamics of X-ray binary jets, where the jet formation occurs on much smaller timescales  (e.g., \cite[]{Tetarenko2017}).

Spectral line VLBI observations at 86\,GHz can also be used to study SiO maser emission near stars. In stars on the asymptotic giant branch (AGB), SiO maser emission probes the highly dynamical circumstellar gas regions where dust grains are formed and accelerated outwards with the gas, driving the formation of planetary nebulae  (e.g., \cite[]{Dodson_2017}, and references therein). In high-mass star formation regions, imaging SiO maser emission spots can probe the dynamics of the protostellar accretion disk and outflow, shedding more light on the star formation process (e.g., \cite[]{Matthews_2010, Issaoun_2017}).

The planned next-generation Very Large Array (ngVLA) will be able to observe at 86\,GHz with unprecedented sensitivity (\citep[]{Selina2018, mckinnon2019}; see also Section \ref{sec:interoperability}). However, as the ngVLA sites are limited to the American continent, its maximum baseline length and hence angular resolution are significantly smaller than those of the current GMVA. The ngEHT operating at 86\,GHz could provide up to $\sim$3 times longer baselines than the ngVLA while providing significantly more sensitivity than the GMVA due to its increased bandwidth. The ngEHT and ngVLA operating together would form an extremely high-sensitivity and high-resolution array at 86 GHz, providing unprecedented images of AGN jet sources (Section \ref{sec:interoperability}).

\section{Array Specifications and Performance}\label{sec:logistics}

The ngEHT will be comprised of up to 10 new (sub)millimeter radio telescopes distributed around the globe and operating as a VLBI network \citep{Doeleman_2019}.  There are two primary operating modes that the ngEHT is expected to employ.  The first operating mode is a campaign mode, during which the ngEHT dishes will observe alongside the current EHT dishes as part of a large and sensitive array.  Owing to the need to coordinate such observations among many telescopes, a number of which are themselves facility instruments, the campaign mode will likely only operate during a small number of observing windows within any given calendar year.  The second operating mode is a standalone ngEHT mode that will be more versatile and which is expected to operate throughout the year.

\subsection{Array} \label{sec:Array}

\begin{figure}[H]
\includegraphics[width=0.49\textwidth]{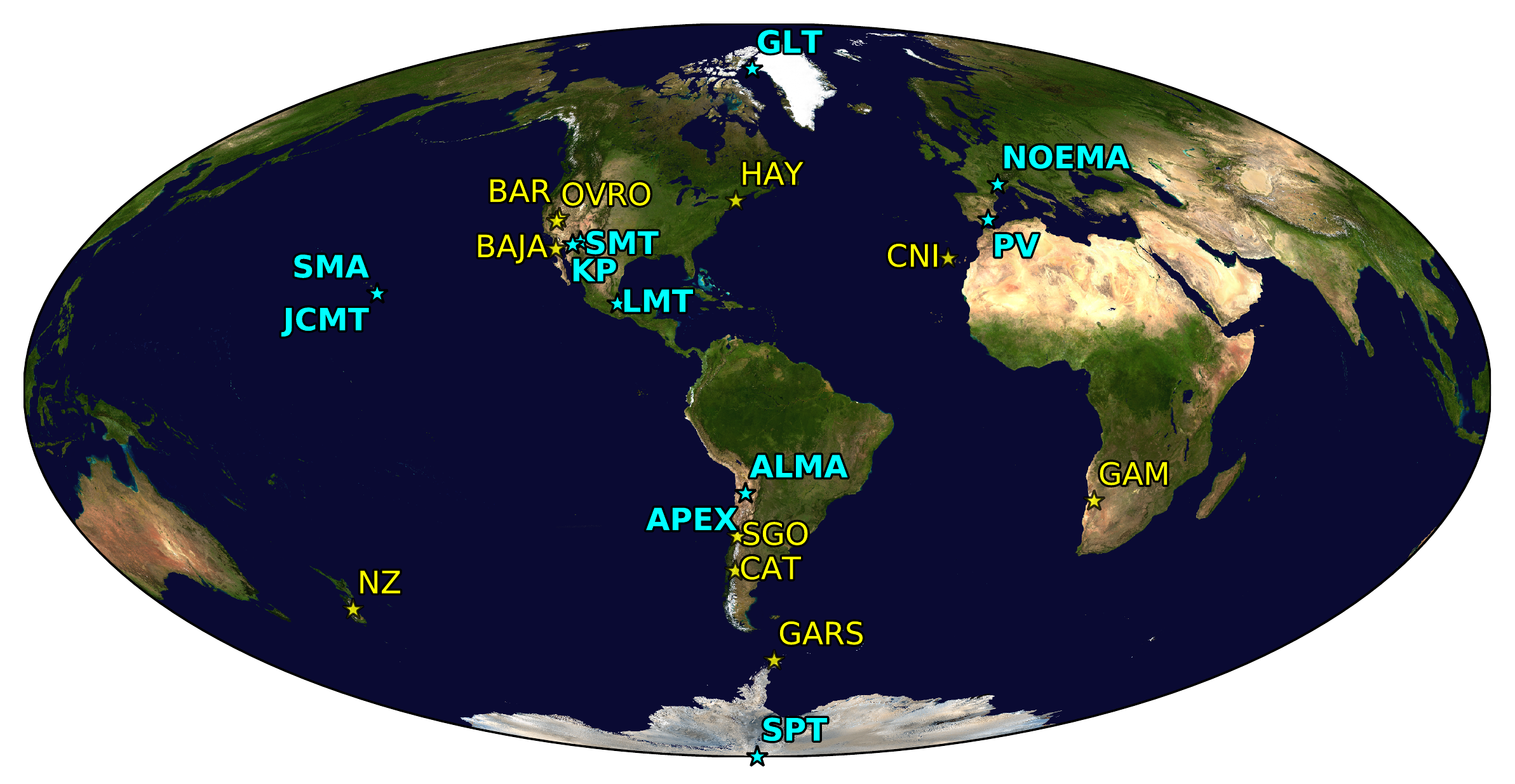}
\includegraphics[width=0.49\textwidth]{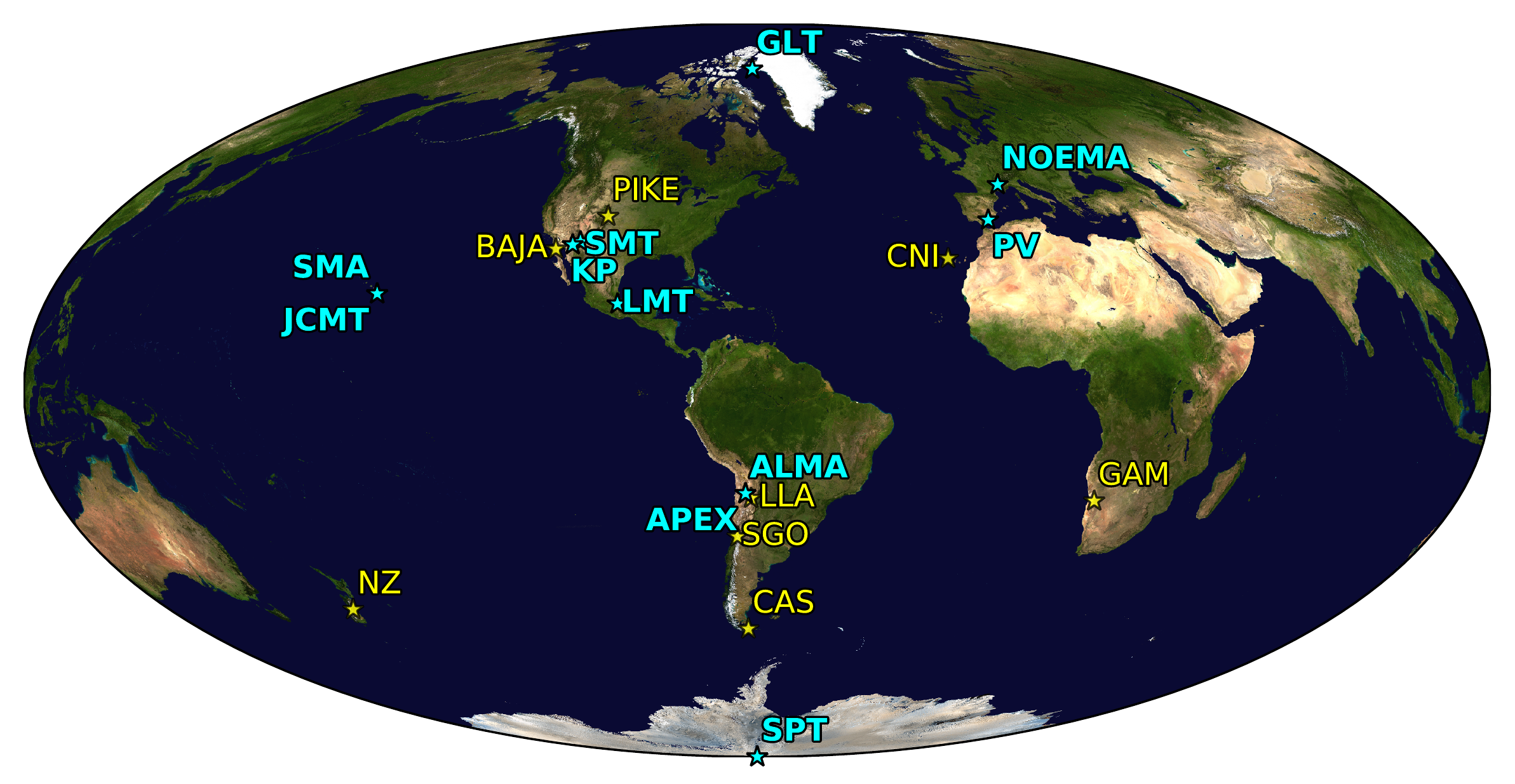}
\caption{Maps of the two ngEHT array configurations, with the current EHT sites shown in cyan and new ngEHT sites in yellow. The ngEHTa configuration with 6\,m dishes is shown on the left, and the ngEHTb configuration with 10\,m dishes is shown on the right. }\label{fig:maps}
\end{figure}

In this article, we consider two different reference array configurations for the ngEHT, which we refer to as the ngEHTa and ngEHTb arrays (Doeleman et~al., in prep; Roelofs et~al. \cite{Roelofs_2023}). The individual sites contained in each of these arrays are listed in Table \ref{tab:vlbi}, and world maps illustrating their global distributions are shown in Figure \ref{fig:maps}.  The ngEHTa configuration contains 10 stations, and the ngEHTb configuration contains 8 stations with a slightly larger average dish diameter.

\begin{table}[H]
     \caption{Station participation in current and future VLBI arrays.}
    \label{tab:vlbi}
    \newcolumntype{C}{>{\centering\arraybackslash}X}
    \begin{tabularx}{\textwidth}{CCCCC}
    \toprule
      \textbf{Station}   & \textbf{EHT}       & \textbf{ngEHTa}    & \textbf{ngEHTb}    & \textbf{GMVA}      \\
      \midrule
      
       ALMA     & {\bf X}   & -         & -         & {\bf X}   \\
       APEX     & {\bf X}   & -         & -         & -         \\
       SMA      & {\bf X}   & -         & -         & -         \\
       JCMT     & {\bf X}   & -         & -         & -         \\
       LMT      & {\bf X}   & -         & -         & {\bf X}   \\
       SMT      & {\bf X}   & -         & -         & -         \\
       KPTO     & {\bf X}   & -         & -         & -         \\
       NOEMA    & {\bf X}   & -         & -         & {\bf X}   \\
       PV       & {\bf X}   & -         & -         & {\bf X}   \\
       SPT      & {\bf X}   & -         & -         & -         \\
       GLT      & {\bf X}   & -         & -         & {\bf X}   \\
       \midrule
       BAJA     & -         & {\bf X}   & {\bf X}   & -         \\
       BAR      & -         & {\bf X}   & -         & -         \\
       CAS      & -         & -         & {\bf X}   & -         \\
       CAT      & -         & {\bf X}   & -         & -         \\
       CNI      & -         & {\bf X}   & {\bf X}   & -         \\
       GAM      & -         & {\bf X}   & {\bf X}   & -         \\
       GARS     & -         & {\bf X}   & -         & -         \\
       HAY      & -         & {\bf X}   & -         & -         \\
       LLA      & -         & -         & {\bf X}   & -         \\
       NZ       & -         & {\bf X}   & {\bf X}   & -         \\
       OVRO     & -         & {\bf X}   & -         & -         \\
       PIKE     & -         & -         & {\bf X}   & -         \\
       SGO      & -         & {\bf X}   & {\bf X}   & -         \\
       \midrule
       GBT      & -         & -         & -         & {\bf X}   \\
       BR       & -         & -         & -         & {\bf X}   \\
       FD       & -         & -         & -         & {\bf X}   \\
       KP       & -         & -         & -         & {\bf X}   \\
       LA       & -         & -         & -         & {\bf X}   \\
       MK       & -         & -         & -         & {\bf X}   \\
       NL       & -         & -         & -         & {\bf X}   \\
       OV       & -         & -         & -         & {\bf X}   \\
       PT       & -         & -         & -         & {\bf X}   \\
       EF       & -         & -         & -         & {\bf X}   \\
       YS       & -         & -         & -         & {\bf X}   \\
       ONS      & -         & -         & -         & {\bf X}   \\
       MET      & -         & -         & -         & {\bf X}   \\
       KVN      & -         & -         & -         & {\bf X}   \\
       \bottomrule
    \end{tabularx}
  
\end{table}

\begin{figure}[H]
\includegraphics[width=0.49\textwidth]{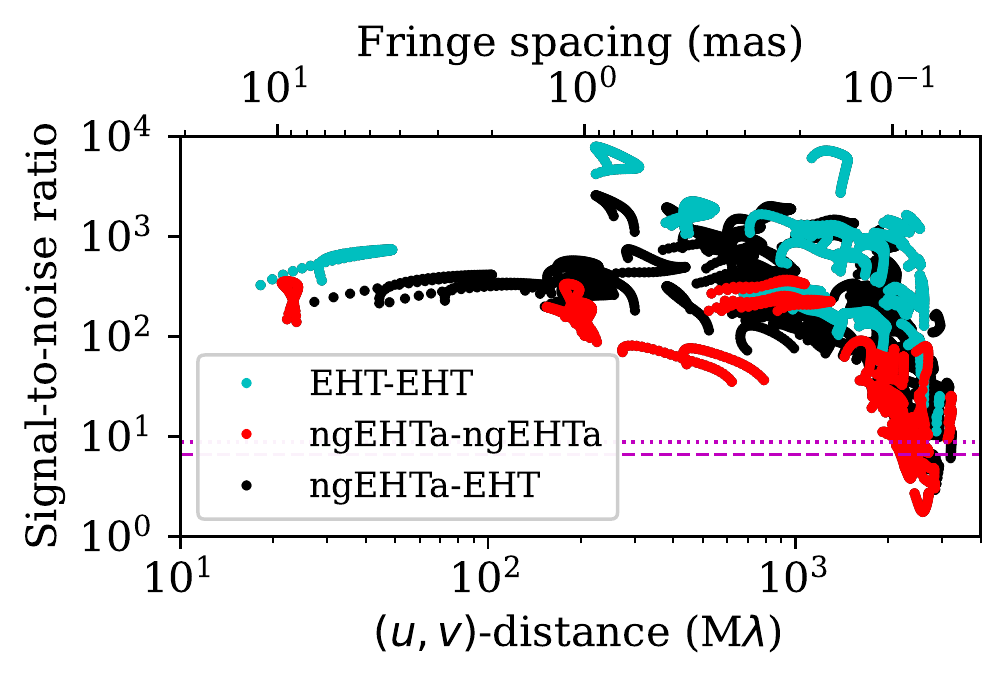}
\includegraphics[width=0.49\textwidth]{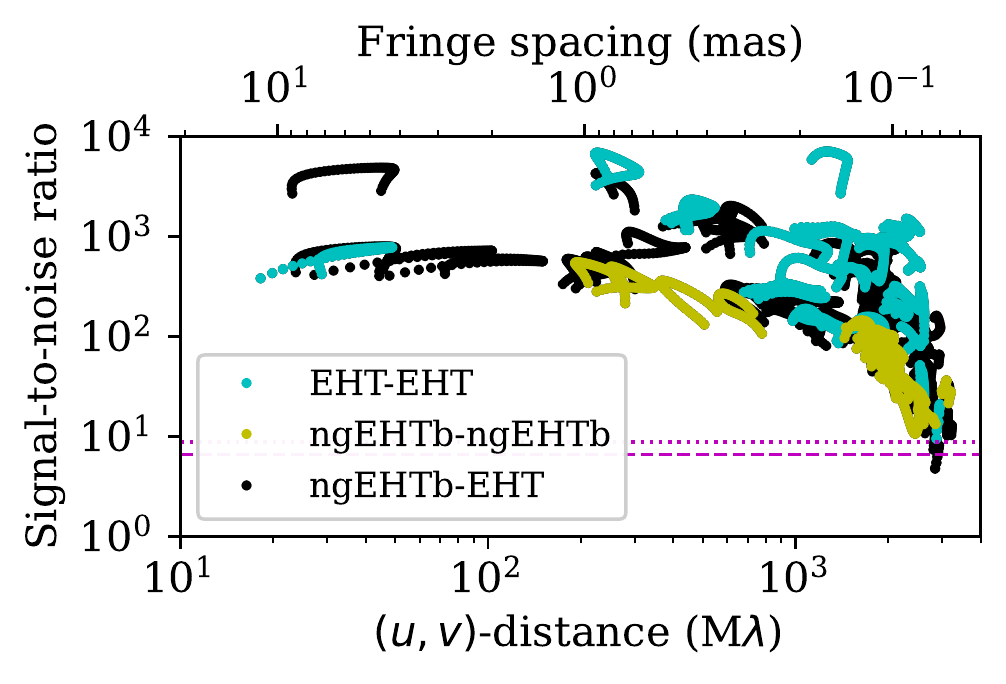}
\includegraphics[width=0.49\textwidth]{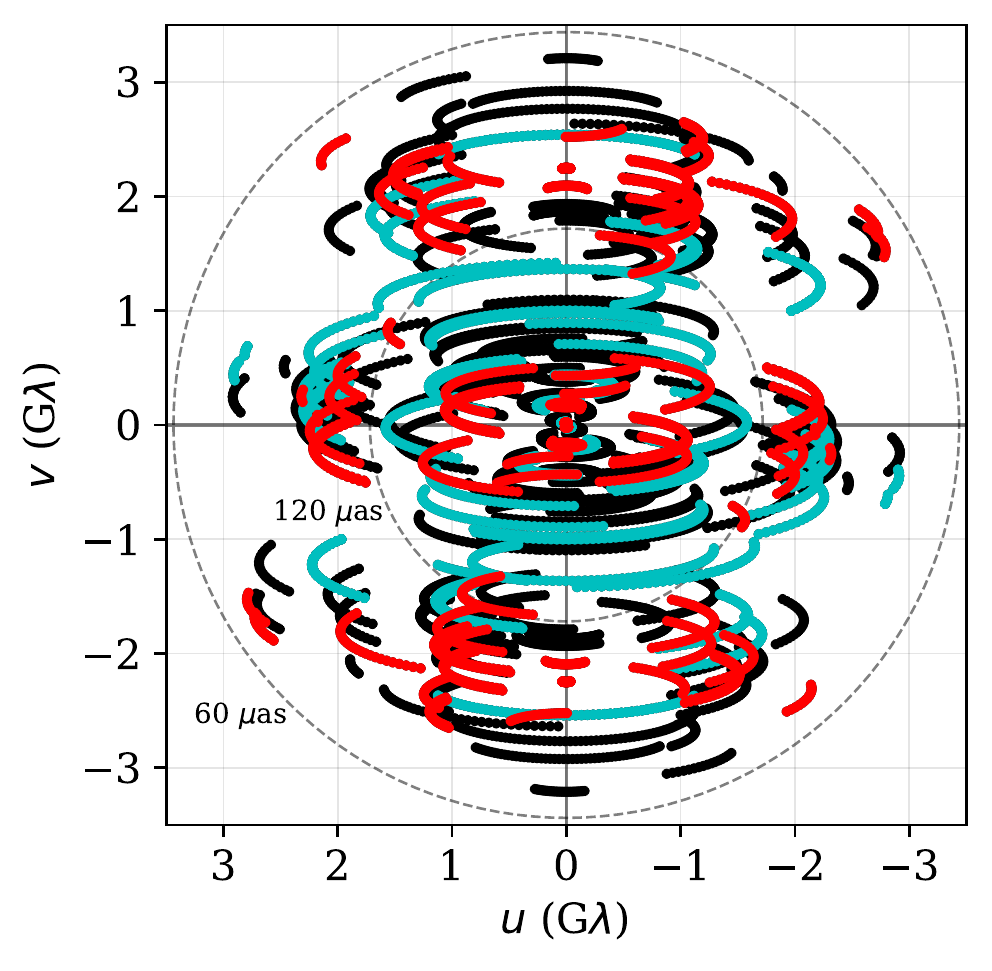}
\includegraphics[width=0.49\textwidth]{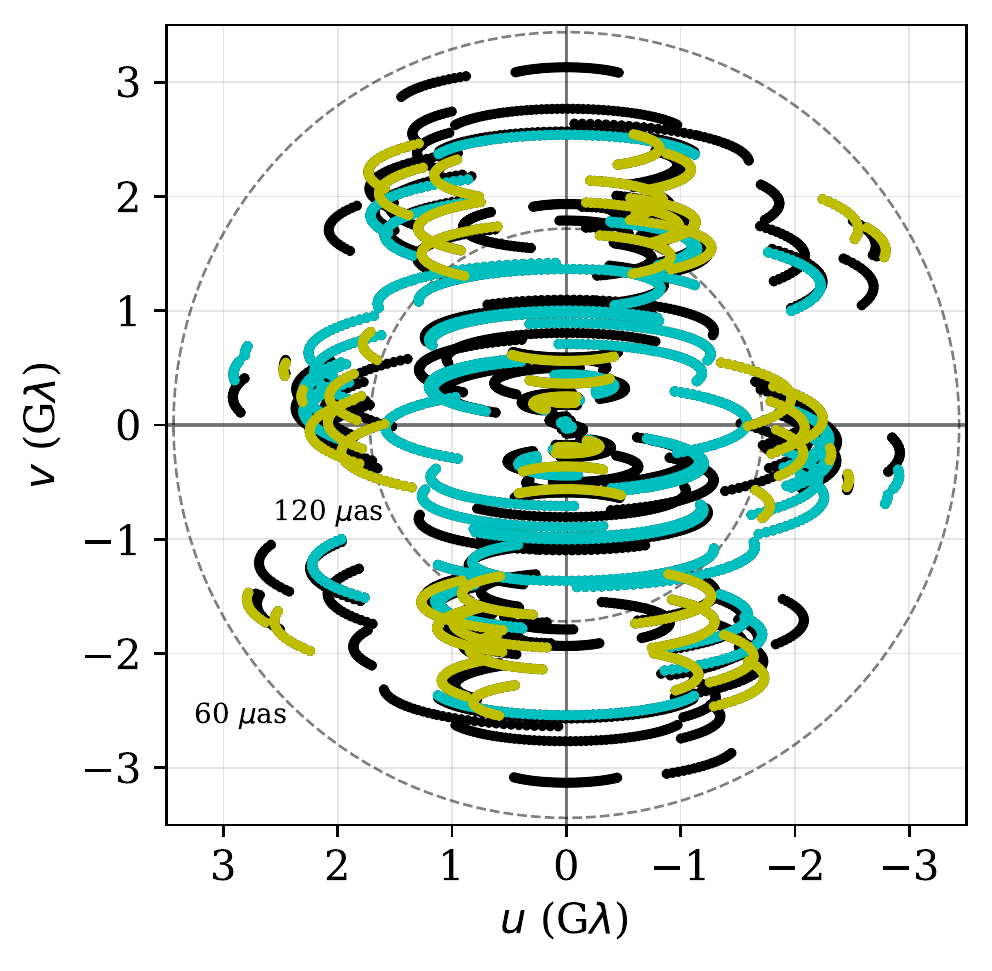}
\caption{Coverage at 86\,GHz $(u,v)$  (bottom row) and signal-to-noise ratio versus $(u,v)$ distance (top row) for the two ngEHT array configurations, as viewed from M87.  The left panels show baselines between EHT stations in cyan, baselines between ngEHTa stations in red, and baselines between ngEHTa and EHT stations in black.  The right panels show baselines between EHT stations in cyan, baselines between ngEHTb stations in yellow, and baselines between ngEHTb and EHT stations in black.  The horizontal dashed and dotted lines in the top panels show the 86\,GHz S/N levels necessary to achieve 90\% phase coherence at 230 and 345\,GHz, respectively.}\label{fig:coverages}
\end{figure}

For the ngEHTa reference configuration, 7 of the 10 new sites were assumed to be equipped with 6-meter dishes, and the remaining three sites were the Haystack Observatory (HAY; 37-meter dish), the Gamsberg mountain (GAM; the primary candidate site for the Africa Millimetre Telescope, which will refurbish the 15-meter SEST dish currently in Chile), and Owens Valley Radio Observatory (OVRO; 10-meter dish).  For the ngEHTb configuration, 7 out of 8 new sites were assumed to be equipped with 10-meter dishes, and the remaining site was GAM with a 15-meter dish.  The bottom row of panels in Figure \ref{fig:coverages} shows the $(u,v)$ coverage for both of these arrays as seen from M87, and the top row of panels shows the expected signal-to-noise ratio as a function of baseline length.

\begin{figure}[H]
\begin{center}
\includegraphics[width=\textwidth]{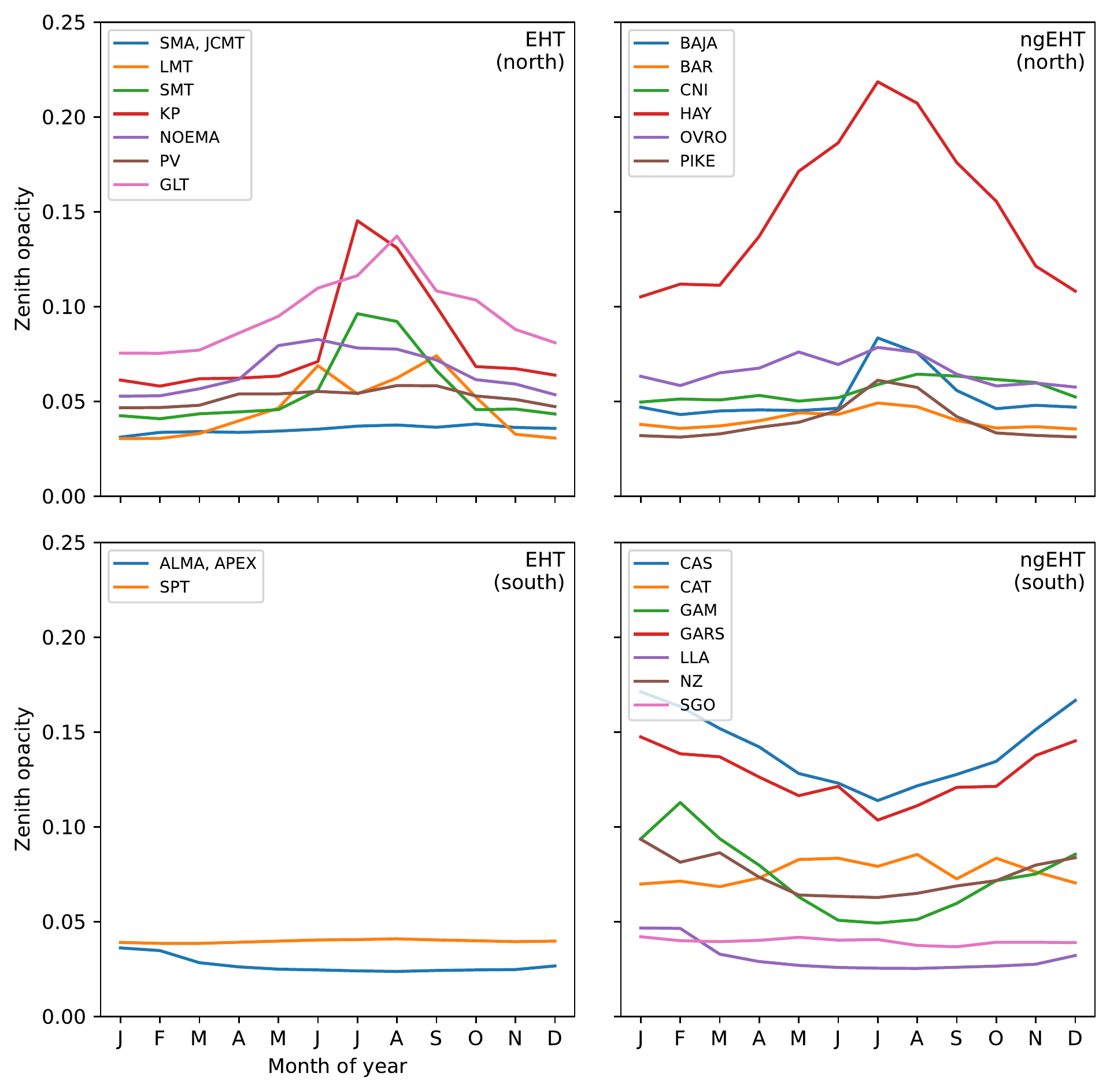}
\end{center}
\caption{Median-zenith atmospheric opacity at 86\,GHz as a function of time at each of the sites in the EHT (\textbf{left panels}) and ngEHT (\textbf{right panels}) arrays; the sites are split by hemisphere. The top row shows northern hemisphere sites, and the bottom row shows southern hemisphere sites.  We note that for frequencies below $\sim$130\,GHz, the atmosphere over Chajnantor is typically more transparent even than that over the South Pole (e.g., \cite[][]{Kovac_2007}).}\label{fig:opacities_86GHz}
\end{figure}

The various new ngEHT sites were selected primarily because of their suitability for observations at 230 and 345\,GHz \citep{Raymond_2021}, and as a result, they tend to have excellent prospects for 86\,GHz observations.  Figure \ref{fig:opacities_86GHz} shows the median 86\,GHz zenith opacities at each of the EHT and ngEHT sites as a function of month during the year.  We can see that the majority of the sites exhibit median 86\,GHz opacities less than 0.1 throughout the entire year, corresponding to $\gtrsim$90\% atmospheric transmission.

\subsection{Synthetic Data} \label{sec:SynthData}

For the various explorations carried out in this article, we have generated synthetic interferometric datasets using the \texttt{eht-imaging} library \citep{Chael_2016,Chael_2018}.  As the input source structure for M87, we used images generated from the GRMHD simulations described in \cite{Chael_2019} and ray-traced at observer frequencies of 86, 230, and 345\,GHz.  Figure \ref{fig:m87model} shows the images of the source at these three frequencies.

During synthetic data generation, SEFDs for each of the stations have been determined following the procedure and atmospheric parameters from \cite{Raymond_2021}.  We carried out a Monte Carlo weather sampling procedure, whereby 100 versions of each synthetic dataset were generated using independent instantiations of the atmospheric temperature and zenith opacity at every site.  All results presented in this article were then computed using the statistics of these 100 samples for each synthetic observation.

\subsection{Performance Metrics} \label{sec:PerformanceMetrics}

We employed two different metrics to assess the performance of a particular array.  Our selected metrics can be computed directly from visibility measurements, so as to be independent of the various specific algorithmic and procedural choices that go into \mbox{image reconstruction.}

Our first metric is the "point source sensitivity," or PSS, which is a measure of overall array sensitivity.  For a set of $N$ complex visibilities with thermal noises $\sigma_i$, the PSS is \mbox{given by}

\begin{equation}
\text{PSS} = \left( \sum_{i=1}^N \frac{1}{\sigma_i^2} \right)^{-1/2} .
\end{equation}

\noindent The PSS has units of flux density, and for a perfectly-calibrated array, it would be equal to the measurement uncertainty in the flux density of an observed point source.  A smaller value for the PSS thus indicates a more sensitive array.

Our second metric is the "$(u,v)$-filling fraction," or FF, which was developed by \cite{Palumbo_2019} and is a measure of how completely-filled the Fourier coverage is.  Computation of the FF depends not only on the $(u,v)$ coverage, but also on the specifications of two additional values: an angular resolution and a field of view.  Given a circle in the $(u,v)$ plane with radius determined by the specified angular resolution, the FF metric value is taken to be the fraction of this circle's area that is occupied by the $(u,v)$ coverage after convolution with a circular tophat function with a radius determined by the specified field of view.  For all FF calculations in this article, we specified a 56.4\,$\upmu$as angular resolution (equal to that of an Earth-diameter baseline observed at 86\,GHz), and we specified a field of view of 1\,mas for M87 observations.  The FF metric value is normalized to fall between zero and one: a value of zero indicates no coverage, and a value of one indicates a fully covered \mbox{Fourier plane.}

\begin{figure}[H]
\includegraphics[width=0.7\columnwidth]{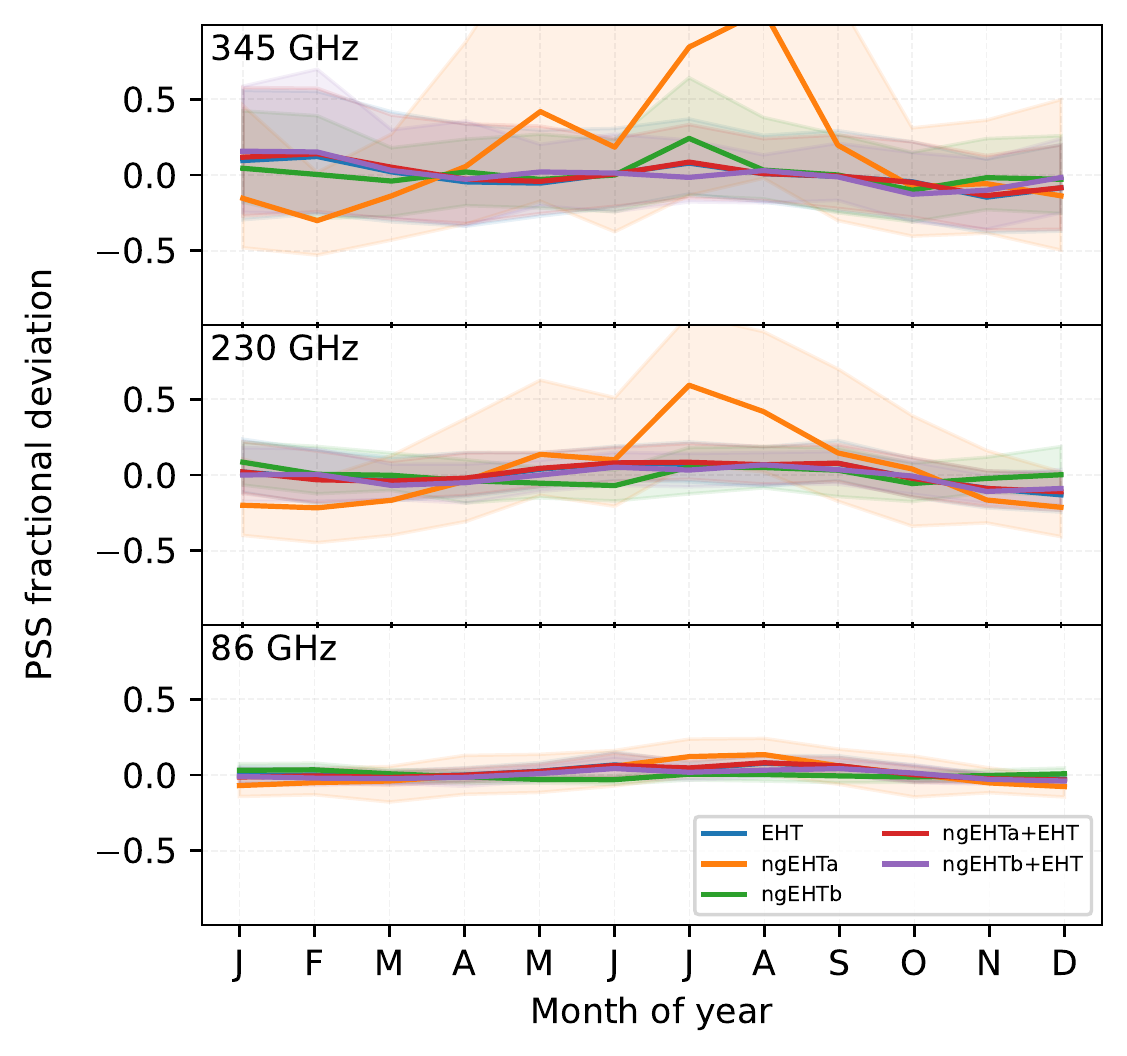}
\includegraphics[width=0.7\columnwidth]{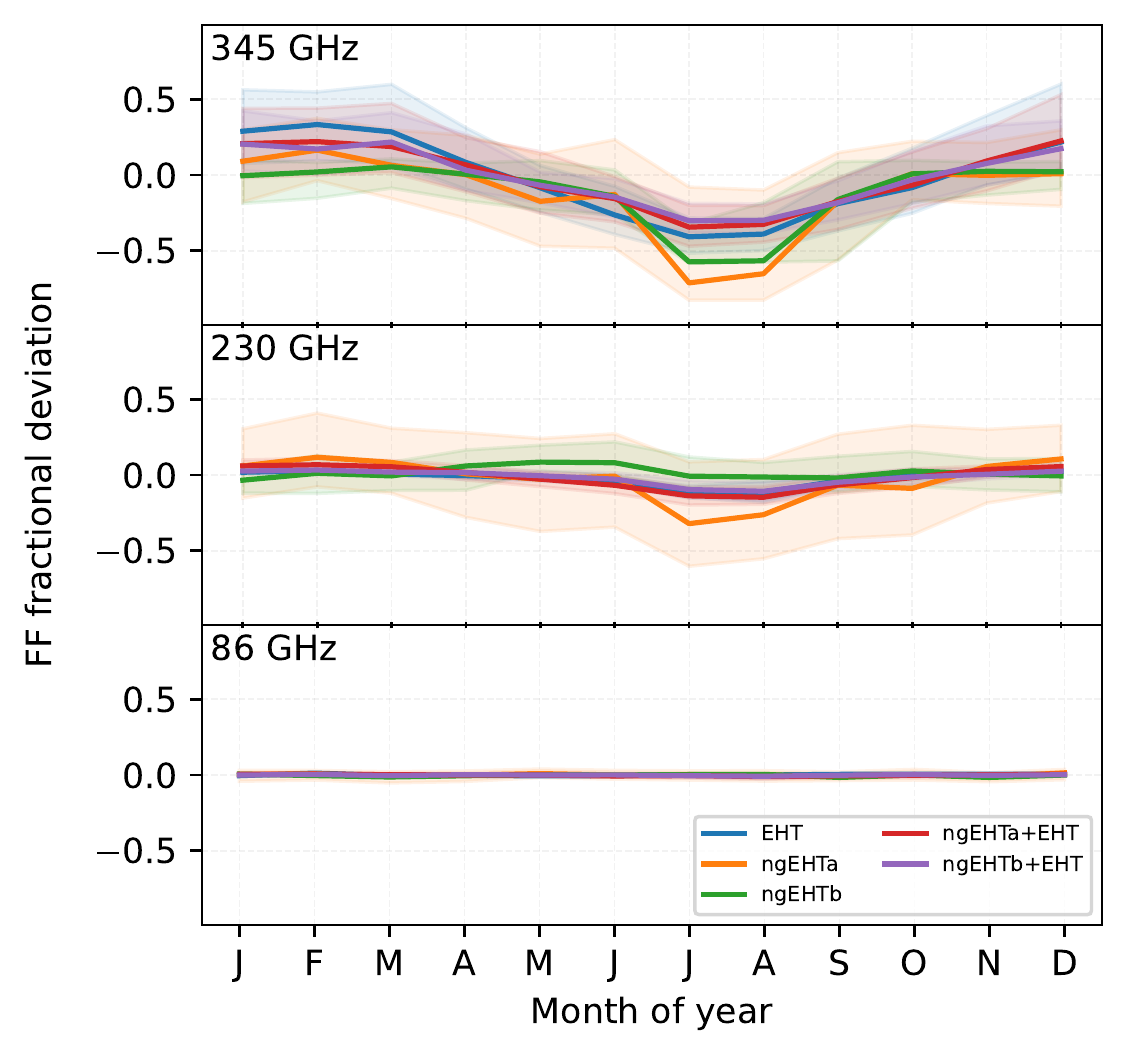}
\caption{Deviation of the PSS (top panel) and FF (bottom panel) metrics from their yearly median values, each plotted as a fraction of that median value versus time for five different array configurations. The EHT, ngEHTa, ngEHTb, ngEHTa+EHT, and ngEHTb+EHT arrays are each plotted in a different color, as indicated in the legend.  The top, middle, and bottom rows in each panel correspond to observing frequencies of 345, 230, and 86\,GHz, respectively.  The light shaded region around each line encompasses the inter-quartile range determined by the weather Monte Carlo procedure.  A target field-of-view of 1000\,$\mu$as has been assumed for all $(u,v)$-filling fraction computations.}\label{fig:compare_metrics}
\end{figure}

Figure \ref{fig:compare_metrics} shows the PSS and FF metric behavior for the EHT, ngEHT, and composite arrays throughout the year, relative to their median values.  We see that for both the PSS and FF metrics, the 86\,GHz behavior is substantially more stable in time than the corresponding metrics at 230\,GHz or 345\,GHz.  The arrays may suffer from substantial performance degradation when observing at 230\,GHz or 345\,GHz in the northern summer relative to the northern winter, but observations at 86\,GHz should not be significantly impacted.

\subsection{Multi-Frequency Calibration Techniques}\label{sec:phase_trans}

One promising calibration technique made possible by the addition of 86\,GHz capabilities is FPT, in which atmospheric phase fluctuations are tracked at 86\,GHz and then transferred to the higher-frequency bands. FPT results in increased coherence time at the higher frequencies. Bona fide astrometry can be added using SFPR, by interleaving observations of a second source to remove the remaining FPT dispersive residual terms. A comprehensive error analysis formulation of FPT and SFPR was initially presented in \citet{Rioja_2011}.  Overviews and discussions can be found in Dodson et al. \cite{Dodson_2017} and \citet{Rioja_2020}.

FPT requires observing a source simultaneously in (at least) two different frequencies.  The key assumption underlying FPT is that phase tracking and calibration at one of the frequencies---almost always taken to be the lower frequency---is easier than at the other frequency.  There are at least three reasons for why phase calibration is easier at \mbox{lower frequencies:}
\begin{enumerate}
    \item At (sub)millimeter observing wavelengths, the most rapidly-fluctuating contribution to the visibility phase comes from the troposphere, whose timescale typically decreases with increasing observing frequency and whose magnitude is proportional to $\nu$.
    \item Atmospheric absorption and receiver noise temperatures are lower at 86 GHz than they are at higher frequencies, essentially making each telescope more sensitive and permitting higher S/N to be achieved within any given integration time.
    \item Dimensionless baseline lengths are proportional to $\nu$, meaning that the spatial scales probed by any given baseline are larger when observing at lower frequencies.  As many VLBI sources (e.g., AGN) are resolved at (sub)millimeter observing wavelengths, shorter baselines typically have higher correlated flux densities at lower frequencies in this regime, again permitting higher S/N to be achieved within any given \mbox{integration time.}
\end{enumerate}

To successfully carry out FPT, the source must be detectable at the lower frequency within a timescale that is approximately equal to the phase coherence timescale at the higher frequency, such that the phase variations can be tracked over time.  As the tropospheric term dominates these phase variations, and because the magnitude of the tropospheric variations is proportional to $\nu$, we can apply a frequency-scaled version of the lower frequency phase solution to the higher-frequency data.  For periods of time over which the intrinsic source phases are only slowly varying, the removal of the dominant phase corruption permits substantially increased integration times.

FPT imposes more demanding S/N requirements for the low-frequency detection than would typically be necessary for single-frequency phase calibration.  The S/N of a detection is related to the RMS phase fluctuations $\sigma_{\phi}$ by
\begin{equation}
\text{S/N} \approx \frac{1}{\sigma_{\phi}} .
\end{equation}
\noindent The RMS phase fluctuations in turn determine the coherence, $\eta$, which for Gaussian variations is given by
\begin{equation}
\eta = e^{-\sigma_{\phi}^2 / 2} = e^{-1 / 2 (\text{S/N})^2} .
\end{equation}
\noindent The RMS phase fluctuations at the lower frequency are also scaled by the frequency ratio when transferred to the higher frequency, meaning that the effective S/N at the higher frequency is smaller by the same factor.  Thus, achieving a coherence of $\eta \geq 0.9$ at a frequency of 345\,GHz requires $\text{S/N} \gtrsim 2.2$.  However, if the phase at 345\,GHz is being determined by FPT from 86\,GHz, then the equivalent S/N at 86\,GHz must be $\text{S/N} \gtrsim 8.8$ to achieve the same 345\,GHz coherence.  The horizontal dashed and dotted lines in the top panels of \mbox{Figure \ref{fig:coverages}} show the 86\,GHz S/N levels necessary to achieve $\eta \geq 0.9$ at 230\,GHz and \mbox{345\,GHz.}

\begin{figure}[H]
\includegraphics[width=0.98\columnwidth]{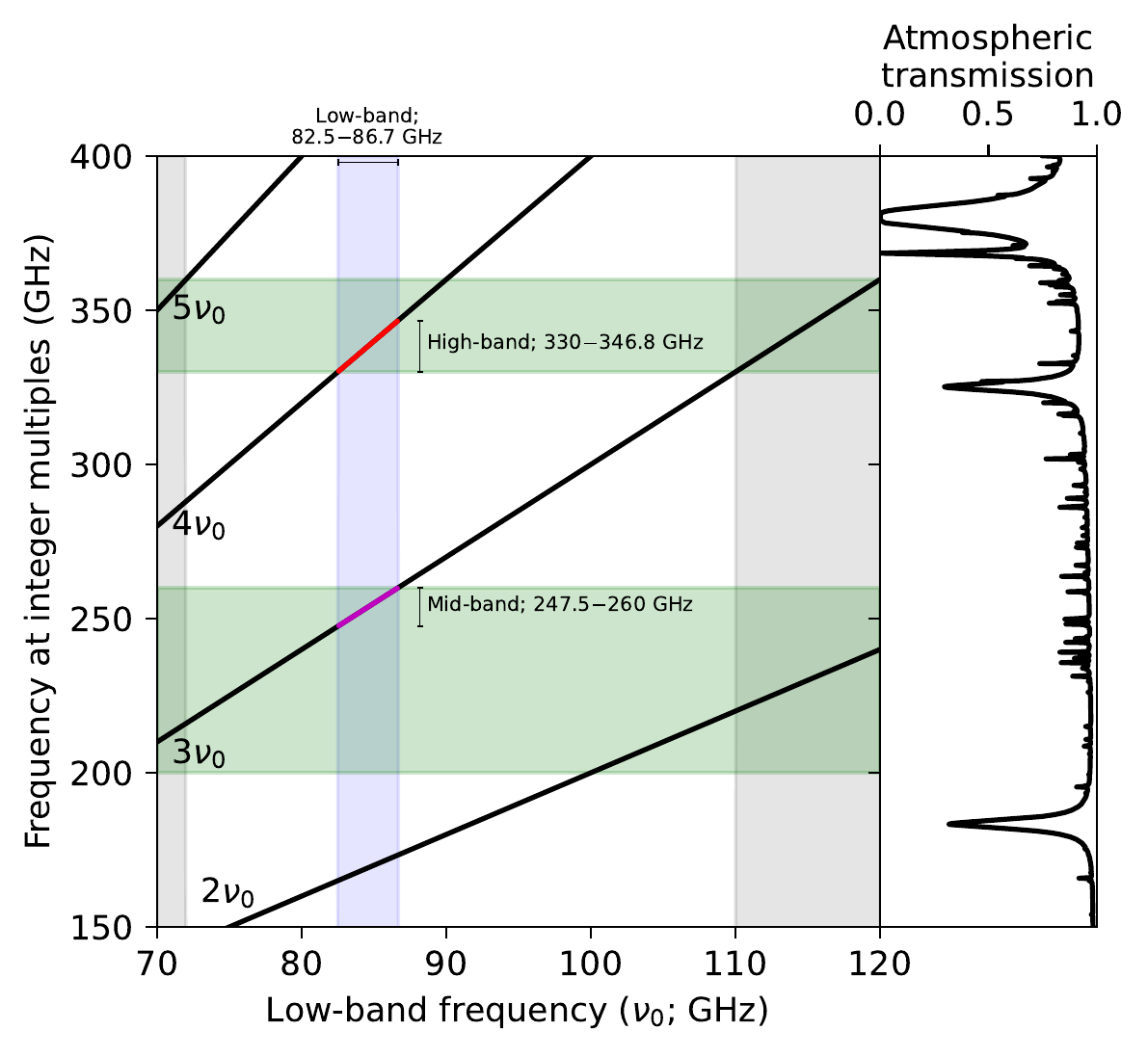}
\caption{Frequency
 coverage constraints imposed by the desire to transfer phase information from a low-frequency band ("low-band," around $\sim$86\,GHz) to two different higher-frequency bands ("mid-band" around $\sim$230\,GHz and "high-band" around $\sim$345\,GHz).  The black curves in the left panel show integer multiples of the low-band frequency, and the green shaded regions indicate approximate available spectral windows for the mid- and high-frequency receivers.  The three vertical shaded regions indicate low-band frequency ranges where an integer multiple of that frequency passes through both the mid- and high-band spectral windows.  The middle vertical shaded region (highlighted in blue) corresponds to our proposed spectral arrangement; the available frequency ranges for each of the three bands are labeled, and the corresponding segments of the black curves are highlighted.  For reference, the right panel shows the atmospheric transmission as a function \mbox{of frequency.}}\label{fig:frequency_coverage}
\end{figure}

The scaling of the phase variations at the lower frequency before applying them to the higher frequency can also result in phase-wrapping ambiguities.  Such ambiguities are avoided only if the frequency ratio between the lower and higher frequencies is an integer~\citep{Dodson_2014}.  For the more general non-integer case, these ambiguities can introduce seemingly random phase jumps whenever the lower frequency phase wraps.  Attempting to "unwrap" the phases prior to transferring can improve the performance, but this is an imperfect solution that will perform increasingly poorly as SNR decreases.  It is thus preferable to maintain an integer value $R$ when employing the FPT technique, which motivates particular choices of frequency bands.  Figure \ref{fig:frequency_coverage} illustrates how such constraints manifest for the proposed ngEHT frequency configuration containing three bands.  One "optimal" arrangement is highlighted in blue and corresponds to a $\sim$4\,GHz bandwidth in the lowest frequency band (spanning $\sim$82.5--86.5\,GHz), a $\sim$12\,GHz bandwidth in the middle frequency band (spanning $\sim$248--260\,GHz), and a $\sim$16\,GHz bandwidth in the highest frequency band (spanning $\sim$330--346\,GHz).

Applications at lower frequencies have demonstrated that FPT-aided coherence times can extend to tens of minutes \citep{Rioja_2014}, and adding in a third frequency to remove residual ionospheric phase fluctuations can potentially extend the coherence times to multiple hours \citep{Zhao_2018}.  Integration times of minutes have also been achieved by the EHT using phase stabilization at 230\,GHz alone for sources with $\gtrsim$Jy-level flux densities \citep{M87PaperIII}.  An FPT from 86\,GHz would enable similarly increased coherence times for substantially weaker sources than would otherwise be observable with the ngEHT.

\section{Technical Interoperability}\label{sec:interoperability}

In the previous section we have shown that the ngEHT stations are able to observe at 86\,GHz for the entire year. This offers great flexibility to enhance observing time at the higher frequencies, but also to provide stand-alone 86\,GHz observing time with the array when observation conditions at the higher frequencies are poor. 

The addition of 86\,GHz capabilities to the ngEHT telescopes creates an opportunity for interoperability with major current and upcoming facilities. The new ngEHT dishes will most likely have small ($\leq$10\,m) diameters, meaning that the sensitivities of baselines between ngEHT dishes will be comparatively modest relative to, e.g., many EHT baselines. Substantial increases in both sensitivity and coverage at 86\,GHz could be achieved by jointly observing with the ngEHT and one or more other arrays.
In this section, we explore the capabilities of the ngEHT stations on their own and in combination with two external facilities (see Tables \ref{tab:vlbi} and \ref{tab:array}): the Global Millimeter VLBI Array (GMVA), the current leading 86\,GHz VLBI array, and the next generation Very Large Array (ngVLA), a future major facility that is expected to become the most sensitive array observing at 86\,GHz.

\begin{table}[H]
   \caption{Station overview and receiver capabilities, showing which sites are capable (or expected to be capable) of observing in which frequency bands.}
    \label{tab:array}
    \newcolumntype{C}{>{\centering\arraybackslash}X}
    \begin{tabularx}{\textwidth}{CCCC}
    \toprule
      \textbf{Station}   & \textbf{86 GHz} & \textbf{230 GHz} & \textbf{345 GHz}  \\
      \midrule

       ALMA  & {\bf X} & {\bf X} & {\bf X} \\
       APEX  & - & {\bf X} & {\bf X} \\
       SMA  & - & {\bf X} & {\bf X} \\
       JCMT  & {\bf X} & {\bf X} & {\bf X} \\
       LMT  & {\bf X} & {\bf X} & planned \\
       SMT  & - & {\bf X} & {\bf X} \\
       KPTO  & - & {\bf X} & - \\
       NOEMA  & {\bf X} & {\bf X} & {\bf X} \\
       PV  & {\bf X} & {\bf X} & {\bf X} \\
       SPT  & - & {\bf X} & {\bf X} \\
       GLT  & {\bf X} & {\bf X} & {\bf X} \\
       \hline
       BAJA  & {\bf X} & {\bf X} & {\bf X} \\
       BAR  & {\bf X} & {\bf X} & {\bf X} \\
       CAS  & {\bf X} & {\bf X} & {\bf X} \\
       CAT  & {\bf X} & {\bf X} & {\bf X} \\
       CNI  & {\bf X} & {\bf X} & {\bf X} \\
       GAM  & {\bf X} & {\bf X} & -\\
       GARS  & {\bf X} & {\bf X} & {\bf X} \\
       HAY  & {\bf X} & {\bf X} & {\bf X} \\
       LLA  & {\bf X} & {\bf X} & {\bf X} \\
       NZ  & {\bf X} & {\bf X} & {\bf X} \\
       OVRO  & {\bf X} & {\bf X} & {\bf X} \\
       PIKE  & {\bf X} & {\bf X} & {\bf X} \\
       SGO  & {\bf X} & {\bf X} & {\bf X} \\
       ngVLA & {\bf X} & - & - \\
       \midrule
       GBT  & {\bf X} & - & - \\
       BR  & {\bf X} & - & - \\
       FD  & {\bf X} & - & - \\
       KP  & {\bf X} & - & - \\
       LA  & {\bf X} & - & - \\
       MK  & {\bf X} & - & - \\
       NL  & {\bf X} & - & - \\
       OV  & {\bf X} & - & - \\
       PT  & {\bf X} & - & - \\
       EF  & {\bf X} & - & - \\
       YS  & {\bf X} & - & - \\
       ONS  & {\bf X} & - & - \\
       MET  & {\bf X} & - & - \\
       KVN  & {\bf X} & {\bf X} & - \\
       \bottomrule
    \end{tabularx}
   
\end{table}

The baseline sensitivity of the GMVA is currently limited by the recording bandwidth of the VLBA array (4\,Gbps), and the ngEHT is planning to operate with a bandwidth of 256\,Gbps. Currently, EHT sites with 86\,GHz receivers are able to observe, as part of the GMVA, by only correlating a fraction of the observed frequency band. They have potential applications also for sub-arraying stations that are able to record at a higher rate. An alternative to the GMVA would be to make use of the high bandwidth of the ngEHT+EHT 86\,GHz sites. This alternative offers a significant increase in baseline sensitivity due to the higher recording rate (especially valuable for weak polarization signals) and comparable coverage and point-source sensitivity to the GMVA.  However, to more accurately reflect the capabilities of the current GMVA, all synthetic datasets labeled "GMVA" in this paper are limited to 512\,MHz.  The synthetic datasets for the other arrays use the bandwidths specified in Figure \ref{fig:frequency_coverage}.

\begin{figure}[H]
\includegraphics[width=0.82\textwidth]{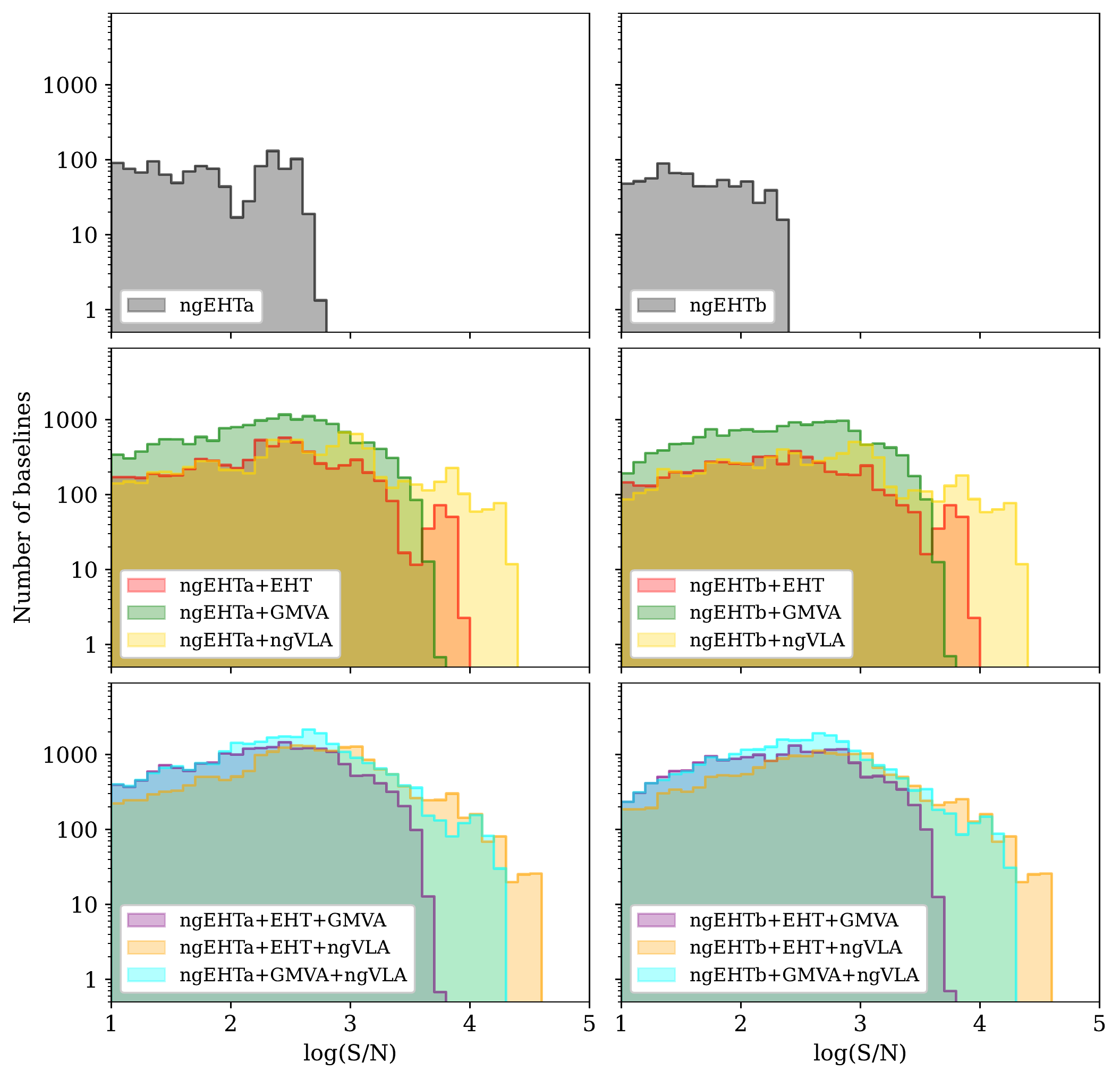}
\caption{Histograms of baseline signal-to-noise ratios for the different arrays observing M87 at 86\,GHz.  Each histogram is the result of averaging over 100 Monte Carlo realizations of weather at every site, so the resulting histogram bins do not necessarily contain integer numbers of baselines.}\label{fig:SNR_average}
\end{figure}

\begin{figure}[H]
\includegraphics[width=0.82\columnwidth]{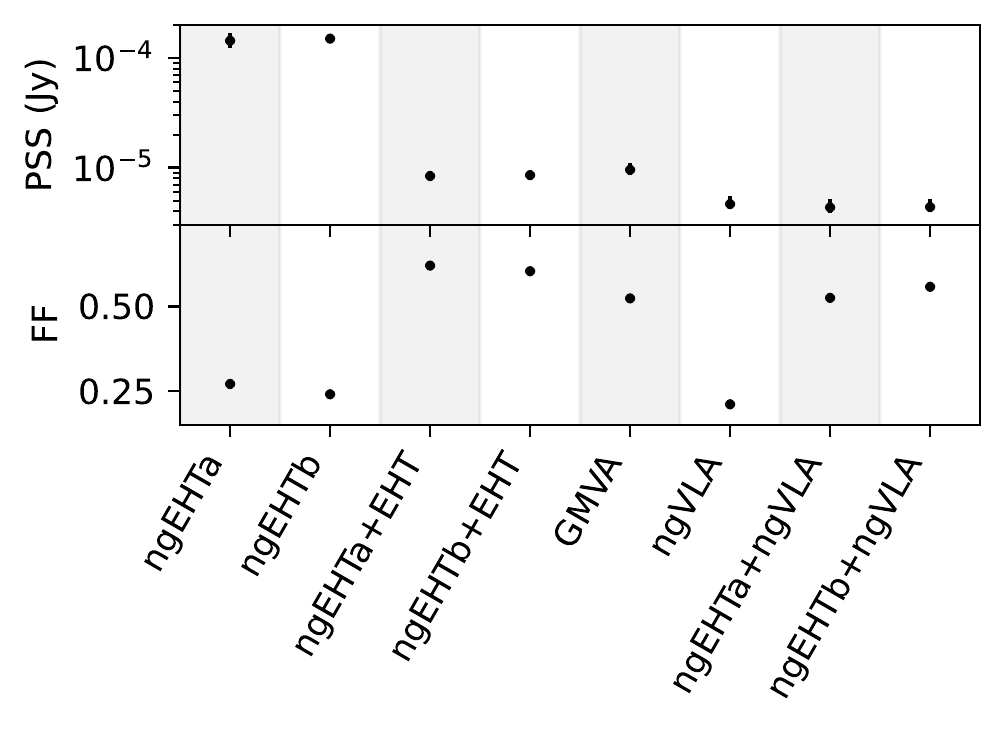}
\caption{Comparison of the point source sensitivity (PSS, top panel, in units of Jy) and $(u,v)$-filling fraction (FF, bottom panel, unitless) metrics for several different array configurations. Each point shows the median metric value determined for 86\,GHz M87 observations across 100 Monte Carlo realizations of weather at every site and across each month of the year, and the errorbars indicate the 16th to 84th percentile ranges.}\label{fig:metric_comparison}
\end{figure}

Figure \ref{fig:SNR_average} shows histograms of the baseline signal-to-noise ratio (S/N) for a number of potential array combinations.  When observing as a standalone array, the ngEHT achieves a typical baseline S/N in tens or hundreds.  By jointly observing with the EHT, GMVA, and/or ngVLA, a typical baseline S/N of hundreds or thousands is achieved; some baselines have an S/N in excess of $10^4$.  A measure of the total array sensitivity is captured by the PSS metric plotted in Figure \ref{fig:metric_comparison}, which improves by more than an order of magnitude when observing with the EHT, GMVA, and/or ngVLA alongside the ngEHT.  Figure \ref{fig:metric_comparison} also shows the improvement in the FF metric that is achieved by joint observations; we can see that joint observations substantially improve the FF to make it superior to that of the standalone ngEHT and the standalone ngVLA.

\begin{figure}[H]
    \begin{adjustwidth}{-\extralength}{0cm}
\centering
        \begin{tabular}{cccc}
             Ground Truth & \hspace{-1cm} ngVLA & \hspace{-1cm} ngEHTa + EHT & \hspace{-1cm} ngEHTa + EHT + ngVLA \vspace{0.1cm}\\
 \includegraphics[width=0.235\linewidth]{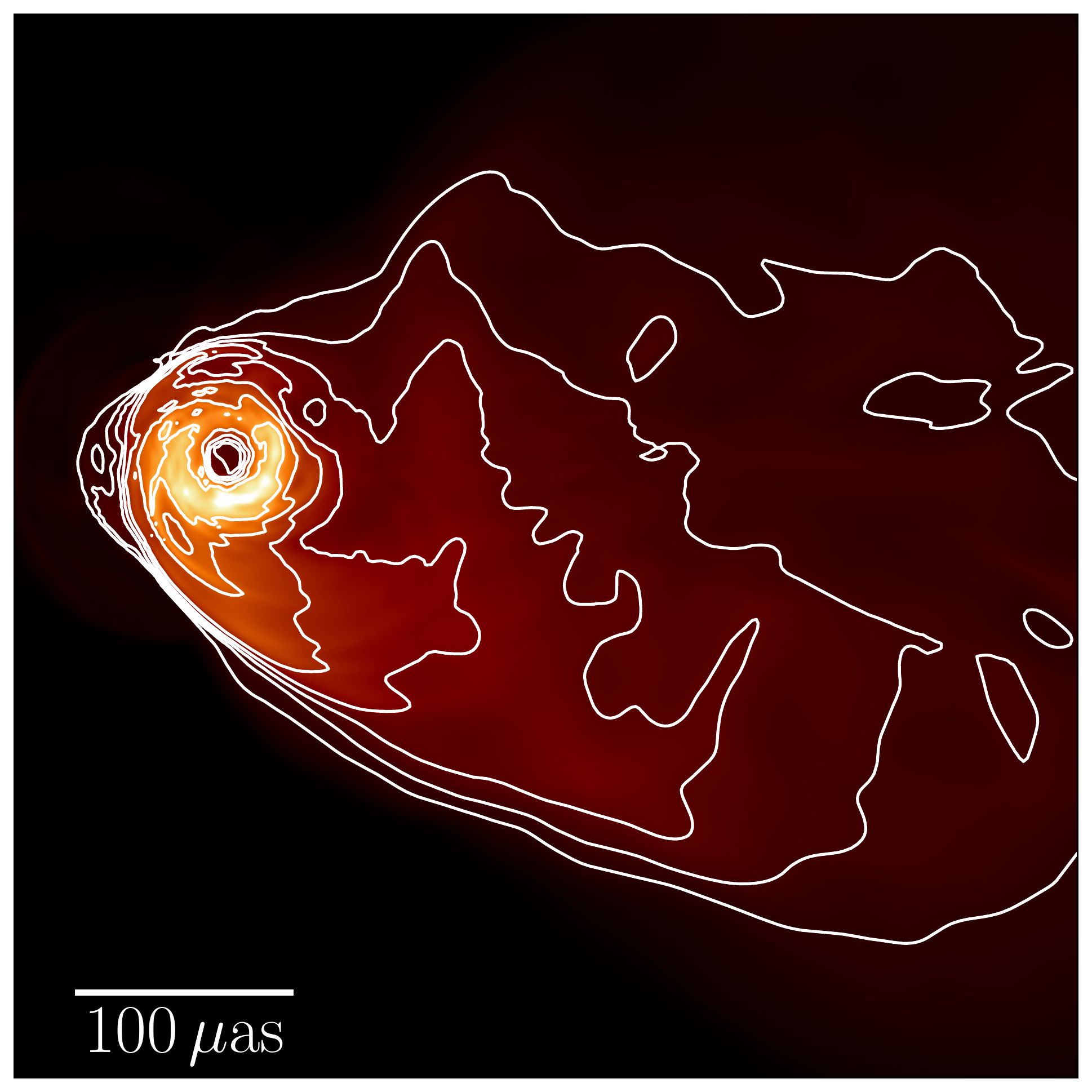} & \hspace{-0.7cm} \includegraphics[width=0.235\linewidth]{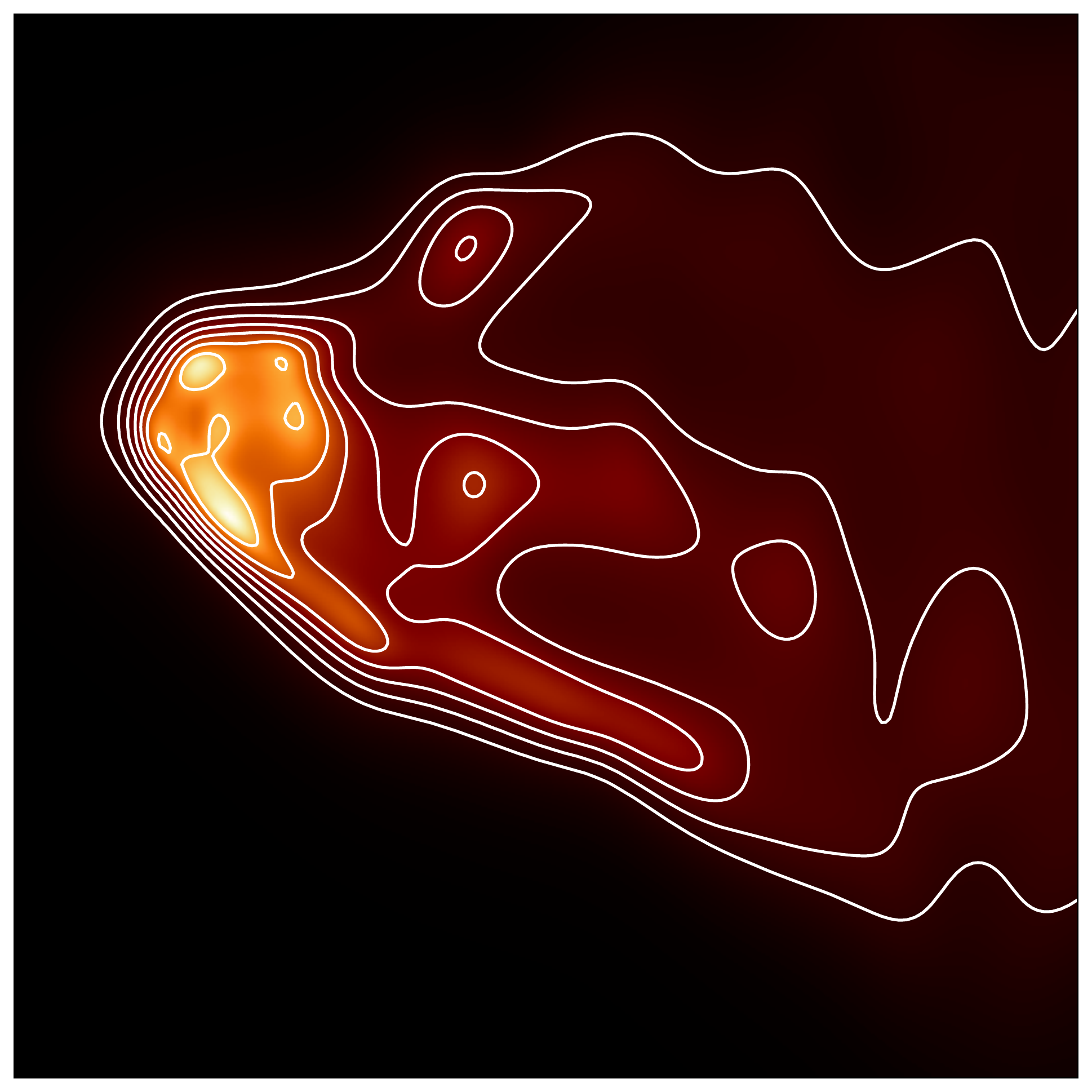} & \hspace{-0.7cm}
 \includegraphics[width=0.235\linewidth]{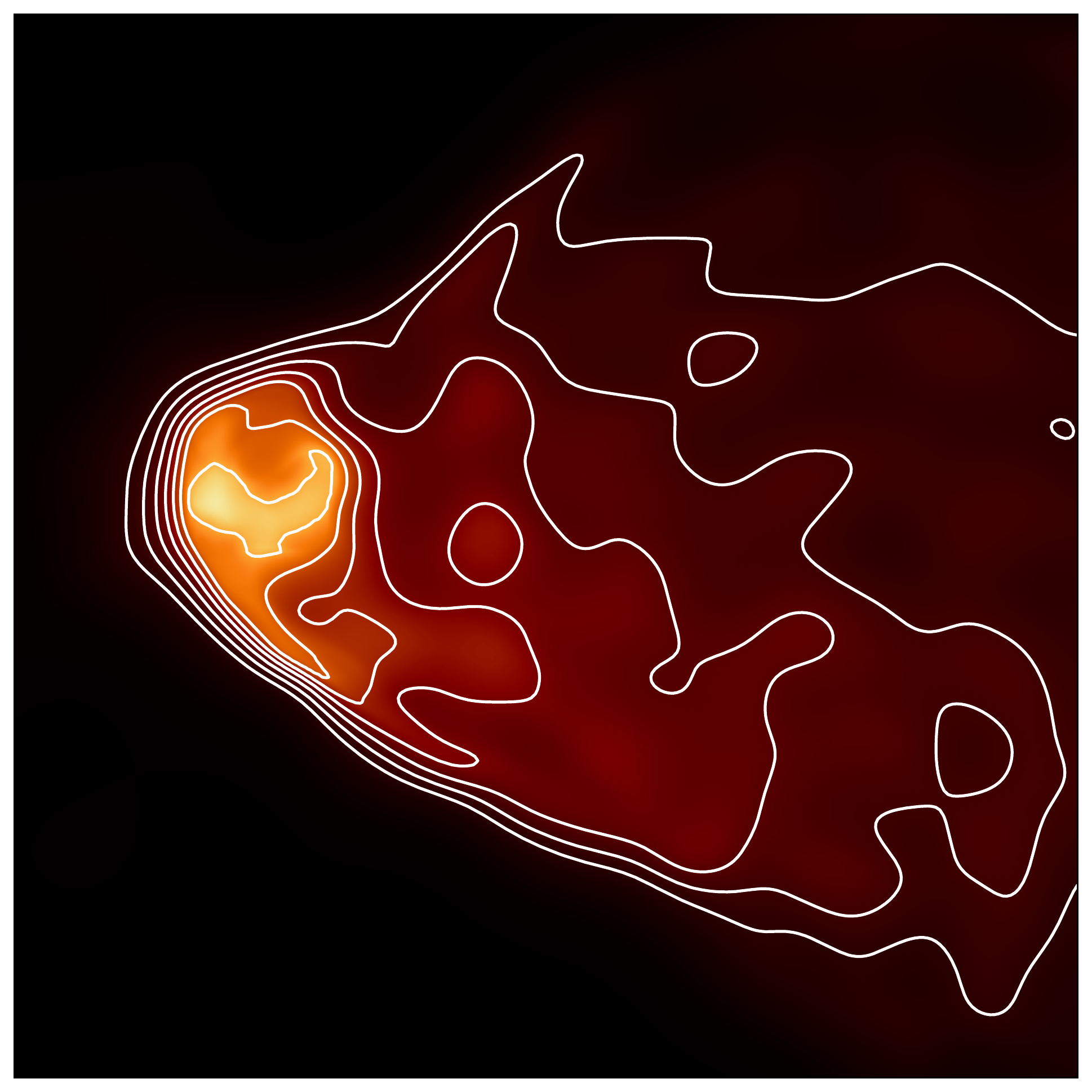} &\hspace{-0.47cm}
    \raisebox{-0.01\linewidth}[0pt][0pt]{\includegraphics[width=0.28\linewidth]{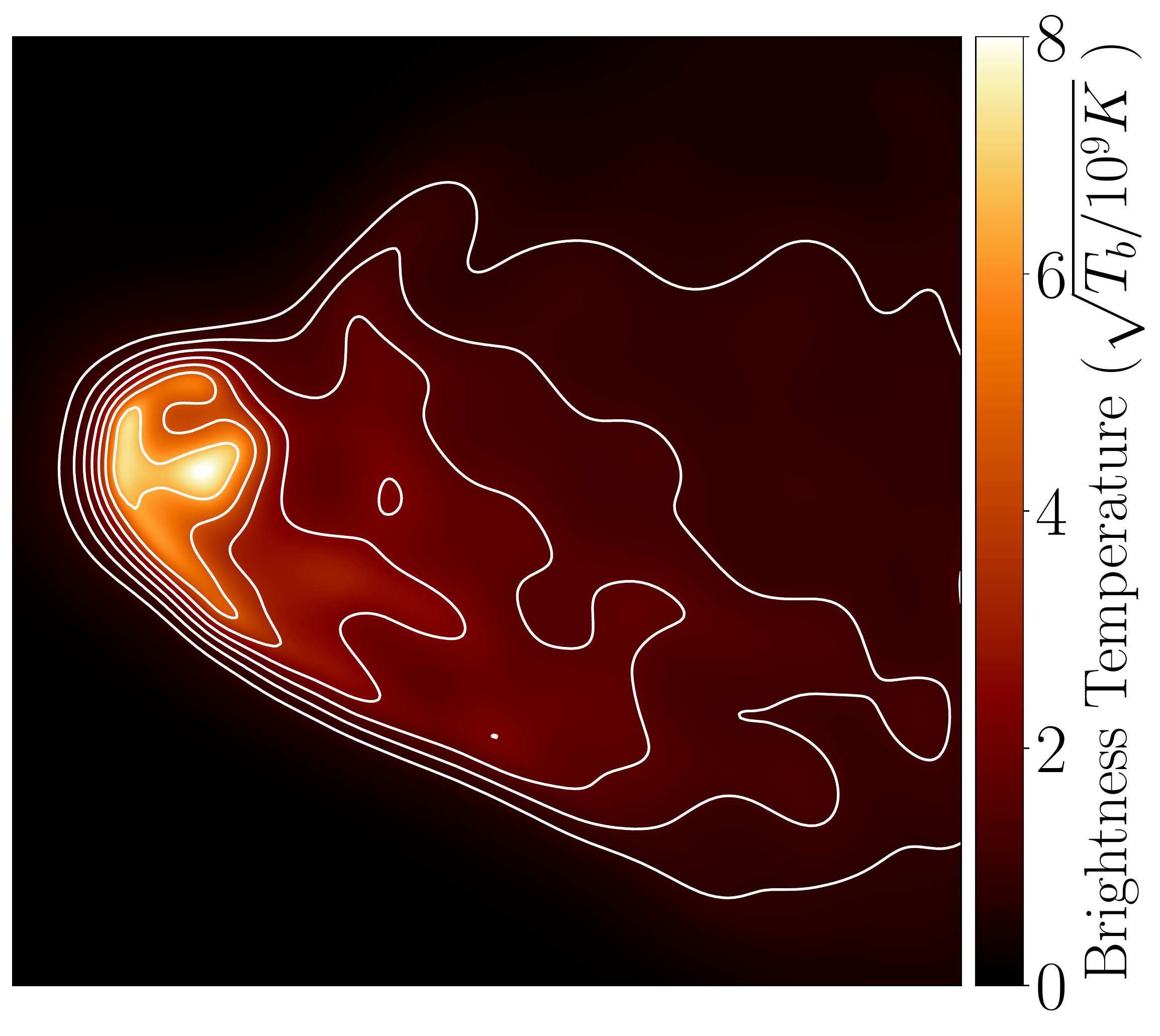}} \\
    \includegraphics[width=0.235\linewidth]{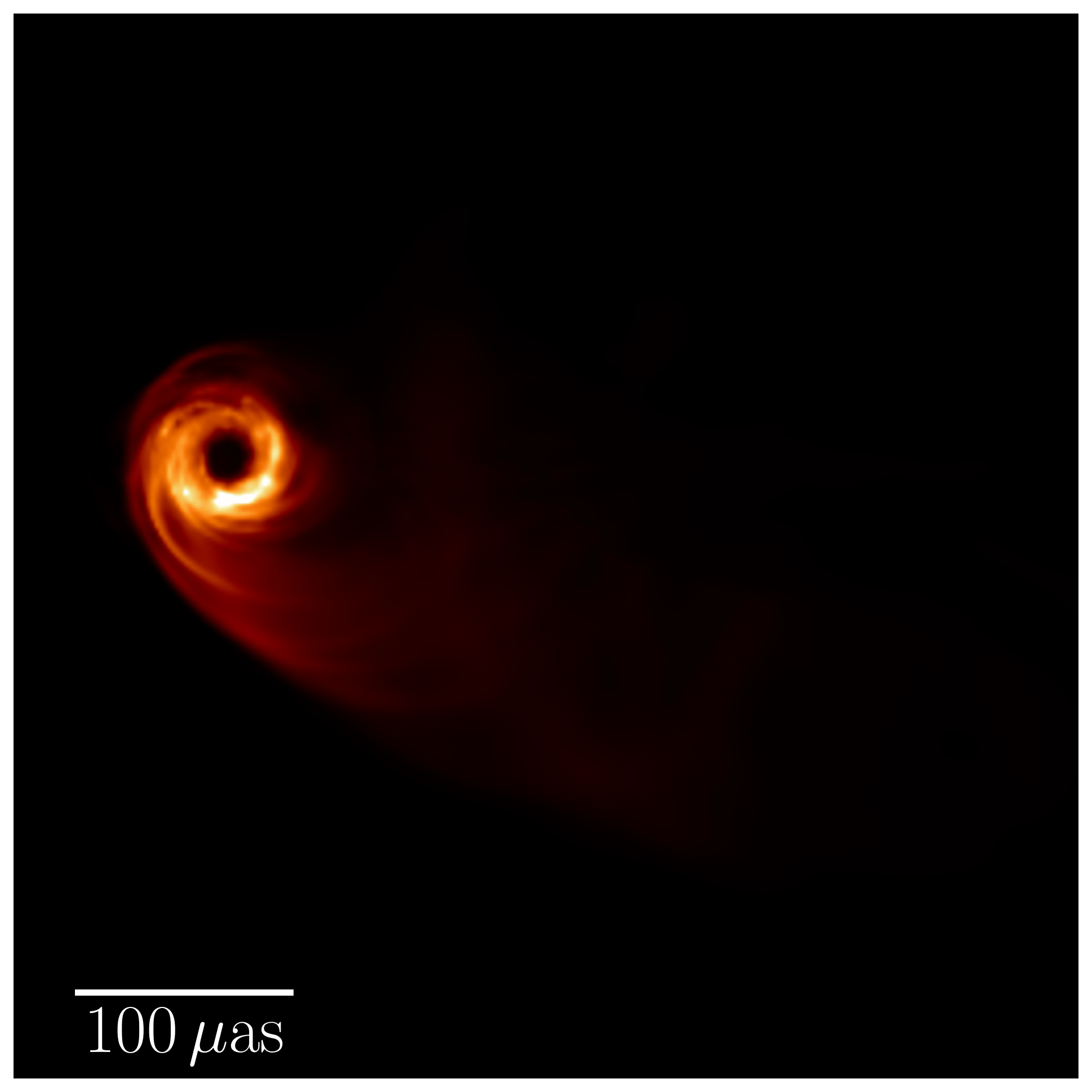} & \hspace{-0.7cm} \includegraphics[width=0.235\linewidth]{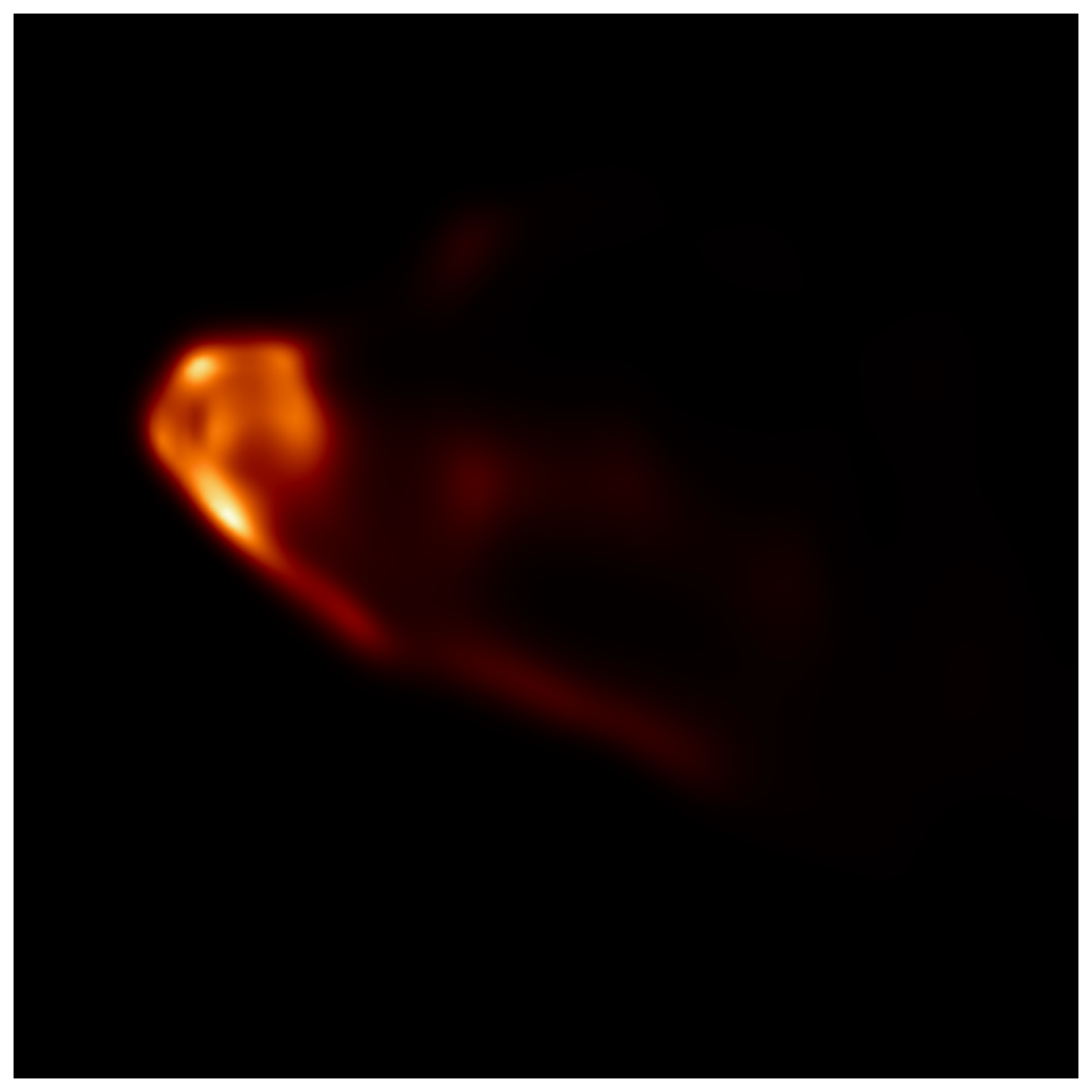} & \hspace{-0.7cm}
 \includegraphics[width=0.235\linewidth]{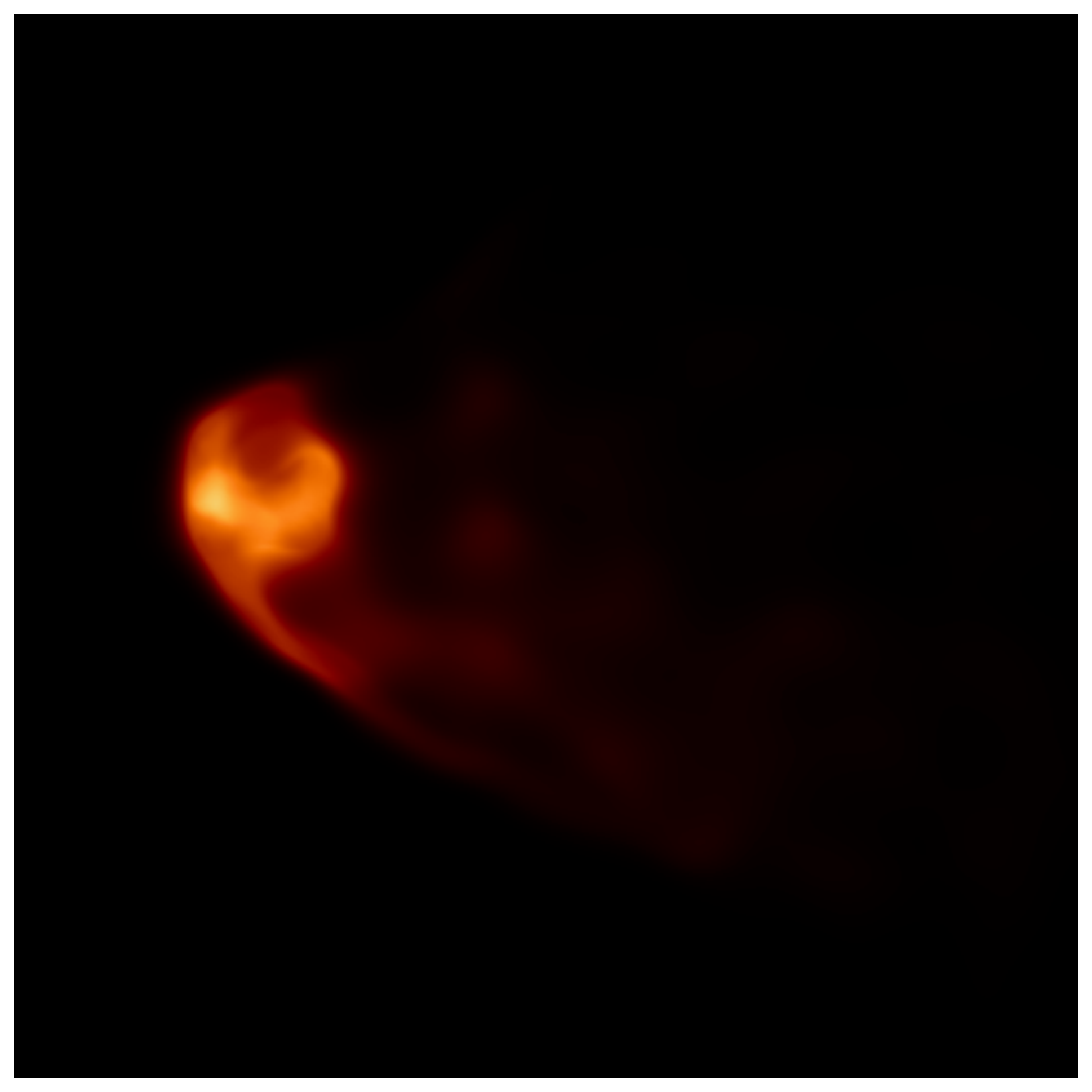} &\hspace{-0.47cm}
    \raisebox{-0.01\linewidth}[0pt][0pt]{\includegraphics[width=0.285\linewidth]{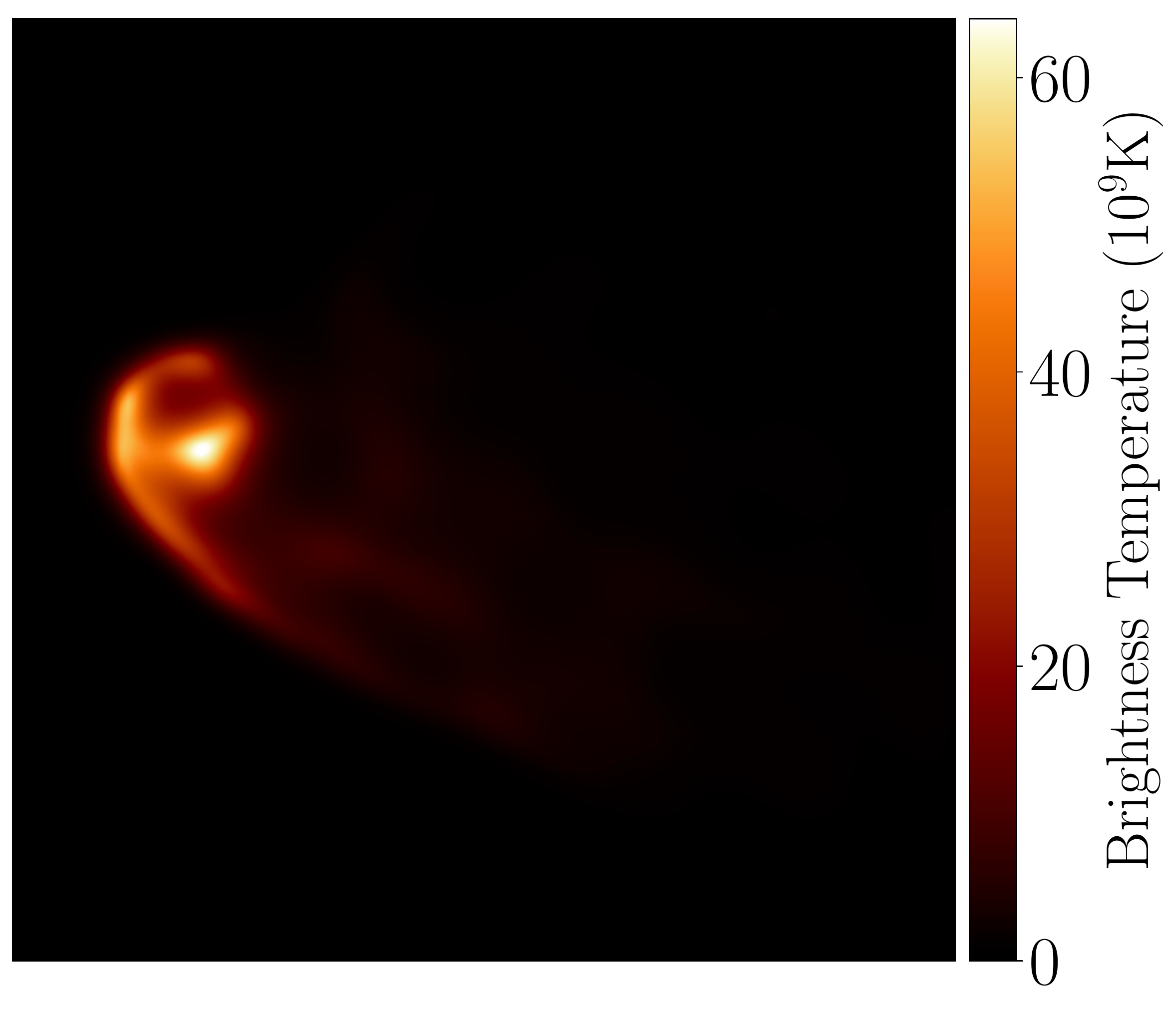}} 
    \end{tabular}\end{adjustwidth}
    \caption{Demonstration of interoperability with ngVLA using simulated observations of the M87 jet at 86\,GHz. The columns from left to right are: the ground truth image; the reconstruction with the ngVLA only; the reconstruction with the full ngEHT array; and the reconstruction with the full ngEHT array combined with the ngVLA. Top: Images plotted in square-root scale to emphasize the jet reconstruction; the contours are spaced logarithmically, starting at 1\% of the peak value and increasing by factors of 2. Bottom: Images on a linear scale to emphasize the shadow reconstruction. The sensitivity of the ngVLA improves the dynamic range of the ngEHT, and the ngEHT boosts the resolution of the ngVLA, enabling imaging of both the shadow and the jet at 86\,GHz with \mbox{high fidelity.}}
    \label{fig:ngVLAdemo}
\end{figure}

We also prepared a demonstration of the imaging capabilities with the ngVLA, which is shown in Figure \ref{fig:ngVLAdemo}. We used the long-baseline ngVLA array configuration (LBA), which consists of 30 18-meter dishes on 10 sites across the United States. Co-located stations were modeled as a single site with properly scaled sensitivity, and in addition, we modeled the ngVLA core in New Mexico as a single highly sensitive site (with an SEFD of 10 Jy). We used the underlying model for M87 shown in Section \ref{sec:drivers}, and obtained reconstructions at 86\,GHz with the ngVLA alone, with the ngEHT+EHT combined array (here acting as a high-sensitivity alternative to the GMVA), and with the ngEHT+EHT+ngVLA combination. We notice that the ngVLA greatly improves the dynamic range for higher-fidelity reconstructions of the jet, whereas the ngEHT+EHT increases the resolution of the reconstruction and enables the imaging of the central brightness depression related to the black-hole shadow. Owing to the increased optical depth at 86 GHz compared to, e.g., 230 and 345 GHz (see also \mbox{Figure \ref{fig:m87model}}), the appearance of this central brightness depression probes the "inner shadow" rather than the photon ring, which provides opportunities for measuring accretion flow and black hole properties \citep{Chael2021}. While the optical depth of M87 at 86\,GHz is uncertain, our simulation results inform the potential for long-term monitoring with the ngEHT+EHT+ngVLA at 86\,GHz to study dynamics of both the (inner) shadow and the jet. 

\vspace{-6pt}

\section{Summary and Conclusions}\label{sec:Summary}

We argue that supplementing the ngEHT with 86\,GHz observing capabilities---particularly simultaneous multi-frequency capabilities at 86, 230, and 345\,GHz---would improve the overall performance and flexibility of the array, helping it to achieve its primary scientific goals of high-dynamic-range images and movies of the M87 jet base region and the evolving accretion flow around Sgr A*.

One of the main benefits afforded by the addition of 86\,GHz capabilities to the ngEHT would be the ability to carry out agile (i.e., rapid response or target of opportunity) and/or year-round observations.  The primary observing frequencies of the ngEHT (230 and 345\,GHz) require very stable atmospheric conditions, which renders consistent and agile observing difficult.  We have shown that observing at 86\,GHz should be possible year-round, as there are reliable weather conditions at all sites, allowing for more flexibility to observe transient sources and other targets.

Through multi-frequency imaging, 86\,GHz observations can be combined with 230/345 GHz observations to permit imaging of the M87 jet structure with high fidelity.  A standalone ngEHT array (i.e., without the addition of EHT or other sites) can only reliably reconstruct the M87 shadow and jet if 86\,GHz information is present in multi-frequency imaging.  Observations at 86\,GHz would thus open up a significant fraction of time in which some or all of the EHT sites may not be available, but core ngEHT science \mbox{remains achievable.}

Simultaneous observations at 86\,GHz, together with 230/345\,GHz, are a practical requirement for absolute phase calibration and astrometry.  Observing at 86\,GHz would permit the ngEHT to connect with a well-established astrometric network of facilities around the world.  Furthermore, simultaneous multi-frequency observing capabilities would enable full-array enhancements in calibration and sensitivity via multi-frequency phase reference techniques.  By transferring the simultaneously measured phases at 86\,GHz to the higher frequencies, the effective coherence times at 230 and 345\,GHz could be increased from a few seconds to tens of minutes (and perhaps even longer).

In addition to simultaneous observations with the higher frequency bands, standalone 86\,GHz capabilities of the ngEHT could also be used communally and to connect with other facilities, such as the currently operating GMVA and the upcoming ngVLA.  Joint observations with the ngEHT and ngVLA at 86\,GHz would leverage their substantial complementarity; the ngVLA provides sensitivity and the ngEHT provides angular resolution.  We demonstrate that if the accretion flow in M87 is sufficiently optically thin at 86\,GHz, then joint observations with the ngEHT and ngVLA could image both the horizon-scale shadow structure and the extended jet emission.
\vspace{6pt}

\authorcontributions{Conceptualization, S.S.D., S.I., D.W.P., and F.R.; methodology, R.A., L.B., A.C., R.D., S.I., M.D.J., D.W.P., M.J.R., and F.R.; software, R.A., A.C., S.I., D.W.P., A.W.R., and F.R.; writing---original draft preparation, A.C., S.I., D.W.P., and F.R.; writing---review and editing, K.A., R.A., L.B., A.C., R.D., and S.S.D., V.L.F., G.F., S.I., M.D.J., G.N., D.W.P., A.W.R., M.J.R., F.R., and  R.P.J.T.; visualization, A.C., S.I., D.W.P., and  F.R.; supervision, S.S.D.; project administration, S.S.D. and  G.F. All authors have read and agreed to the published version of the manuscript.}

\funding{Support for this work was provided by the NSF through grants AST-1440254, AST-1935980, and AST-2034306; and by the Gordon and Betty Moore Foundation through grant GBMF-10423. This work has been supported in part by the Black Hole Initiative at Harvard University, which is funded by grants from the John Templeton Foundation and the Gordon and Betty Moore Foundation to Harvard University. S.I. is supported by the NASA Hubble Fellowship grant HST-HF2-51482.001-A awarded by the Space Telescope Science Institute, which is operated by the Association of Universities for Research in Astronomy, Inc., for NASA, under contract NAS5-26555. A.C. was supported by the NASA Hubble Fellowship grant HST-HF2-51431.001-A awarded by the Space Telescope Science Institute, which is operated by the Association of Universities for Research in Astronomy, Inc., for NASA, under contract NAS5-26555. K.A. has been also financially supported by following NSF grants: OMA-2029670, AST-1614868, AST-2107681, AST-2132700.}

\dataavailability{ No observational data were used in this publication.} 
\conflictsofinterest{The authors declare no conflict of interest.}
\begin{adjustwidth}{-\extralength}{0cm}

\reftitle{References}

\PublishersNote{}
\end{adjustwidth}

\begin{thebibliography}{999}

\bibitem[{Event Horizon Telescope Collaboration}
  \em{et~al.}(2019{\natexlab{a}}){Event Horizon Telescope Collaboration},
  {Akiyama}, {Alberdi}, {Alef}, {Asada}, {Azulay}, {Baczko}, {Ball},
  {Balokovi{\'c}}, {Barrett}, and et~al.]{M87PaperI}
{Event Horizon Telescope Collaboration}, et~al.
\newblock {First M87 Event Horizon Telescope Results. I. The Shadow of the
  Supermassive Black Hole}.
\newblock {\em  Astrophys. J. Lett.} {\bf 2019}, {\em 875},~L1.
\newblock {{https://doi.org/10.3847/2041-8213/ab0ec7}}.

\bibitem[{Event Horizon Telescope Collaboration}
  \em{et~al.}(2019{\natexlab{b}}){Event Horizon Telescope Collaboration},
  {Akiyama}, {Alberdi}, {Alef}, {Asada}, {Azulay}, {Baczko}, {Ball},
  {Balokovi{\'c}}, {Barrett}, and et~al.]{M87PaperII}
{Event Horizon Telescope Collaboration}, et~al.
\newblock {First M87 Event Horizon Telescope Results. II. Array and
  Instrumentation}.
\newblock {\em Astrophys. J. Lett.} {\bf 2019}, {\em 875},~L2.
\newblock {{https://doi.org/10.3847/2041-8213/ab0c96}}.

\bibitem[{Event Horizon Telescope Collaboration}
  \em{et~al.}(2019{\natexlab{c}}){Event Horizon Telescope Collaboration},
  {Akiyama}, {Alberdi}, {Alef}, {Asada}, {Azulay}, {Baczko}, {Ball},
  {Balokovi{\'c}}, {Barrett}, and et~al.]{M87PaperIII}
{Event Horizon Telescope Collaboration}, et~al.
\newblock {First M87 Event Horizon Telescope Results. III. Data Processing and
  Calibration}.
\newblock {\em Astrophys. J. Lett.} {\bf 2019}, {\em 875},~L3.
\newblock {{https://doi.org/10.3847/2041-8213/ab0c57}}.

\bibitem[{Event Horizon Telescope Collaboration}
  \em{et~al.}(2019{\natexlab{d}}){Event Horizon Telescope Collaboration},
  {Akiyama}, {Alberdi}, {Alef}, {Asada}, {Azulay}, {Baczko}, {Ball},
  {Balokovi{\'c}}, {Barrett}, and et~al.]{M87PaperIV}
{Event Horizon Telescope Collaboration}, et~al.
\newblock {First M87 Event Horizon Telescope Results. IV. Imaging the Central
  Supermassive Black Hole}.
\newblock {\em Astrophys. J. Lett.} {\bf 2019}, {\em 875},~L4.
\newblock {{https://doi.org/10.3847/2041-8213/ab0e85}}.

\bibitem[{Event Horizon Telescope Collaboration}
  \em{et~al.}(2019{\natexlab{e}}){Event Horizon Telescope Collaboration},
  {Akiyama}, {Alberdi}, {Alef}, {Asada}, {Azulay}, {Baczko}, {Ball},
  {Balokovi{\'c}}, {Barrett}, and et~al.]{M87PaperV}
{Event Horizon Telescope Collaboration}, et~al. 
\newblock {First M87 Event Horizon Telescope Results. V. Physical Origin of the
  Asymmetric Ring}.
\newblock {\em Astrophys. J. Lett.} {\bf 2019}, {\em 875},~L5.
\newblock {{https://doi.org/10.3847/2041-8213/ab0f43}}.

\bibitem[{Event Horizon Telescope Collaboration}
  \em{et~al.}(2019{\natexlab{f}}){Event Horizon Telescope Collaboration},
  {Akiyama}, {Alberdi}, {Alef}, {Asada}, {Azulay}, {Baczko}, {Ball},
  {Balokovi{\'c}}, {Barrett}, and et~al.]{M87PaperVI}
{Event Horizon Telescope Collaboration}, et~al.
\newblock {First M87 Event Horizon Telescope Results. VI. The Shadow and Mass
  of the Central Black Hole}.
\newblock {\em Astrophys. J. Lett.} {\bf 2019}, {\em 875},~L6.
\newblock {{https://doi.org/10.3847/2041-8213/ab1141}}.

\bibitem[{Event Horizon Telescope Collaboration}
  \em{et~al.}(2021{\natexlab{a}}){Event Horizon Telescope Collaboration},
  {Akiyama}, {Algaba}, {Alberdi}, {Alef}, {Anantua}, {Asada}, {Azulay},
  {Baczko}, {Ball}, {Balokovi{\'c}}, {Barrett}, {Benson}, {Bintley},
  {Blackburn}, {Blundell}, {Boland}, {Bouman}, {Bower}, {Boyce}, {Bremer},
  {Brinkerink}, {Brissenden}, {Britzen}, {Broderick}, {Broguiere}, {Bronzwaer},
  {Byun}, {Carlstrom}, {Chael}, {Chan}, {Chatterjee}, {Chatterjee}, {Chen},
  {Chen}, {Chesler}, {Cho}, {Christian}, {Conway}, {Cordes}, {Crawford},
  {Crew}, {Cruz-Osorio}, {Cui}, {Davelaar}, {De Laurentis}, {Deane}, {Dempsey},
  {Desvignes}, {Dexter}, {Doeleman}, {Eatough}, {Falcke}, {Farah}, {Fish},
  {Fomalont}, {Ford}, {Fraga-Encinas}, {Freeman}, {Friberg}, {Fromm},
  {Fuentes}, {Galison}, {Gammie}, {Garc{\'\i}a}, {Gentaz}, {Georgiev}, {Goddi},
  {Gold}, {G{\'o}mez}, {G{\'o}mez-Ruiz}, {Gu}, {Gurwell}, {Hada}, {Haggard},
  {Hecht}, {Hesper}, {Ho}, {Ho}, {Honma}, {Huang}, {Huang}, {Hughes}, {Ikeda},
  {Inoue}, {Issaoun}, {James}, {Jannuzi}, {Janssen}, {Jeter}, {Jiang},
  {Jimenez-Rosales}, {Johnson}, {Jorstad}, {Jung}, {Karami}, {Karuppusamy},
  {Kawashima}, {Keating}, {Kettenis}, {Kim}, {Kim}, {Kim}, {Kim}, {Kino},
  {Koay}, {Kofuji}, {Koch}, {Koyama}, {Kramer}, {Kramer}, {Krichbaum}, {Kuo},
  {Lauer}, {Lee}, {Levis}, {Li}, {Li}, {Lindqvist}, {Lico}, {Lindahl}, {Liu},
  {Liu}, {Liuzzo}, {Lo}, {Lobanov}, {Loinard}, {Lonsdale}, {Lu}, {MacDonald},
  {Mao}, {Marchili}, {Markoff}, {Marrone}, {Marscher}, {Mart{\'\i}-Vidal},
  {Matsushita}, {Matthews}, {Medeiros}, {Menten}, {Mizuno}, {Mizuno}, {Moran},
  {Moriyama}, {Moscibrodzka}, {M{\"u}ller}, {Musoke}, {Mej{\'\i}as},
  {Michalik}, {Nadolski}, {Nagai}, {Nagar}, {Nakamura}, {Narayan}, {Narayanan},
  {Natarajan}, {Nathanail}, {Neilsen}, {Neri}, {Ni}, {Noutsos}, {Nowak},
  {Okino}, {Olivares}, {Ortiz-Le{\'o}n}, {Oyama}, {{\"O}zel}, {Palumbo},
  {Park}, {Patel}, {Pen}, {Pesce}, {Pi{\'e}tu}, {Plambeck}, {PopStefanija},
  {Porth}, {P{\"o}tzl}, {Prather}, {Preciado-L{\'o}pez}, {Psaltis}, {Pu},
  {Ramakrishnan}, {Rao}, {Rawlings}, {Raymond}, {Rezzolla}, {Ricarte},
  {Ripperda}, {Roelofs}, {Rogers}, {Ros}, {Rose}, {Roshanineshat}, {Rottmann},
  {Roy}, {Ruszczyk}, {Rygl}, {S{\'a}nchez}, {S{\'a}nchez-Arguelles}, {Sasada},
  {Savolainen}, {Schloerb}, {Schuster}, {Shao}, {Shen}, {Small}, {Sohn},
  {SooHoo}, {Sun}, {Tazaki}, {Tetarenko}, {Tiede}, {Tilanus}, {Titus}, {Toma},
  {Torne}, {Trent}, {Traianou}, {Trippe}, {van Bemmel}, {van Langevelde}, {van
  Rossum}, {Wagner}, {Ward-Thompson}, {Wardle}, {Weintroub}, {Wex}, {Wharton},
  {Wielgus}, {Wong}, {Wu}, {Yoon}, {Young}, {Young}, {Younsi}, {Yuan}, {Yuan},
  {Zensus}, {Zhao}, and {Zhao}]{M87PaperVII}
{Event Horizon Telescope Collaboration}, et~al. 
\newblock {First M87 Event Horizon Telescope Results. VII. Polarization of the
  Ring}.
\newblock {\em Astrophys. J. Lett.} {\bf 2021}, {\em 910},~L12.
\newblock {{https://doi.org/10.3847/2041-8213/abe71d}}.

\bibitem[{Event Horizon Telescope Collaboration}
  \em{et~al.}(2021{\natexlab{b}}){Event Horizon Telescope Collaboration},
  {Akiyama}, {Algaba}, {Alberdi}, {Alef}, {Anantua}, {Asada}, {Azulay},
  {Baczko}, {Ball}, {Balokovi{\'c}}, {Barrett}, {Benson}, {Bintley},
  {Blackburn}, {Blundell}, {Boland}, {Bouman}, {Bower}, {Boyce}, {Bremer},
  {Brinkerink}, {Brissenden}, {Britzen}, {Broderick}, {Broguiere}, {Bronzwaer},
  {Byun}, {Carlstrom}, {Chael}, {Chan}, {Chatterjee}, {Chatterjee}, {Chen},
  {Chen}, {Chesler}, {Cho}, {Christian}, {Conway}, {Cordes}, {Crawford},
  {Crew}, {Cruz-Osorio}, {Cui}, {Davelaar}, {De Laurentis}, {Deane}, {Dempsey},
  {Desvignes}, {Dexter}, {Doeleman}, {Eatough}, {Falcke}, {Farah}, {Fish},
  {Fomalont}, {Ford}, {Fraga-Encinas}, {Friberg}, {Fromm}, {Fuentes},
  {Galison}, {Gammie}, {Garc{\'\i}a}, {Gelles}, {Gentaz}, {Georgiev}, {Goddi},
  {Gold}, {G{\'o}mez}, {G{\'o}mez-Ruiz}, {Gu}, {Gurwell}, {Hada}, {Haggard},
  {Hecht}, {Hesper}, {Himwich}, {Ho}, {Ho}, {Honma}, {Huang}, {Huang},
  {Hughes}, {Ikeda}, {Inoue}, {Issaoun}, {James}, {Jannuzi}, {Janssen},
  {Jeter}, {Jiang}, {Jimenez-Rosales}, {Johnson}, {Jorstad}, {Jung}, {Karami},
  {Karuppusamy}, {Kawashima}, {Keating}, {Kettenis}, {Kim}, {Kim}, {Kim},
  {Kim}, {Kino}, {Koay}, {Kofuji}, {Koch}, {Koyama}, {Kramer}, {Kramer},
  {Krichbaum}, {Kuo}, {Lauer}, {Lee}, {Levis}, {Li}, {Li}, {Lindqvist}, {Lico},
  {Lindahl}, {Liu}, {Liu}, {Liuzzo}, {Lo}, {Lobanov}, {Loinard}, {Lonsdale},
  {Lu}, {MacDonald}, {Mao}, {Marchili}, {Markoff}, {Marrone}, {Marscher},
  {Mart{\'\i}-Vidal}, {Matsushita}, {Matthews}, {Medeiros}, {Menten}, {Mizuno},
  {Mizuno}, {Moran}, {Moriyama}, {Moscibrodzka}, {M{\"u}ller}, {Musoke}, {Mus
  Mej{\'\i}as}, {Michalik}, {Nadolski}, {Nagai}, {Nagar}, {Nakamura},
  {Narayan}, {Narayanan}, {Natarajan}, {Nathanail}, {Neilsen}, {Neri}, {Ni},
  {Noutsos}, {Nowak}, {Okino}, {Olivares}, {Ortiz-Le{\'o}n}, {Oyama},
  {{\"O}zel}, {Palumbo}, {Park}, {Patel}, {Pen}, {Pesce}, {Pi{\'e}tu},
  {Plambeck}, {PopStefanija}, {Porth}, {P{\"o}tzl}, {Prather},
  {Preciado-L{\'o}pez}, {Psaltis}, {Pu}, {Ramakrishnan}, {Rao}, {Rawlings},
  {Raymond}, {Rezzolla}, {Ricarte}, {Ripperda}, {Roelofs}, {Rogers}, {Ros},
  {Rose}, {Roshanineshat}, {Rottmann}, {Roy}, {Ruszczyk}, {Rygl},
  {S{\'a}nchez}, {S{\'a}nchez-Arguelles}, {Sasada}, {Savolainen}, {Schloerb},
  {Schuster}, {Shao}, {Shen}, {Small}, {Sohn}, {SooHoo}, {Sun}, {Tazaki},
  {Tetarenko}, {Tiede}, {Tilanus}, {Titus}, {Toma}, {Torne}, {Trent},
  {Traianou}, {Trippe}, {van Bemmel}, {van Langevelde}, {van Rossum}, {Wagner},
  {Ward-Thompson}, {Wardle}, {Weintroub}, {Wex}, {Wharton}, {Wielgus}, {Wong},
  {Wu}, {Yoon}, {Young}, {Young}, {Younsi}, {Yuan}, {Yuan}, {Zensus}, {Zhao},
  and {Zhao}]{M87PaperVIII}
{Event Horizon Telescope Collaboration}, et~al.
\newblock {First M87 Event Horizon Telescope Results. VIII. Magnetic Field
  Structure near The Event Horizon}.
\newblock {\em Astrophys. J. Lett.} {\bf 2021}, {\em 910},~L13.
\newblock {{https://doi.org/10.3847/2041-8213/abe4de}}.

\bibitem[{Event Horizon Telescope Collaboration}
  \em{et~al.}(2022{\natexlab{a}}){Event Horizon Telescope Collaboration},
  {Akiyama}, {Alberdi}, {Alef}, {Algaba}, {Anantua}, {Asada}, {Azulay}, {Bach},
  {Baczko}, {Ball}, {Balokovi{\'c}}, {Barrett}, {Baub{\"o}ck}, {Benson},
  {Bintley}, {Blackburn}, {Blundell}, {Bouman}, {Bower}, {Boyce}, {Bremer},
  {Brinkerink}, {Brissenden}, {Britzen}, {Broderick}, {Broguiere}, {Bronzwaer},
  {Bustamante}, {Byun}, {Carlstrom}, {Ceccobello}, {Chael}, {Chan},
  {Chatterjee}, {Chatterjee}, {Chen}, {Chen}, {Cheng}, {Cho}, {Christian},
  {Conroy}, {Conway}, {Cordes}, {Crawford}, {Crew}, {Cruz-Osorio}, {Cui},
  {Davelaar}, {Laurentis}, {Deane}, {Dempsey}, {Desvignes}, {Dexter}, {Dhruv},
  {Doeleman}, {Dougal}, {Dzib}, {Eatough}, {Emami}, {Falcke}, {Farah}, {Fish},
  {Fomalont}, {Ford}, {Fraga-Encinas}, {Freeman}, {Friberg}, {Fromm},
  {Fuentes}, {Galison}, {Gammie}, {Garc{\'\i}a}, {Gentaz}, {Georgiev}, {Goddi},
  {Gold}, {G{\'o}mez-Ruiz}, {G{\'o}mez}, {Gu}, {Gurwell}, {Hada}, {Haggard},
  {Haworth}, {Hecht}, {Hesper}, {Heumann}, {Ho}, {Ho}, {Honma}, {Huang},
  {Huang}, {Hughes}, {Ikeda}, {Impellizzeri}, {Inoue}, {Issaoun}, {James},
  {Jannuzi}, {Janssen}, {Jeter}, {Jiang}, {Jim{\'e}nez-Rosales}, {Johnson},
  {Jorstad}, {Joshi}, {Jung}, {Karami}, {Karuppusamy}, {Kawashima}, {Keating},
  {Kettenis}, {Kim}, {Kim}, {Kim}, {Kim}, {Kino}, {Koay}, {Kocherlakota},
  {Kofuji}, {Koch}, {Koyama}, {Kramer}, {Kramer}, {Krichbaum}, {Kuo}, {Bella},
  {Lauer}, {Lee}, {Lee}, {Leung}, {Levis}, {Li}, {Lico}, {Lindahl},
  {Lindqvist}, {Lisakov}, {Liu}, {Liu}, {Liuzzo}, {Lo}, {Lobanov}, {Loinard},
  {Lonsdale}, {Lu}, {Mao}, {Marchili}, {Markoff}, {Marrone}, {Marscher},
  {Mart{\'\i}-Vidal}, {Matsushita}, {Matthews}, {Medeiros}, {Menten},
  {Michalik}, {Mizuno}, {Mizuno}, {Moran}, {Moriyama}, {Moscibrodzka},
  {M{\"u}ller}, {Mus}, {Musoke}, {Myserlis}, {Nadolski}, {Nagai}, {Nagar},
  {Nakamura}, {Narayan}, {Narayanan}, {Natarajan}, {Nathanail}, {Fuentes},
  {Neilsen}, {Neri}, {Ni}, {Noutsos}, {Nowak}, {Oh}, {Okino}, {Olivares},
  {Ortiz-Le{\'o}n}, {Oyama}, {{\"O}zel}, {Palumbo}, {Paraschos}, {Park},
  {Parsons}, {Patel}, {Pen}, {Pesce}, {Pi{\'e}tu}, {Plambeck}, {PopStefanija},
  {Porth}, {P{\"o}tzl}, {Prather}, {Preciado-L{\'o}pez}, {Psaltis}, {Pu},
  {Ramakrishnan}, {Rao}, {Rawlings}, {Raymond}, {Rezzolla}, {Ricarte},
  {Ripperda}, {Roelofs}, {Rogers}, {Ros}, {Romero-Ca{\~n}izales},
  {Roshanineshat}, {Rottmann}, {Roy}, {Ruiz}, {Ruszczyk}, {Rygl},
  {S{\'a}nchez}, {S{\'a}nchez-Arg{\"u}elles}, {S{\'a}nchez-Portal}, {Sasada},
  {Satapathy}, {Savolainen}, {Schloerb}, {Schonfeld}, {Schuster}, {Shao},
  {Shen}, {Small}, {Sohn}, {SooHoo}, {Souccar}, {Sun}, {Tazaki}, {Tetarenko},
  {Tiede}, {Tilanus}, {Titus}, {Torne}, {Traianou}, {Trent}, {Trippe}, {Turk},
  {van Bemmel}, {van Langevelde}, {van Rossum}, {Vos}, {Wagner},
  {Ward-Thompson}, {Wardle}, {Weintroub}, {Wex}, {Wharton}, {Wielgus}, {Wiik},
  {Witzel}, {Wondrak}, {Wong}, {Wu}, {Yamaguchi}, {Yoon}, {Young}, {Young},
  {Younsi}, {Yuan}, {Yuan}, {Zensus}, {Zhang}, {Zhao}, {Zhao}, {Agurto},
  {Allardi}, {Amestica}, {Araneda}, {Arriagada}, {Berghuis}, {Bertarini},
  {Berthold}, {Blanchard}, {Brown}, {C{\'a}rdenas}, {Cantzler}, {Caro},
  {Castillo-Dom{\'\i}nguez}, {Chan}, {Chang}, {Chang}, {Chang}, {Chang},
  {Chen}, {Chilson}, {Chuter}, {Ciechanowicz}, {Colin-Beltran}, {Coulson},
  {Crowley}, {Degenaar}, {Dornbusch}, {Dur{\'a}n}, {Everett}, {Faber},
  {Forster}, {Fuchs}, {Gale}, {Geertsema}, {Gonz{\'a}lez}, {Graham}, {Gueth},
  {Halverson}, {Han}, {Han}, {Hasegawa}, {Hern{\'a}ndez-Rebollar}, {Herrera},
  {Herrero-Illana}, {Heyminck}, {Hirota}, {Hoge}, {Hostler Schimpf}, {Howie},
  {Huang}, {Jiang}, {Jinchi}, {John}, {Kimura}, {Klein}, {Kubo}, {Kuroda},
  {Kwon}, {Lacasse}, {Laing}, {Leitch}, {Li}, {Liu}, {Liu}, {Lin}, {Lu},
  {Mac-Auliffe}, {Martin-Cocher}, {Matulonis}, {Maute}, {Messias},
  {Meyer-Zhao}, {Monta{\~n}a}, {Montenegro-Montes}, {Montgomerie}, {Moreno
  Nolasco}, {Muders}, {Nishioka}, {Norton}, {Nystrom}, {Ogawa}, {Olivares},
  {Oshiro}, {P{\'e}rez-Beaupuits}, {Parra}, {Phillips}, {Poirier}, {Pradel},
  {Qiu}, {Raffin}, {Rahlin}, {Ram{\'\i}rez}, {Ressler}, {Reynolds},
  {Rodr{\'\i}guez-Montoya}, {Saez-Madain}, {Santana}, {Shaw}, {Shirkey},
  {Silva}, {Snow}, {Sousa}, {Sridharan}, {Stahm}, {Stark}, {Test},
  {Torstensson}, {Venegas}, {Walther}, {Wei}, {White}, {Wieching}, {Wijnands},
  {Wouterloot}, {Yu}, {Yu (于威)}, and {Zeballos}]{SgrAPaperI}
{Event Horizon Telescope Collaboration}, et~al. 
\newblock {First Sagittarius A* Event Horizon Telescope Results. I. The Shadow
  of the Supermassive Black Hole in the Center of the Milky Way}.
\newblock {\em Astrophys. J. Lett.} {\bf 2022}, {\em 930},~L12.
\newblock {{https://doi.org/10.3847/2041-8213/ac6674}}.

\bibitem[{Event Horizon Telescope Collaboration}
  \em{et~al.}(2022{\natexlab{b}}){Event Horizon Telescope Collaboration},
  {Akiyama}, {Alberdi}, {Alef}, {Algaba}, {Anantua}, {Asada}, {Azulay}, {Bach},
  {Baczko}, {Ball}, {Balokovi{\'c}}, {Barrett}, {Baub{\"o}ck}, {Benson},
  {Bintley}, {Blackburn}, {Blundell}, {Bouman}, {Bower}, {Boyce}, {Bremer},
  {Brinkerink}, {Brissenden}, {Britzen}, {Broderick}, {Broguiere}, {Bronzwaer},
  {Bustamante}, {Byun}, {Carlstrom}, {Ceccobello}, {Chael}, {Chan},
  {Chatterjee}, {Chatterjee}, {Chen}, {Chen}, {Cheng}, {Cho}, {Christian},
  {Conroy}, {Conway}, {Cordes}, {Crawford}, {Crew}, {Cruz-Osorio}, {Cui},
  {Davelaar}, {De Laurentis}, {Deane}, {Dempsey}, {Desvignes}, {Dexter},
  {Dhruv}, {Doeleman}, {Dougal}, {Dzib}, {Eatough}, {Emami}, {Falcke}, {Farah},
  {Fish}, {Fomalont}, {Ford}, {Fraga-Encinas}, {Freeman}, {Friberg}, {Fromm},
  {Fuentes}, {Galison}, {Gammie}, {Garc{\'\i}a}, {Gentaz}, {Georgiev}, {Goddi},
  {Gold}, {G{\'o}mez-Ruiz}, {G{\'o}mez}, {Gu}, {Gurwell}, {Hada}, {Haggard},
  {Haworth}, {Hecht}, {Hesper}, {Heumann}, {Ho}, {Ho}, {Honma}, {Huang},
  {Huang}, {Hughes}, {Ikeda}, {Impellizzeri}, {Inoue}, {Issaoun}, {James},
  {Jannuzi}, {Janssen}, {Jeter}, {Jiang}, {Jim{\'e}nez-Rosales}, {Johnson},
  {Jorstad}, {Joshi}, {Jung}, {Karami}, {Karuppusamy}, {Kawashima}, {Keating},
  {Kettenis}, {Kim}, {Kim}, {Kim}, {Kim}, {Kino}, {Koay}, {Kocherlakota},
  {Kofuji}, {Koch}, {Koyama}, {Kramer}, {Kramer}, {Krichbaum}, {Kuo}, {Bella},
  {Lauer}, {Lee}, {Lee}, {Leung}, {Levis}, {Li}, {Lico}, {Lindahl},
  {Lindqvist}, {Lisakov}, {Liu}, {Liu}, {Liuzzo}, {Lo}, {Lobanov}, {Loinard},
  {Lonsdale}, {Lu}, {Mao}, {Marchili}, {Markoff}, {Marrone}, {Marscher},
  {Mart{\'\i}-Vidal}, {Matsushita}, {Matthews}, {Medeiros}, {Menten},
  {Michalik}, {Mizuno}, {Mizuno}, {Moran}, {Moriyama}, {Moscibrodzka},
  {M{\"u}ller}, {Mus}, {Musoke}, {Myserlis}, {Nadolski}, {Nagai}, {Nagar},
  {Nakamura}, {Narayan}, {Narayanan}, {Natarajan}, {Nathanail}, {Fuentes},
  {Neilsen}, {Neri}, {Ni}, {Noutsos}, {Nowak}, {Oh}, {Okino}, {Olivares},
  {Ortiz-Le{\'o}n}, {Oyama}, {{\"O}zel}, {Palumbo}, {Paraschos}, {Park},
  {Parsons}, {Patel}, {Pen}, {Pesce}, {Pi{\'e}tu}, {Plambeck}, {PopStefanija},
  {Porth}, {P{\"o}tzl}, {Prather}, {Preciado-L{\'o}pez}, {Psaltis}, {Pu},
  {Ramakrishnan}, {Rao}, {Rawlings}, {Raymond}, {Rezzolla}, {Ricarte},
  {Ripperda}, {Roelofs}, {Rogers}, {Ros}, {Romero-Ca{\~n}izales},
  {Roshanineshat}, {Rottmann}, {Roy}, {Ruiz}, {Ruszczyk}, {Rygl},
  {S{\'a}nchez}, {S{\'a}nchez-Arg{\"u}elles}, {S{\'a}nchez-Portal}, {Sasada},
  {Satapathy}, {Savolainen}, {Schloerb}, {Schonfeld}, {Schuster}, {Shao},
  {Shen}, {Small}, {Sohn}, {SooHoo}, {Souccar}, {Sun}, {Tazaki}, {Tetarenko},
  {Tiede}, {Tilanus}, {Titus}, {Torne}, {Traianou}, {Trent}, {Trippe}, {Turk},
  {van Bemmel}, {van Langevelde}, {van Rossum}, {Vos}, {Wagner},
  {Ward-Thompson}, {Wardle}, {Weintroub}, {Wex}, {Wharton}, {Wielgus}, {Wiik},
  {Witzel}, {Wondrak}, {Wong}, {Wu}, {Yamaguchi}, {Yoon}, {Young}, {Young},
  {Younsi}, {Yuan}, {Yuan}, {Zensus}, {Zhang}, {Zhao}, {Zhao}, {Agurto},
  {Araneda}, {Arriagada}, {Bertarini}, {Berthold}, {Blanchard}, {Brown},
  {C{\'a}rdenas}, {Cantzler}, {Caro}, {Chuter}, {Ciechanowicz}, {Coulson},
  {Crowley}, {Degenaar}, {Dornbusch}, {Dur{\'a}n}, {Forster}, {Geertsema},
  {Gonz{\'a}lez}, {Graham}, {Gueth}, {Han}, {Herrera}, {Herrero-Illana},
  {Heyminck}, {Hoge}, {Huang}, {Jiang}, {John}, {Klein}, {Kubo}, {Kuroda},
  {Kwon}, {Laing}, {Liu}, {Liu}, {Mac-Auliffe}, {Martin-Cocher}, {Matulonis},
  {Messias}, {Meyer-Zhao}, {Montenegro-Montes}, {Montgomerie}, {Muders},
  {Nishioka}, {Norton}, {Olivares}, {P{\'e}rez-Beaupuits}, {Parra}, {Poirier},
  {Pradel}, {Raffin}, {Ram{\'\i}rez}, {Reynolds}, {Saez-Madain}, {Santana},
  {Silva}, {Sousa}, {Stahm}, {Torstensson}, {Venegas}, {Walther}, {Wieching},
  {Wijnands}, and {Wouterloot}]{SgrAPaperII}
{Event Horizon Telescope Collaboration}, et~al.
\newblock {First Sagittarius A* Event Horizon Telescope Results. II. EHT and
  Multiwavelength Observations, Data Processing, and Calibration}.
\newblock {\em Astrophys. J. Lett.} {\bf 2022}, {\em 930},~L13.
\newblock {{https://doi.org/10.3847/2041-8213/ac6675}}.

\bibitem[{Event Horizon Telescope Collaboration}
  \em{et~al.}(2022{\natexlab{c}}){Event Horizon Telescope Collaboration},
  {Akiyama}, {Alberdi}, {Alef}, {Algaba}, {Anantua}, {Asada}, {Azulay}, {Bach},
  {Baczko}, {Ball}, {Balokovi{\'c}}, {Barrett}, {Baub{\"o}ck}, {Benson},
  {Bintley}, {Blackburn}, {Blundell}, {Bouman}, {Bower}, {Boyce}, {Bremer},
  {Brinkerink}, {Brissenden}, {Britzen}, {Broderick}, {Broguiere}, {Bronzwaer},
  {Bustamante}, {Byun}, {Carlstrom}, {Ceccobello}, {Chael}, {Chan},
  {Chatterjee}, {Chatterjee}, {Chen}, {Chen}, {Cheng}, {Cho}, {Christian},
  {Conroy}, {Conway}, {Cordes}, {Crawford}, {Crew}, {Cruz-Osorio}, {Cui},
  {Davelaar}, {De Laurentis}, {Deane}, {Dempsey}, {Desvignes}, {Dexter},
  {Dhruv}, {Doeleman}, {Dougal}, {Dzib}, {Eatough}, {Emami}, {Falcke}, {Farah},
  {Fish}, {Fomalont}, {Ford}, {Fraga-Encinas}, {Freeman}, {Friberg}, {Fromm},
  {Fuentes}, {Galison}, {Gammie}, {Garc{\'\i}a}, {Gentaz}, {Georgiev}, {Goddi},
  {Gold}, {G{\'o}mez-Ruiz}, {G{\'o}mez}, {Gu}, {Gurwell}, {Hada}, {Haggard},
  {Haworth}, {Hecht}, {Hesper}, {Heumann}, {Ho}, {Ho}, {Honma}, {Huang},
  {Huang}, {Hughes}, {Ikeda}, {Impellizzeri}, {Inoue}, {Issaoun}, {James},
  {Jannuzi}, {Janssen}, {Jeter}, {Jiang}, {Jim{\'e}nez-Rosales}, {Johnson},
  {Jorstad}, {Joshi}, {Jung}, {Karami}, {Karuppusamy}, {Kawashima}, {Keating},
  {Kettenis}, {Kim}, {Kim}, {Kim}, {Kim}, {Kino}, {Koay}, {Kocherlakota},
  {Kofuji}, {Koch}, {Koyama}, {Kramer}, {Kramer}, {Krichbaum}, {Kuo}, {Bella},
  {Lauer}, {Lee}, {Lee}, {Leung}, {Levis}, {Li}, {Lico}, {Lindahl},
  {Lindqvist}, {Lisakov}, {Liu}, {Liu}, {Liuzzo}, {Lo}, {Lobanov}, {Loinard},
  {Lonsdale}, {Lu}, {Mao}, {Marchili}, {Markoff}, {Marrone}, {Marscher},
  {Mart{\'\i}-Vidal}, {Matsushita}, {Matthews}, {Medeiros}, {Menten},
  {Michalik}, {Mizuno}, {Mizuno}, {Moran}, {Moriyama}, {Moscibrodzka},
  {M{\"u}ller}, {Mus}, {Musoke}, {Myserlis}, {Nadolski}, {Nagai}, {Nagar},
  {Nakamura}, {Narayan}, {Narayanan}, {Natarajan}, {Nathanail}, {Fuentes},
  {Neilsen}, {Neri}, {Ni}, {Noutsos}, {Nowak}, {Oh}, {Okino}, {Olivares},
  {Ortiz-Le{\'o}n}, {Oyama}, {{\"O}zel}, {Palumbo}, {Paraschos}, {Park},
  {Parsons}, {Patel}, {Pen}, {Pesce}, {Pi{\'e}tu}, {Plambeck}, {PopStefanija},
  {Porth}, {P{\"o}tzl}, {Prather}, {Preciado-L{\'o}pez}, {Psaltis}, {Pu},
  {Ramakrishnan}, {Rao}, {Rawlings}, {Raymond}, {Rezzolla}, {Ricarte},
  {Ripperda}, {Roelofs}, {Rogers}, {Ros}, {Romero-Ca{\~n}izales},
  {Roshanineshat}, {Rottmann}, {Roy}, {Ruiz}, {Ruszczyk}, {Rygl},
  {S{\'a}nchez}, {S{\'a}nchez-Arg{\"u}elles}, {S{\'a}nchez-Portal}, {Sasada},
  {Satapathy}, {Savolainen}, {Schloerb}, {Schonfeld}, {Schuster}, {Shao},
  {Shen}, {Small}, {Sohn}, {SooHoo}, {Souccar}, {Sun}, {Tazaki}, {Tetarenko},
  {Tiede}, {Tilanus}, {Titus}, {Torne}, {Traianou}, {Trent}, {Trippe}, {Turk},
  {van Bemmel}, {van Langevelde}, {van Rossum}, {Vos}, {Wagner},
  {Ward-Thompson}, {Wardle}, {Weintroub}, {Wex}, {Wharton}, {Wielgus}, {Wiik},
  {Witzel}, {Wondrak}, {Wong}, {Wu}, {Yamaguchi}, {Yoon}, {Young}, {Young},
  {Younsi}, {Yuan}, {Yuan}, {Zensus}, {Zhang}, {Zhao}, and
  {Zhao}]{SgrAPaperIII}
{Event Horizon Telescope Collaboration}, et~al.
\newblock {First Sagittarius A* Event Horizon Telescope Results. III. Imaging
  of the Galactic Center Supermassive Black Hole}.
\newblock {\em Astrophys. J. Lett.} {\bf 2022}, {\em 930},~L14.
\newblock {{https://doi.org/10.3847/2041-8213/ac6429}}.

\bibitem[{Event Horizon Telescope Collaboration}
  \em{et~al.}(2022{\natexlab{d}}){Event Horizon Telescope Collaboration},
  {Akiyama}, {Alberdi}, {Alef}, {Algaba}, {Anantua}, {Asada}, {Azulay}, {Bach},
  {Baczko}, {Ball}, {Balokovi{\'c}}, {Barrett}, {Baub{\"o}ck}, {Benson},
  {Bintley}, {Blackburn}, {Blundell}, {Bouman}, {Bower}, {Boyce}, {Bremer},
  {Brinkerink}, {Brissenden}, {Britzen}, {Broderick}, {Broguiere}, {Bronzwaer},
  {Bustamante}, {Byun}, {Carlstrom}, {Ceccobello}, {Chael}, {Chan},
  {Chatterjee}, {Chatterjee}, {Chen}, {Chen}, {Cheng}, {Cho}, {Christian},
  {Conroy}, {Conway}, {Cordes}, {Crawford}, {Crew}, {Cruz-Osorio}, {Cui},
  {Davelaar}, {Laurentis}, {Deane}, {Dempsey}, {Desvignes}, {Dexter}, {Dhruv},
  {Doeleman}, {Dougal}, {Dzib}, {Eatough}, {Emami}, {Falcke}, {Farah}, {Fish},
  {Fomalont}, {Ford}, {Fraga-Encinas}, {Freeman}, {Friberg}, {Fromm},
  {Fuentes}, {Galison}, {Gammie}, {Garc{\'\i}a}, {Gentaz}, {Georgiev}, {Goddi},
  {Gold}, {G{\'o}mez-Ruiz}, {G{\'o}mez}, {Gu}, {Gurwell}, {Hada}, {Haggard},
  {Haworth}, {Hecht}, {Hesper}, {Heumann}, {Ho}, {Ho}, {Honma}, {Huang},
  {Huang}, {Hughes}, {Ikeda}, {Impellizzeri}, {Inoue}, {Issaoun}, {James},
  {Jannuzi}, {Janssen}, {Jeter}, {Jiang}, {Jim{\'e}nez-Rosales}, {Johnson},
  {Jorstad}, {Joshi}, {Jung}, {Karami}, {Karuppusamy}, {Kawashima}, {Keating},
  {Kettenis}, {Kim}, {Kim}, {Kim}, {Kim}, {Kino}, {Koay}, {Kocherlakota},
  {Kofuji}, {Koch}, {Koyama}, {Kramer}, {Kramer}, {Krichbaum}, {Kuo}, {Bella},
  {Lauer}, {Lee}, {Lee}, {Leung}, {Levis}, {Li}, {Lico}, {Lindahl},
  {Lindqvist}, {Lisakov}, {Liu}, {Liu}, {Liuzzo}, {Lo}, {Lobanov}, {Loinard},
  {Lonsdale}, {Lu}, {Mao}, {Marchili}, {Markoff}, {Marrone}, {Marscher},
  {Mart{\'\i}-Vidal}, {Matsushita}, {Matthews}, {Medeiros}, {Menten},
  {Michalik}, {Mizuno}, {Mizuno}, {Moran}, {Moriyama}, {Moscibrodzka},
  {M{\"u}ller}, {Mus}, {Musoke}, {Myserlis}, {Nadolski}, {Nagai}, {Nagar},
  {Nakamura}, {Narayan}, {Narayanan}, {Natarajan}, {Nathanail}, {Fuentes},
  {Neilsen}, {Neri}, {Ni}, {Noutsos}, {Nowak}, {Oh}, {Okino}, {Olivares},
  {Ortiz-Le{\'o}n}, {Oyama}, {Palumbo}, {Paraschos}, {Park}, {Parsons},
  {Patel}, {Pen}, {Pesce}, {Pi{\'e}tu}, {Plambeck}, {PopStefanija}, {Porth},
  {P{\"o}tzl}, {Prather}, {Preciado-L{\'o}pez}, {Pu}, {Ramakrishnan}, {Rao},
  {Rawlings}, {Raymond}, {Rezzolla}, {Ricarte}, {Ripperda}, {Roelofs},
  {Rogers}, {Ros}, {Romero-Ca{\~n}izales}, {Roshanineshat}, {Rottmann}, {Roy},
  {Ruiz}, {Ruszczyk}, {Rygl}, {S{\'a}nchez}, {S{\'a}nchez-Arg{\"u}elles},
  {S{\'a}nchez-Portal}, {Sasada}, {Satapathy}, {Savolainen}, {Schloerb},
  {Schonfeld}, {Schuster}, {Shao}, {Shen}, {Small}, {Sohn}, {SooHoo},
  {Souccar}, {Sun}, {Tazaki}, {Tetarenko}, {Tiede}, {Tilanus}, {Titus},
  {Torne}, {Traianou}, {Trent}, {Trippe}, {Turk}, {van Bemmel}, {van
  Langevelde}, {van Rossum}, {Vos}, {Wagner}, {Ward-Thompson}, {Wardle},
  {Weintroub}, {Wex}, {Wharton}, {Wielgus}, {Wiik}, {Witzel}, {Wondrak},
  {Wong}, {Wu}, {Yamaguchi}, {Yoon}, {Young}, {Young}, {Younsi}, {Yuan},
  {Yuan}, {Zensus}, {Zhang}, {Zhao}, {Zhao}, and {Chang}]{SgrAPaperIV}
{Event Horizon Telescope Collaboration}, et~al.
\newblock {First Sagittarius A* Event Horizon Telescope Results. IV.
  Variability, Morphology, and Black Hole Mass}.
\newblock {\em Astrophys. J. Lett.} {\bf 2022}, {\em 930},~L15.
\newblock {{https://doi.org/10.3847/2041-8213/ac6736}}.

\bibitem[{Event Horizon Telescope Collaboration}
  \em{et~al.}(2022{\natexlab{e}}){Event Horizon Telescope Collaboration},
  {Akiyama}, {Alberdi}, {Alef}, {Algaba}, {Anantua}, {Asada}, {Azulay}, {Bach},
  {Baczko}, {Ball}, {Balokovi{\'c}}, {Barrett}, {Baub{\"o}ck}, {Benson},
  {Bintley}, {Blackburn}, {Blundell}, {Bouman}, {Bower}, {Boyce}, {Bremer},
  {Brinkerink}, {Brissenden}, {Britzen}, {Broderick}, {Broguiere}, {Bronzwaer},
  {Bustamante}, {Byun}, {Carlstrom}, {Ceccobello}, {Chael}, {Chan},
  {Chatterjee}, {Chatterjee}, {Chen}, {Chen}, {Cheng}, {Cho}, {Christian},
  {Conroy}, {Conway}, {Cordes}, {Crawford}, {Crew}, {Cruz-Osorio}, {Cui},
  {Davelaar}, {De Laurentis}, {Deane}, {Dempsey}, {Desvignes}, {Dexter},
  {Dhruv}, {Doeleman}, {Dougal}, {Dzib}, {Eatough}, {Emami}, {Falcke}, {Farah},
  {Fish}, {Fomalont}, {Ford}, {Fraga-Encinas}, {Freeman}, {Friberg}, {Fromm},
  {Fuentes}, {Galison}, {Gammie}, {Garc{\'\i}a}, {Gentaz}, {Georgiev}, {Goddi},
  {Gold}, {G{\'o}mez-Ruiz}, {G{\'o}mez}, {Gu}, {Gurwell}, {Hada}, {Haggard},
  {Haworth}, {Hecht}, {Hesper}, {Heumann}, {Ho}, {Ho}, {Honma}, {Huang},
  {Huang}, {Hughes}, {Ikeda}, {Violette Impellizzeri}, {Inoue}, {Issaoun},
  {James}, {Jannuzi}, {Janssen}, {Jeter}, {Jiang}, {Jim{\'e}nez-Rosales},
  {Johnson}, {Jorstad}, {Joshi}, {Jung}, {Karami}, {Karuppusamy}, {Kawashima},
  {Keating}, {Kettenis}, {Kim}, {Kim}, {Kim}, {Kim}, {Kino}, {Koay},
  {Kocherlakota}, {Kofuji}, {Koch}, {Koyama}, {Kramer}, {Kramer}, {Krichbaum},
  {Kuo}, {Bella}, {Lauer}, {Lee}, {Lee}, {Leung}, {Levis}, {Li}, {Lico},
  {Lindahl}, {Lindqvist}, {Lisakov}, {Liu}, {Liu}, {Liuzzo}, {Lo}, {Lobanov},
  {Loinard}, {Lonsdale}, {Lu}, {Mao}, {Marchili}, {Markoff}, {Marrone},
  {Marscher}, {Mart{\'\i}-Vidal}, {Matsushita}, {Matthews}, {Medeiros},
  {Menten}, {Michalik}, {Mizuno}, {Mizuno}, {Moran}, {Moriyama},
  {Moscibrodzka}, {M{\"u}ller}, {Mus}, {Musoke}, {Myserlis}, {Nadolski},
  {Nagai}, {Nagar}, {Nakamura}, {Narayan}, {Narayanan}, {Natarajan},
  {Nathanail}, {Navarro Fuentes}, {Neilsen}, {Neri}, {Ni}, {Noutsos}, {Nowak},
  {Oh}, {Okino}, {Olivares}, {Ortiz-Le{\'o}n}, {Oyama}, {{\"O}zel}, {Palumbo},
  {Filippos Paraschos}, {Park}, {Parsons}, {Patel}, {Pen}, {Pesce},
  {Pi{\'e}tu}, {Plambeck}, {PopStefanija}, {Porth}, {P{\"o}tzl}, {Prather},
  {Preciado-L{\'o}pez}, {Psaltis}, {Pu}, {Ramakrishnan}, {Rao}, {Rawlings},
  {Raymond}, {Rezzolla}, {Ricarte}, {Ripperda}, {Roelofs}, {Rogers}, {Ros},
  {Romero-Ca{\~n}izales}, {Roshanineshat}, {Rottmann}, {Roy}, {Ruiz},
  {Ruszczyk}, {Rygl}, {S{\'a}nchez}, {S{\'a}nchez-Arg{\"u}elles},
  {S{\'a}nchez-Portal}, {Sasada}, {Satapathy}, {Savolainen}, {Schloerb},
  {Schonfeld}, {Schuster}, {Shao}, {Shen}, {Small}, {Sohn}, {SooHoo},
  {Souccar}, {Sun}, {Tazaki}, {Tetarenko}, {Tiede}, {Tilanus}, {Titus},
  {Torne}, {Traianou}, {Trent}, {Trippe}, {Turk}, {van Bemmel}, {van
  Langevelde}, {van Rossum}, {Vos}, {Wagner}, {Ward-Thompson}, {Wardle},
  {Weintroub}, {Wex}, {Wharton}, {Wielgus}, {Wiik}, {Witzel}, {Wondrak},
  {Wong}, {Wu}, {Yamaguchi}, {Yoon}, {Young}, {Young}, {Younsi}, {Yuan},
  {Yuan}, {Zensus}, {Zhang}, {Zhao}, {Zhao}, {Chan}, {Qiu}, {Ressler}, and
  {White}]{SgrAPaperV}
{Event Horizon Telescope Collaboration}, et~al.
\newblock {First Sagittarius A* Event Horizon Telescope Results. V. Testing
  Astrophysical Models of the Galactic Center Black Hole}.
\newblock {\em Astrophys. J. Lett.} {\bf 2022}, {\em 930},~L16.
\newblock {{https://doi.org/10.3847/2041-8213/ac6672}}.

\bibitem[{Event Horizon Telescope Collaboration}
  \em{et~al.}(2022{\natexlab{f}}){Event Horizon Telescope Collaboration},
  {Akiyama}, {Alberdi}, {Alef}, {Algaba}, {Anantua}, {Asada}, {Azulay}, {Bach},
  {Baczko}, {Ball}, {Balokovi{\'c}}, {Barrett}, {Baub{\"o}ck}, {Benson},
  {Bintley}, {Blackburn}, {Blundell}, {Bouman}, {Bower}, {Boyce}, {Bremer},
  {Brinkerink}, {Brissenden}, {Britzen}, {Broderick}, {Broguiere}, {Bronzwaer},
  {Bustamante}, {Byun}, {Carlstrom}, {Ceccobello}, {Chael}, {Chan},
  {Chatterjee}, {Chatterjee}, {Chen}, {Chen}, {Cheng}, {Cho}, {Christian},
  {Conroy}, {Conway}, {Cordes}, {Crawford}, {Crew}, {Cruz-Osorio}, {Cui},
  {Davelaar}, {De Laurentis}, {Deane}, {Dempsey}, {Desvignes}, {Dexter},
  {Dhruv}, {Doeleman}, {Dougal}, {Dzib}, {Eatough}, {Emami}, {Falcke}, {Farah},
  {Fish}, {Fomalont}, {Ford}, {Fraga-Encinas}, {Freeman}, {Friberg}, {Fromm},
  {Fuentes}, {Galison}, {Gammie}, {Garc{\'\i}a}, {Gentaz}, {Georgiev}, {Goddi},
  {Gold}, {G{\'o}mez-Ruiz}, {G{\'o}mez}, {Gu}, {Gurwell}, {Hada}, {Haggard},
  {Haworth}, {Hecht}, {Hesper}, {Heumann}, {Ho}, {Ho}, {Honma}, {Huang},
  {Huang}, {Hughes}, {Ikeda}, {Impellizzeri}, {Inoue}, {Issaoun}, {James},
  {Jannuzi}, {Janssen}, {Jeter}, {Jiang}, {Jim{\'e}nez-Rosales}, {Johnson},
  {Jorstad}, {Joshi}, {Jung}, {Karami}, {Karuppusamy}, {Kawashima}, {Keating},
  {Kettenis}, {Kim}, {Kim}, {Kim}, {Kim}, {Kino}, {Koay}, {Kocherlakota},
  {Kofuji}, {Koch}, {Koyama}, {Kramer}, {Kramer}, {Krichbaum}, {Kuo}, {Bella},
  {Lauer}, {Lee}, {Lee}, {Leung}, {Levis}, {Li}, {Lico}, {Lindahl},
  {Lindqvist}, {Lisakov}, {Liu}, {Liu}, {Liuzzo}, {Lo}, {Lobanov}, {Loinard},
  {Lonsdale}, {Lu}, {Mao}, {Marchili}, {Markoff}, {Marrone}, {Marscher},
  {Mart{\'\i}-Vidal}, {Matsushita}, {Matthews}, {Medeiros}, {Menten},
  {Michalik}, {Mizuno}, {Mizuno}, {Moran}, {Moriyama}, {Moscibrodzka},
  {M{\"u}ller}, {Mus}, {Musoke}, {Myserlis}, {Nadolski}, {Nagai}, {Nagar},
  {Nakamura}, {Narayan}, {Narayanan}, {Natarajan}, {Nathanail}, {Fuentes},
  {Neilsen}, {Neri}, {Ni}, {Noutsos}, {Nowak}, {Oh}, {Okino}, {Olivares},
  {Ortiz-Le{\'o}n}, {Oyama}, {{\"O}zel}, {Palumbo}, {Paraschos}, {Park},
  {Parsons}, {Patel}, {Pen}, {Pesce}, {Pi{\'e}tu}, {Plambeck}, {PopStefanija},
  {Porth}, {P{\"o}tzl}, {Prather}, {Preciado-L{\'o}pez}, {Psaltis}, {Pu},
  {Ramakrishnan}, {Rao}, {Rawlings}, {Raymond}, {Rezzolla}, {Ricarte},
  {Ripperda}, {Roelofs}, {Rogers}, {Ros}, {Romero-Ca{\~n}izales},
  {Roshanineshat}, {Rottmann}, {Roy}, {Ruiz}, {Ruszczyk}, {Rygl},
  {S{\'a}nchez}, {S{\'a}nchez-Arg{\"u}elles}, {S{\'a}nchez-Portal}, {Sasada},
  {Satapathy}, {Savolainen}, {Schloerb}, {Schonfeld}, {Schuster}, {Shao},
  {Shen}, {Small}, {Sohn}, {SooHoo}, {Souccar}, {Sun}, {Tazaki}, {Tetarenko},
  {Tiede}, {Tilanus}, {Titus}, {Torne}, {Traianou}, {Trent}, {Trippe}, {Turk},
  {van Bemmel}, {van Langevelde}, {van Rossum}, {Vos}, {Wagner},
  {Ward-Thompson}, {Wardle}, {Weintroub}, {Wex}, {Wharton}, {Wielgus}, {Wiik},
  {Witzel}, {Wondrak}, {Wong}, {Wu}, {Yamaguchi}, {Yoon}, {Young}, {Young},
  {Younsi}, {Yuan}, {Yuan}, {Zensus}, {Zhang}, {Zhao}, and {Zhao}]{SgrAPaperVI}
{Event Horizon Telescope Collaboration}, et~al.
\newblock {First Sagittarius A* Event Horizon Telescope Results. VI. Testing
  the Black Hole Metric}.
\newblock {\em Astrophys. J. Lett.} {\bf 2022}, {\em 930},~L17.
\newblock {{https://doi.org/10.3847/2041-8213/ac6756}}.

\bibitem[{Doeleman} \em{et~al.}(2019){Doeleman}, {Blackburn}, {Dexter},
  {Gomez}, {Johnson}, {Palumbo}, {Weintroub}, {Farah}, {Fish}, {Loinard},
  {Lonsdale}, {Narayanan}, {Patel}, {Pesce}, {Raymond}, {Tilanus}, {Wielgus},
  {Akiyama}, {Bower}, {Broderick}, {Deane}, {Fromm}, {Gammie}, {Gold},
  {Janssen}, {Kawashima}, {Krichbaum}, {Marrone}, {Matthews}, {Mizuno},
  {Rezzolla}, {Roelofs}, {Ros}, {Savolainen}, {Yuan}, {Zhao}, {Blackburn},
  {Doeleman}, {Dexter}, {Gomez}, {Johnson}, {Palumbo}, {Weintroub}, {Farah},
  {Fish}, {Loinard}, {Lonsdale}, {Narayanan}, {Patel}, {Pesce}, {Raymond},
  {Tilanus}, {Wielgus}, {Akiyama}, {Bower}, {Broderick}, {Deane}, {Fromm},
  {Gammie}, {Gold}, {Janssen}, {Kawashima}, {Krichbaum}, {Marrone}, {Matthews},
  {Mizuno}, {Rezzolla}, {Roelofs}, {Ros}, {Savolainen}, {Yuan}, and
  {Zhao}]{Doeleman_2019}
{Doeleman}, S.; {Blackburn}, L.; {Dexter}, J.; {Gomez}, J.L.; {Johnson}, M.D.;
  {Palumbo}, D.C.; {Weintroub}, J.; {Farah}, J.R.; {Fish}, V.; {Loinard}, L.;
  et~al.
\newblock {Studying Black Holes on Horizon Scales with VLBI Ground Arrays}.
\newblock  \emph{Bull. Am. Astron. Soc.},
  \textbf{2019}, \emph{51}, {256}.

\bibitem[Bustamante \em{et~al.}(2023)Bustamante, Blackburn, Narayanan,
  Schloerb, and Hughes]{Bustamante_2023}
Bustamante, S.; Blackburn, L.; Narayanan, G.; Schloerb, F.P.; Hughes, D.
\newblock The Role of the Large Millimeter Telescope in Black Hole Science with
  the Next-Generation Event Horizon Telescope.
\newblock {\em Galaxies} {\bf 2023}, {\em 11}, 2.
\newblock {{https://doi.org/10.3390/galaxies11010002}}.

\bibitem[{Chael} \em{et~al.}(2022){Chael}, {Issaoun}, {Pesce}, {Johnson},
  {Ricarte}, {Fromm}, and {Mizuno}]{Chael_2022}
{Chael}, A.; {Issaoun}, S.; {Pesce}, D.w.; {Johnson}, M.D.; {Ricarte}, A.;
  {Fromm}, C.M.; {Mizuno}, Y.
\newblock {Multi-frequency Black Hole Imaging for the Next-Generation Event
  Horizon Telescope}.
\newblock {\em arXiv} {\bf 2022}, arXiv:2210.12226.

\bibitem[{Issaoun} \em{et~al.}(2019){Issaoun}, {Johnson}, {Blackburn},
  {Brinkerink}, {Mo{\'s}cibrodzka}, {Chael}, {Goddi}, {Mart{\'{\i}}-Vidal},
  {Wagner}, {Doeleman}, {Falcke}, {Krichbaum}, {Akiyama}, {Bach}, {Bouman},
  {Bower}, {Broderick}, {Cho}, {Crew}, {Dexter}, {Fish}, {Gold}, {G{\'o}mez},
  {Hada}, {Hern{\'a}ndez-G{\'o}mez}, {Jan{\ss}en}, {Kino}, {Kramer}, {Loinard},
  {Lu}, {Markoff}, {Marrone}, {Matthews}, {Moran}, {M{\"u}ller}, {Roelofs},
  {Ros}, {Rottmann}, {Sanchez}, {Tilanus}, {de Vicente}, {Wielgus}, {Zensus},
  and {Zhao}]{Issaoun_2019}
{Issaoun}, S.; {Johnson}, M.D.; {Blackburn}, L.; {Brinkerink}, C.D.;
  {Mo{\'s}cibrodzka}, M.; {Chael}, A.; {Goddi}, C.; {Mart{\'{\i}}-Vidal}, I.;
  {Wagner}, J.; {Doeleman}, S.S.;  et~al.
\newblock {The Size, Shape, and Scattering of Sagittarius A* at 86 GHz: First
  VLBI with ALMA}.
\newblock {\em ApJ} {\bf 2019}, {\em 871},~30.
\newblock {{https://doi.org/10.3847/1538-4357/aaf732}}.

\bibitem[{Issaoun} \em{et~al.}(2021){Issaoun}, {Johnson}, {Blackburn},
  {Broderick}, {Tiede}, {Wielgus}, {Doeleman}, {Falcke}, {Akiyama}, {Bower},
  {Brinkerink}, {Chael}, {Cho}, {G{\'o}mez}, {Hern{\'a}ndez-G{\'o}mez},
  {Hughes}, {Kino}, {Krichbaum}, {Liuzzo}, {Loinard}, {Markoff}, {Marrone},
  {Mizuno}, {Moran}, {Pidopryhora}, {Ros}, {Rygl}, {Shen}, and
  {Wagner}]{Issaoun_2021}
{Issaoun}, S.; {Johnson}, M.D.; {Blackburn}, L.; {Broderick}, A.; {Tiede}, P.;
  {Wielgus}, M.; {Doeleman}, S.S.; {Falcke}, H.; {Akiyama}, K.; {Bower}, G.C.;
  et~al.
\newblock {Persistent Non--Gaussian Structure in the Image of Sagittarius A* at
  86 GHz}.
\newblock {\em ApJ} {\bf 2021}, {\em 915},~99.
\newblock {{https://doi.org/10.3847/1538-4357/ac00b0}}.

\bibitem[{Gwinn} \em{et~al.}(2014){Gwinn}, {Kovalev}, {Johnson}, and
  {Soglasnov}]{Gwinn_2014}
{Gwinn}, C.R.; {Kovalev}, Y.Y.; {Johnson}, M.D.; {Soglasnov}, V.A.
\newblock {Discovery of Substructure in the Scatter-broadened Image of Sgr A*}.
\newblock {\em ApJL} {\bf 2014}, {\em 794},~L14.
\newblock {{https://doi.org/10.1088/2041-8205/794/1/L14}}.

\bibitem[{Psaltis} \em{et~al.}(2018){Psaltis}, {Johnson}, {Narayan},
  {Medeiros}, {Blackburn}, and {Bower}]{Psaltis_2018}
{Psaltis}, D.; {Johnson}, M.; {Narayan}, R.; {Medeiros}, L.; {Blackburn}, L.;
  {Bower}, G.
\newblock {A Model for Anisotropic Interstellar Scattering and its Application
  to Sgr A*}.
\newblock {\em arXiv} {\bf 2018}, arXiv:1805.01242.

\bibitem[{Chael} \em{et~al.}(2019){Chael}, {Narayan}, and
  {Johnson}]{Chael_2019}
{Chael}, A.; {Narayan}, R.; {Johnson}, M.D.
\newblock {Two-temperature, Magnetically Arrested Disc simulations of the jet
  from the supermassive black hole in M87}.
\newblock {\em MNRAS} {\bf 2019}, {\em 486},~2873--2895.
\newblock {{https://doi.org/10.1093/mnras/stz988}}.

\bibitem[{Porcas} and {Rioja}(2002)]{Porcas_2002}
{Porcas}, R.W.; {Rioja}, M.J.
\newblock {VLBI phase-reference investigations at 86 GHz}.
\newblock In Proceedings of the 6th EVN Symposium,  Bonn, Germany, 25--28 June 2002;
  p.~65.

\bibitem[{Asaki} \em{et~al.}(1996){Asaki}, {Saito}, {Kawabe}, {Morita}, and
  {Sasao}]{Asaki_1996}
{Asaki}, Y.; {Saito}, M.; {Kawabe}, R.; {Morita}, K.I.; {Sasao}, T.
\newblock {Phase compensation experiments with the paired antennas method}.
\newblock {\em Radio Sci.} {\bf 1996}, {\em 31},~1615--1626.
\newblock {{https://doi.org/10.1029/96RS02629}}.

\bibitem[{Middelberg} \em{et~al.}(2005){Middelberg}, {Roy}, {Walker}, and
  {Falcke}]{Middelberg_2005}
{Middelberg}, E.; {Roy}, A.L.; {Walker}, R.C.; {Falcke}, H.
\newblock {VLBI observations of weak sources using fast frequency switching}.
\newblock {\em A\&A} {\bf 2005}, {\em 433},~897--909.
\newblock {{https://doi.org/10.1051/0004-6361:20042078}}.

\bibitem[{Dodson} and {Rioja}(2009)]{Dodson_2009}
{Dodson}, R.; {Rioja}, M.J.
\newblock {VLBA Scientific Memorandum n. 31: Astrometric calibration of mm-VLBI
  using ``Source/Frequency Phase Referenced'' observations}.
\newblock {\em arXiv} {\bf 2009}, arXiv:0910.1159.

\bibitem[{Rioja} and {Dodson}(2011)]{Rioja_2011}
{Rioja}, M.; {Dodson}, R.
\newblock {High-precision Astrometric Millimeter Very Long Baseline
  Interferometry Using a New Method for Atmospheric Calibration}.
\newblock {\em AJ} {\bf 2011}, {\em 141},~114.
\newblock {{https://doi.org/10.1088/0004-6256/141/4/114}}.

\bibitem[{Han} \em{et~al.}(2013){Han}, {Lee}, {Kang}, {Oh}, {Byun}, {Je},
  {Chung}, {Wi}, {Song}, {Kang}, {Lee}, {Kim}, {Sasao}, {Goldsmith}, and
  {Wylde}]{Han_2013}
{Han}, S.T.; {Lee}, J.W.; {Kang}, J.; {Oh}, C.S.; {Byun}, D.Y.; {Je}, D.H.;
  {Chung}, M.H.; {Wi}, S.O.; {Song}, M.; {Kang}, Y.W.;  et~al.
\newblock {Korean VLBI Network Receiver Optics for Simultaneous Multifrequency
  Observation: Evaluation}.
\newblock {\em PASP} {\bf 2013}, {\em 125},~539.
\newblock {{https://doi.org/10.1086/671125}}.

\bibitem[{Rioja} \em{et~al.}(2015){Rioja}, {Dodson}, {Jung}, and
  {Sohn}]{Rioja_2015}
{Rioja}, M.J.; {Dodson}, R.; {Jung}, T.; {Sohn}, B.W.
\newblock {The Power of Simultaneous Multifrequency Observations for mm-VLBI:
  Astrometry up to 130 GHz with the KVN}.
\newblock {\em AJ} {\bf 2015}, {\em 150},~202.
\newblock {{https://doi.org/10.1088/0004-6256/150/6/202}}.

\bibitem[{Algaba} \em{et~al.}(2015){Algaba}, {Zhao}, {Lee}, {Byun}, {Kang},
  {Kim}, {Kim}, {Kim}, {Kim}, {Kino}, {Miyazaki}, {Park}, {Trippe}, and
  {Wajima}]{Algaba2015}
{Algaba}, J.C.; {Zhao}, G.Y.; {Lee}, S.S.; {Byun}, D.Y.; {Kang}, S.C.; {Kim},
  D.W.; {Kim}, J.Y.; {Kim}, J.S.; {Kim}, S.W.; {Kino}, M.;  et~al.
\newblock {Interferometric Monitoring of GAMMA{\textendash}RAY Bright Active
  Galactic Nuclei II: Frequency Phase Transfer}.
\newblock {\em J. Korean Astron. Soc.} {\bf 2015}, {\em
  48},~237--255.
\newblock {{https://doi.org/10.5303/JKAS.2015.48.5.237}}.

\bibitem[{Zhao} \em{et~al.}(2019){Zhao}, {Jung}, {Sohn}, {Kino}, {Honma},
  {Dodson}, {Rioja}, {Han}, {Shibata}, {Byun}, {Akiyama}, {Algaba}, {An},
  {Cheng}, {Cho}, {Cui}, {Hada}, {Hodgson}, {Jiang}, {Lee}, {Lee}, {Niinuma},
  {Park}, {Ro}, {Sawada-Satoh}, {Shen}, {Tazaki}, {Trippe}, {Wajima}, and
  {Zhang}]{Zhao2019}
{Zhao}, G.Y.; {Jung}, T.; {Sohn}, B.W.; {Kino}, M.; {Honma}, M.; {Dodson}, R.;
  {Rioja}, M.; {Han}, S.T.; {Shibata}, K.; {Byun}, D.Y.;  et~al.
\newblock {Source-Frequency Phase-Referencing Observation of AGNS with KAVA
  Using Simultaneous Dual-Frequency Receiving}.
\newblock {\em J. Korean Astron. Soc.} {\bf 2019}, {\em
  52},~23--30.
\newblock {{https://doi.org/10.5303/JKAS.2019.52.1.23}}.

\bibitem[{Zhao} \em{et~al.}(2018){Zhao}, {Algaba}, {Lee}, {Jung}, {Dodson},
  {Rioja}, {Byun}, {Hodgson}, {Kang}, {Kim}, {Kim}, {Kim}, {Kim}, {Kino},
  {Miyazaki}, {Park}, {Trippe}, and {Wajima}]{Zhao_2018}
{Zhao}, G.Y.; {Algaba}, J.C.; {Lee}, S.S.; {Jung}, T.; {Dodson}, R.; {Rioja},
  M.; {Byun}, D.Y.; {Hodgson}, J.; {Kang}, S.; {Kim}, D.W.;  et~al.
\newblock {The Power of Simultaneous Multi-frequency Observations for mm-VLBI:
  Beyond Frequency Phase Transfer}.
\newblock {\em AJ} {\bf 2018}, {\em 155},~26.
\newblock {{https://doi.org/10.3847/1538-3881/aa99e0}}.

\bibitem[{Lee} \em{et~al.}(2008){Lee}, {Lobanov}, {Krichbaum}, {Witzel},
  {Zensus}, {Bremer}, {Greve}, and {Grewing}]{Lee_2008}
{Lee}, S.S.; {Lobanov}, A.P.; {Krichbaum}, T.P.; {Witzel}, A.; {Zensus}, A.;
  {Bremer}, M.; {Greve}, A.; {Grewing}, M.
\newblock {A Global 86 GHz VLBI Survey of Compact Radio Sources}.
\newblock {\em AJ} {\bf 2008}, {\em 136},~159--180.
\newblock
  {{https://doi.org/10.1088/0004-6256/136/1/15910.48550/arXiv.0803.4035}}.

\bibitem[{Hada} \em{et~al.}(2011){Hada}, {Doi}, {Kino}, {Nagai}, {Hagiwara},
  and {Kawaguchi}]{Hada_2011}
{Hada}, K.; {Doi}, A.; {Kino}, M.; {Nagai}, H.; {Hagiwara}, Y.; {Kawaguchi}, N.
\newblock {An origin of the radio jet in M87 at the location of the central
  black hole}.
\newblock {\em Nature} {\bf 2011}, {\em 477},~185--187.
\newblock {{https://doi.org/10.1038/nature10387}}.

\bibitem[{Hada} \em{et~al.}(2016){Hada}, {Kino}, {Doi}, {Nagai}, {Honma},
  {Akiyama}, {Tazaki}, {Lico}, {Giroletti}, {Giovannini}, {Orienti}, and
  {Hagiwara}]{Hada_2016}
{Hada}, K.; {Kino}, M.; {Doi}, A.; {Nagai}, H.; {Honma}, M.; {Akiyama}, K.;
  {Tazaki}, F.; {Lico}, R.; {Giroletti}, M.; {Giovannini}, G.;  et~al.
\newblock {High-sensitivity 86 GHz (3.5 mm) VLBI Observations of M87: Deep
  Imaging of the Jet Base at a Resolution of 10 Schwarzschild Radii}.
\newblock {\em ApJ} {\bf 2016}, {\em 817},~131.
\newblock
  {{https://doi.org/10.3847/0004-637X/817/2/13110.48550/arXiv.1512.03783}}.

\bibitem[{Kim} \em{et~al.}(2018){Kim}, {Krichbaum}, {Lu}, {Ros}, {Bach},
  {Bremer}, {de Vicente}, {Lindqvist}, and {Zensus}]{Kim_2018}
{Kim}, J.Y.; {Krichbaum}, T.P.; {Lu}, R.S.; {Ros}, E.; {Bach}, U.; {Bremer},
  M.; {de Vicente}, P.; {Lindqvist}, M.; {Zensus}, J.A.
\newblock {The limb-brightened jet of M87 down to the 7 Schwarzschild radii
  scale}.
\newblock {\em A\&A} {\bf 2018}, {\em 616},~A188.
\newblock
  {{https://doi.org/10.1051/0004-6361/20183292110.48550/arXiv.1805.02478}}.

\bibitem[{Rioja} and {Dodson}(2020)]{Rioja_2020}
{Rioja}, M.J.; {Dodson}, R.
\newblock {Precise radio astrometry and new developments for the
  next-generation of instruments}.
\newblock {\em A\&Ar} {\bf 2020}, {\em 28},~6.
\newblock {{https://doi.org/10.1007/s00159-020-00126-z}}.

\bibitem[Rioja \em{et~al.}(2023)Rioja, Dodson, and Asaki]{Rioja_2023}
Rioja, M.J.; Dodson, R.; Asaki, Y.
\newblock The Transformational Power of Frequency Phase Transfer Methods for
  ngEHT.
\newblock {\em Galaxies} {\bf 2023}, {\em 11}, 16.
\newblock {{https://doi.org/10.3390/galaxies11010016}}.

\bibitem[Jiang \em{et~al.}(2023)Jiang, Zhao, Shen, Rioja, Dodson, Cho, Zhao,
  Eubanks, and Lu]{Jiang_2023}
Jiang, W.; Zhao, G.Y.; Shen, Z.Q.; Rioja, M.J.; Dodson, R.; Cho, I.; Zhao,
  S.S.; Eubanks, M.; Lu, R.S.
\newblock Applications of the Source-Frequency Phase-Referencing Technique for
  ngEHT Observations.
\newblock {\em Galaxies} {\bf 2023}, {\em 11}, 3.
\newblock {{https://doi.org/10.3390/galaxies11010003}}.

\bibitem[{Marcaide} and {Shapiro}(1984)]{Marcaide_1984}
{Marcaide}, J.M.; {Shapiro}, I.I.
\newblock {VLBI study of 1038+528A and B: Discovery of wavelength dependence
  of peak brightness location.}
\newblock {\em ApJ} {\bf 1984}, {\em 276},~56--59.
\newblock {{https://doi.org/10.1086/161592}}.

\bibitem[{Boccardi} \em{et~al.}(2017){Boccardi}, {Krichbaum}, {Ros}, and
  {Zensus}]{Boccardi_2017}
{Boccardi}, B.; {Krichbaum}, T.P.; {Ros}, E.; {Zensus}, J.A.
\newblock {Radio observations of active galactic nuclei with mm-VLBI}.
\newblock {\em A\&Ar} {\bf 2017}, {\em 25},~4.
\newblock {{https://doi.org/10.1007/s00159-017-0105-6}}.

\bibitem[{Kim} \em{et~al.}(2019){Kim}, {Krichbaum}, {Marscher}, {Jorstad},
  {Agudo}, {Thum}, {Hodgson}, {MacDonald}, {Ros}, {Lu}, {Bremer}, {de Vicente},
  {Lindqvist}, {Trippe}, and {Zensus}]{Kim2019}
{Kim}, J.Y.; {Krichbaum}, T.P.; {Marscher}, A.P.; {Jorstad}, S.G.; {Agudo}, I.;
  {Thum}, C.; {Hodgson}, J.A.; {MacDonald}, N.R.; {Ros}, E.; {Lu}, R.S.;
  et~al.
\newblock {Spatially resolved origin of millimeter-wave linear polarization in
  the nuclear region of 3C 84}.
\newblock {\em A\&A} {\bf 2019}, {\em 622},~A196.
\newblock {{https://doi.org/10.1051/0004-6361/201832920}}.

\bibitem[{Paraschos} \em{et~al.}(2021){Paraschos}, {Kim}, {Krichbaum}, and
  {Zensus}]{Paraschos2021}
{Paraschos}, G.F.; {Kim}, J.Y.; {Krichbaum}, T.P.; {Zensus}, J.A.
\newblock {Pinpointing the jet apex of 3C 84}.
\newblock {\em A\&A} {\bf 2021}, {\em 650},~L18.
\newblock {{https://doi.org/10.1051/0004-6361/202140776}}.

\bibitem[{Oh} \em{et~al.}(2022){Oh}, {Hodgson}, {Trippe}, {Krichbaum}, {Kam},
  {Paraschos}, {Kim}, {Rani}, {Sohn}, {Lee}, {Lico}, {Liuzzo}, {Bremer}, and
  {Zensus}]{Oh2022}
{Oh}, J.; {Hodgson}, J.A.; {Trippe}, S.; {Krichbaum}, T.P.; {Kam}, M.;
  {Paraschos}, G.F.; {Kim}, J.Y.; {Rani}, B.; {Sohn}, B.W.; {Lee}, S.S.;
  et~al.
\newblock {A persistent double nuclear structure in 3C 84}.
\newblock {\em MNRAS} {\bf 2022}, {\em 509},~1024--1035.
\newblock {{https://doi.org/10.1093/mnras/stab3056}}.

\bibitem[{G{\'o}mez} \em{et~al.}(2021){G{\'o}mez}, {Traianou}, {Krichbaum},
  {Lobanov}, {Fuentes}, {Lico}, {Zhao}, {Bruni}, {Kovalev}, {Lahteenmaki},
  {Voitsik}, {Lisakov}, {Angelakis}, {Bach}, {Casadio}, {Cho}, {Dey},
  {Gopakumar}, {Gurvits}, {Jorstad}, {Kovalev}, {Lister}, {Marscher},
  {Myserlis}, {Pushkarev}, {Ros}, {Savolainen}, {Tornikoski}, {Valtonen}, and
  {Zensus}]{Gomez2021}
{G{\'o}mez}, J.L.; {Traianou}, E.; {Krichbaum}, T.P.; {Lobanov}, A.; {Fuentes},
  A.; {Lico}, R.; {Zhao}, G.Y.; {Bruni}, G.; {Kovalev}, Y.Y.; {Lahteenmaki},
  A.;  et~al.
\newblock {Probing the innermost regions of AGN jets and their magnetic fields
  with RadioAstron. V. Space and ground millimeter-VLBI imaging of OJ 287}.
\newblock {\em arXiv} {\bf 2021}, arXiv:2111.11200.

\bibitem[{Boccardi} \em{et~al.}(2021){Boccardi}, {Perucho}, {Casadio},
  {Grandi}, {Macconi}, {Torresi}, {Pellegrini}, {Krichbaum}, {Kadler},
  {Giovannini}, {Karamanavis}, {Ricci}, {Madika}, {Bach}, {Ros}, {Giroletti},
  and {Zensus}]{Boccardi2021}
{Boccardi}, B.; {Perucho}, M.; {Casadio}, C.; {Grandi}, P.; {Macconi}, D.;
  {Torresi}, E.; {Pellegrini}, S.; {Krichbaum}, T.P.; {Kadler}, M.;
  {Giovannini}, G.;  et~al.
\newblock {Jet collimation in NGC 315 and other nearby AGN}.
\newblock {\em A\&A} {\bf 2021}, {\em 647},~A67.
\newblock {{https://doi.org/10.1051/0004-6361/202039612}}.

\bibitem[{Casadio} \em{et~al.}(2021){Casadio}, {MacDonald}, {Boccardi},
  {Jorstad}, {Marscher}, {Krichbaum}, {Hodgson}, {Kim}, {Traianou}, {Weaver},
  {G{\'o}mez Garrido}, {Gonz{\'a}lez Garc{\'\i}a}, {Kallunki}, {Lindqvist},
  {S{\'a}nchez}, {Yang}, and {Zensus}]{Casadio2021}
{Casadio}, C.; {MacDonald}, N.R.; {Boccardi}, B.; {Jorstad}, S.G.; {Marscher},
  A.P.; {Krichbaum}, T.P.; {Hodgson}, J.A.; {Kim}, J.Y.; {Traianou}, E.;
  {Weaver}, Z.R.;  et~al.
\newblock {The jet collimation profile at high resolution in BL Lacertae}.
\newblock {\em A\&A} {\bf 2021}, {\em 649},~A153.
\newblock {{https://doi.org/10.1051/0004-6361/202039616}}.

\bibitem[{Nair} \em{et~al.}(2019){Nair}, {Lobanov}, {Krichbaum}, {Ros},
  {Zensus}, {Kovalev}, {Lee}, {Mertens}, {Hagiwara}, {Bremer}, {Lindqvist}, and
  {de Vicente}]{Nair2019}
{Nair}, D.G.; {Lobanov}, A.P.; {Krichbaum}, T.P.; {Ros}, E.; {Zensus}, J.A.;
  {Kovalev}, Y.Y.; {Lee}, S.S.; {Mertens}, F.; {Hagiwara}, Y.; {Bremer}, M.;
  et~al.
\newblock {Global millimeter VLBI array survey of ultracompact extragalactic
  radio sources at 86 GHz}.
\newblock {\em A\&A} {\bf 2019}, {\em 622},~A92.
\newblock {{https://doi.org/10.1051/0004-6361/201833122}}.

\bibitem[{Rani} \em{et~al.}(2014){Rani}, {Krichbaum}, {Marscher}, {Jorstad},
  {Hodgson}, {Fuhrmann}, and {Zensus}]{Rani2014}
{Rani}, B.; {Krichbaum}, T.P.; {Marscher}, A.P.; {Jorstad}, S.G.; {Hodgson},
  J.A.; {Fuhrmann}, L.; {Zensus}, J.A.
\newblock {Jet outflow and gamma-ray emission correlations in S5 0716+714}.
\newblock {\em A\&A} {\bf 2014}, {\em 571},~L2.
\newblock {{https://doi.org/10.1051/0004-6361/201424796}}.

\bibitem[{Rani} \em{et~al.}(2015){Rani}, {Krichbaum}, {Marscher}, {Hodgson},
  {Fuhrmann}, {Angelakis}, {Britzen}, and {Zensus}]{Rani2015}
{Rani}, B.; {Krichbaum}, T.P.; {Marscher}, A.P.; {Hodgson}, J.A.; {Fuhrmann},
  L.; {Angelakis}, E.; {Britzen}, S.; {Zensus}, J.A.
\newblock {Connection between inner jet kinematics and broadband flux
  variability in the BL Lacertae object S5 0716+714}.
\newblock {\em A\&A} {\bf 2015}, {\em 578},~A123.
\newblock {{https://doi.org/10.1051/0004-6361/201525608}}.

\bibitem[{Casadio} \em{et~al.}(2019){Casadio}, {Marscher}, {Jorstad}, {Blinov},
  {MacDonald}, {Krichbaum}, {Boccardi}, {Traianou}, {G{\'o}mez}, {Agudo},
  {Sohn}, {Bremer}, {Hodgson}, {Kallunki}, {Kim}, {Williamson}, and
  {Zensus}]{Casadio2019}
{Casadio}, C.; {Marscher}, A.P.; {Jorstad}, S.G.; {Blinov}, D.A.; {MacDonald},
  N.R.; {Krichbaum}, T.P.; {Boccardi}, B.; {Traianou}, E.; {G{\'o}mez}, J.L.;
  {Agudo}, I.;  et~al.
\newblock {The magnetic field structure in CTA 102 from high-resolution mm-VLBI
  observations during the flaring state in 2016--2017}.
\newblock {\em A\&A} {\bf 2019}, {\em 622},~A158.
\newblock {{https://doi.org/10.1051/0004-6361/201834519}}.

\bibitem[{Schulz} \em{et~al.}(2020){Schulz}, {Kadler}, {Ros}, {Perucho},
  {Krichbaum}, {Agudo}, {Beuchert}, {Lindqvist}, {Mannheim}, {Wilms}, and
  {Zensus}]{Schultz2020}
{Schulz}, R.; {Kadler}, M.; {Ros}, E.; {Perucho}, M.; {Krichbaum}, T.P.;
  {Agudo}, I.; {Beuchert}, T.; {Lindqvist}, M.; {Mannheim}, K.; {Wilms}, J.;
  et~al.
\newblock {Sub-milliarcsecond imaging of a bright flare and ejection event in
  the extragalactic jet 3C 111}.
\newblock {\em A\&A} {\bf 2020}, {\em 644},~A85.
\newblock {{https://doi.org/10.1051/0004-6361/202037737}}.

\bibitem[{Traianou} \em{et~al.}(2020){Traianou}, {Krichbaum}, {Boccardi},
  {Angioni}, {Rani}, {Liu}, {Ros}, {Bach}, {Sokolovsky}, {Lisakov},
  {Kiehlmann}, {Gurwell}, and {Zensus}]{Traianou2020}
{Traianou}, E.; {Krichbaum}, T.P.; {Boccardi}, B.; {Angioni}, R.; {Rani}, B.;
  {Liu}, J.; {Ros}, E.; {Bach}, U.; {Sokolovsky}, K.V.; {Lisakov}, M.M.;
  et~al.
\newblock {Localizing the {\ensuremath{\gamma}}-ray emitting region in the
  blazar TXS 2013+370}.
\newblock {\em A\&A} {\bf 2020}, {\em 634},~A112.
\newblock {{https://doi.org/10.1051/0004-6361/201935756}}.

\bibitem[{Tetarenko} \em{et~al.}(2017){Tetarenko}, {Sivakoff}, {Miller-Jones},
  {Rosolowsky}, {Petitpas}, {Gurwell}, {Wouterloot}, {Fender}, {Heinz},
  {Maitra}, {Markoff}, {Migliari}, {Rupen}, {Rushton}, {Russell}, {Russell},
  and {Sarazin}]{Tetarenko2017}
{Tetarenko}, A.J.; {Sivakoff}, G.R.; {Miller-Jones}, J.C.A.; {Rosolowsky},
  E.W.; {Petitpas}, G.; {Gurwell}, M.; {Wouterloot}, J.; {Fender}, R.; {Heinz},
  S.; {Maitra}, D.;  et~al.
\newblock {Extreme jet ejections from the black hole X-ray binary V404 Cygni}.
\newblock {\em MNRAS} {\bf 2017}, {\em 469},~3141--3162.
\newblock {{https://doi.org/10.1093/mnras/stx1048}}.

\bibitem[{Dodson} \em{et~al.}(2017){Dodson}, {Rioja}, {Jung}, {Gom{\'e}z},
  {Bujarrabal}, {Moscadelli}, {Miller-Jones}, {Tetarenko}, and
  {Sivakoff}]{Dodson_2017}
{Dodson}, R.; {Rioja}, M.J.; {Jung}, T.; {Gom{\'e}z}, J.L.; {Bujarrabal}, V.;
  {Moscadelli}, L.; {Miller-Jones}, J.C.A.; {Tetarenko}, A.J.; {Sivakoff}, G.R.
\newblock {The science case for simultaneous mm-wavelength receivers in radio
  astronomy}.
\newblock {\em New Astron. Rev.} {\bf 2017}, {\em 79},~85--102.
\newblock {{https://doi.org/10.1016/j.newar.2017.09.003}}.

\bibitem[{Matthews} \em{et~al.}(2010){Matthews}, {Greenhill}, {Goddi},
  {Chandler}, {Humphreys}, and {Kunz}]{Matthews_2010}
{Matthews}, L.D.; {Greenhill}, L.J.; {Goddi}, C.; {Chandler}, C.J.;
  {Humphreys}, E.M.L.; {Kunz}, M.W.
\newblock {A Feature Movie of SiO Emission 20-100 AU from the Massive Young
  Stellar Object Orion Source I}.
\newblock {\em ApJ} {\bf 2010}, {\em 708},~80--92.
\newblock {{https://doi.org/10.1088/0004-637X/708/1/80}}.

\bibitem[{Issaoun} \em{et~al.}(2017){Issaoun}, {Goddi}, {Matthews},
  {Greenhill}, {Gray}, {Humphreys}, {Chandler}, {Krumholz}, and
  {Falcke}]{Issaoun_2017}
{Issaoun}, S.; {Goddi}, C.; {Matthews}, L.D.; {Greenhill}, L.J.; {Gray}, M.D.;
  {Humphreys}, E.M.L.; {Chandler}, C.J.; {Krumholz}, M.; {Falcke}, H.
\newblock {VLBA imaging of the 3 mm SiO maser emission in the disk-wind from
  the massive protostellar system Orion Source I}.
\newblock {\em A\&A} {\bf 2017}, {\em 606},~A126.
\newblock {{https://doi.org/10.1051/0004-6361/201731548}}.

\bibitem[{Selina} \em{et~al.}(2018){Selina}, {Murphy}, {McKinnon}, {Beasley},
  {Butler}, {Carilli}, {Clark}, {Durand}, {Erickson}, {Grammer}, {Hiriart},
  {Jackson}, {Kent}, {Mason}, {Morgan}, {Ojeda}, {Rosero}, {Shillue},
  {Sturgis}, and {Urbain}]{Selina2018}
{Selina}, R.J.; {Murphy}, E.J.; {McKinnon}, M.; {Beasley}, A.; {Butler}, B.;
  {Carilli}, C.; {Clark}, B.; {Durand}, S.; {Erickson}, A.; {Grammer}, W.;
  et~al.
\newblock {The ngVLA Reference Design}.
\newblock In {\em Proceedings of the Science with a Next Generation Very Large
  Array}; {Murphy}, E., Ed.; Astronomical Society of the
  Pacific Conference Series; NASA/ADS: Cambridge, MA, USA, 2018; Volume 517, p.~15.

\bibitem[{McKinnon} \em{et~al.}(2019){McKinnon}, {Beasley}, {Murphy}, {Selina},
  {Farnsworth}, and {Walter}]{mckinnon2019}
{McKinnon}, M.; {Beasley}, A.; {Murphy}, E.; {Selina}, R.; {Farnsworth}, R.;
  {Walter}, A.
\newblock {ngVLA: The Next Generation Very Large Array}.
\newblock  \emph{Bull. Am. Astron. Soc.}
  \textbf{2019}, \emph{51}, 81.

\bibitem[{Roelofs} \em{et~al.}(2022){Roelofs}, {Blackburn}, {Lindahl},
  {Doeleman}, {Johnson}, {Arras}, {Chatterjee}, {Emami}, {Fromm}, {Fuentes},
  {Knollmueller}, {Kosogorov}, {Mueller}, {Patel}, {Raymond}, {Tiede},
  {Traianou}, and {Vega}]{Roelofs_2023}
{Roelofs}, F.; {Blackburn}, L.; {Lindahl}, G.; {Doeleman}, S.S.; {Johnson},
  M.D.; {Arras}, P.; {Chatterjee}, K.; {Emami}, R.; {Fromm}, C.; {Fuentes}, A.;
   et~al.
\newblock {The ngEHT Analysis Challenges}.
\newblock {\em arXiv} {\bf 2022}, arXiv:2212.11355.

\bibitem[{Raymond} \em{et~al.}(2021){Raymond}, {Palumbo}, {Paine}, {Blackburn},
  {C{\'o}rdova Rosado}, {Doeleman}, {Farah}, {Johnson}, {Roelofs}, {Tilanus},
  and {Weintroub}]{Raymond_2021}
{Raymond}, A.W.; {Palumbo}, D.; {Paine}, S.N.; {Blackburn}, L.; {C{\'o}rdova
  Rosado}, R.; {Doeleman}, S.S.; {Farah}, J.R.; {Johnson}, M.D.; {Roelofs}, F.;
  {Tilanus}, R.P.J.;  et~al.
\newblock {Evaluation of New Submillimeter VLBI Sites for the Event Horizon
  Telescope}.
\newblock {\em ApJS} {\bf 2021}, {\em 253},~5.
\newblock {{https://doi.org/10.3847/1538-3881/abc3c3}}.

\bibitem[{Chael} \em{et~al.}(2016){Chael}, {Johnson}, {Narayan}, {Doeleman},
  {Wardle}, and {Bouman}]{Chael_2016}
{Chael}, A.A.; {Johnson}, M.D.; {Narayan}, R.; {Doeleman}, S.S.; {Wardle},
  J.F.C.; {Bouman}, K.L.
\newblock {High-resolution Linear Polarimetric Imaging for the Event Horizon
  Telescope}.
\newblock {\em ApJ} {\bf 2016}, {\em 829},~11.
\newblock {{https://doi.org/10.3847/0004-637X/829/1/11}}.

\bibitem[{Chael} \em{et~al.}(2018){Chael}, {Johnson}, {Bouman}, {Blackburn},
  {Akiyama}, and {Narayan}]{Chael_2018}
{Chael}, A.A.; {Johnson}, M.D.; {Bouman}, K.L.; {Blackburn}, L.L.; {Akiyama},
  K.; {Narayan}, R.
\newblock {Interferometric Imaging Directly with Closure Phases and Closure
  Amplitudes}.
\newblock {\em ApJ} {\bf 2018}, {\em 857},~23.
\newblock {{https://doi.org/10.3847/1538-4357/aab6a8}}.

\bibitem[{Kovac} and {Barkats}(2007)]{Kovac_2007}
{Kovac}, J.M.; {Barkats}, D.
\newblock {CMB from the South Pole: Past, Present, and Future}.
\newblock {\em arXiv} {\bf 2007}, arXiv:0707.1075.

\bibitem[{Palumbo} \em{et~al.}(2019){Palumbo}, {Doeleman}, {Johnson}, {Bouman},
  and {Chael}]{Palumbo_2019}
{Palumbo}, D.C.M.; {Doeleman}, S.S.; {Johnson}, M.D.; {Bouman}, K.L.; {Chael},
  A.A.
\newblock {Metrics and Motivations for Earth-Space VLBI: Time-resolving Sgr A*
  with the Event Horizon Telescope}.
\newblock {\em ApJ} {\bf 2019}, {\em 881},~62.
\newblock {{https://doi.org/10.3847/1538-4357/ab2bed}}.

\bibitem[{Dodson} \em{et~al.}(2014){Dodson}, {Rioja}, {Jung}, {Sohn}, {Byun},
  {Cho}, {Lee}, {Kim}, {Kim}, {Oh}, {Han}, {Je}, {Chung}, {Wi}, {Kang}, {Lee},
  {Chung}, {Kim}, {Kim}, {Lee}, {Roh}, {Oh}, {Yeom}, {Song}, and
  {Kang}]{Dodson_2014}
{Dodson}, R.; {Rioja}, M.J.; {Jung}, T.H.; {Sohn}, B.W.; {Byun}, D.Y.; {Cho},
  S.H.; {Lee}, S.S.; {Kim}, J.; {Kim}, K.T.; {Oh}, C.S.;  et~al.
\newblock {Astrometrically Registered Simultaneous Observations of the 22 GHz
  H$_{2}$O and 43 GHz SiO Masers toward R Leonis Minoris Using KVN and
  Source/Frequency Phase Referencing}.
\newblock {\em AJ} {\bf 2014}, {\em 148},~97.
\newblock {{https://doi.org/10.1088/0004-6256/148/5/97}}.

\bibitem[{Rioja} \em{et~al.}(2014){Rioja}, {Dodson}, {Jung}, {Sohn}, {Byun},
  {Agudo}, {Cho}, {Lee}, {Kim}, {Kim}, {Oh}, {Han}, {Je}, {Chung}, {Wi},
  {Kang}, {Lee}, {Chung}, {Ryoung Kim}, {Kim}, {Lee}, {Roh}, {Oh}, {Yeom},
  {Song}, and {Kang}]{Rioja_2014}
{Rioja}, M.J.; {Dodson}, R.; {Jung}, T.; {Sohn}, B.W.; {Byun}, D.Y.; {Agudo},
  I.; {Cho}, S.H.; {Lee}, S.S.; {Kim}, J.; {Kim}, K.T.;  et~al.
\newblock {Verification of the Astrometric Performance of the Korean VLBI
  Network, Using Comparative SFPR Studies with the VLBA at 14/7 mm}.
\newblock {\em AJ} {\bf 2014}, {\em 148},~84.
\newblock {{https://doi.org/10.1088/0004-6256/148/5/84}}.

\bibitem[{Chael} \em{et~al.}(2021){Chael}, {Johnson}, and
  {Lupsasca}]{Chael2021}
{Chael}, A.; {Johnson}, M.D.; {Lupsasca}, A.
\newblock {Observing the Inner Shadow of a Black Hole: A Direct View of the
  Event Horizon}.
\newblock {\em ApJ} {\bf 2021}, {\em 918},~6.
\newblock {{https://doi.org/10.3847/1538-4357/ac09ee}}.

\end{thebibliography}
\end{document}